\documentclass{article}
\usepackage[margin=1in]{geometry}
\usepackage{amsthm,amsmath,amssymb}
\usepackage{graphicx}
\usepackage[hidelinks,breaklinks]{hyperref}
\newtheorem{theorem}{Theorem}[]

\newtheorem{remark1}[theorem]{Remark}
\newenvironment{remark}{\begin{remark1} \rm}{\end{remark1}}

\DeclareMathOperator{\E}{\mathop{}\mathbb{E}}

\newlength{\vertsep}
\newlength{\imsize}
\title{A graphical method of cumulative differences\\between two subpopulations}
\author{Mark Tygert\\{\normalsize Facebook Artificial Intelligence Research}\\
{\normalsize 1 Facebook Way, Menlo Park, CA 94025}\\
{\normalsize Main e-mail address:\ \ {\tt mark\symbol{64}tygert.com}}}

\begin{document}

\maketitle

\begin{abstract}
Comparing the differences in outcomes (that is, in ``dependent variables'')
between two subpopulations
is often most informative when comparing outcomes only for individuals
from the subpopulations who are similar according to ``independent variables.''
The independent variables are generally known as ``scores,''
as in propensity scores for matching or as in the probabilities predicted
by statistical or machine-learned models, for example.
If the outcomes are discrete, then some averaging is necessary
to reduce the noise arising from the outcomes varying randomly
over those discrete values in the observed data.
The traditional method of averaging is to bin the data according to the scores
and plot the average outcome in each bin against the average score in the bin.
However, such binning can be rather arbitrary and yet greatly impacts
the interpretation of displayed deviation between the subpopulations
and assessment of its statistical significance.
Fortunately, such binning is entirely unnecessary in plots
of cumulative differences and in the associated scalar summary metrics that are
analogous to the workhorse statistics of comparing probability distributions
--- those due to Kolmogorov and Smirnov and their refinements due to Kuiper.
The present paper develops such cumulative methods
for the common case in which no score of any member
of the subpopulations being compared is exactly equal to the score
of any other member of either subpopulation.

\bigskip


\end{abstract}

\section{Introduction}
\label{intro}

A fundamental problem in statistics is to compare outcomes
attained by two different subpopulations whose members are matched
via numerical values known as ``scores.''
In this context, the scores are the independent variables,
and the outcomes are the dependent variables.
Propensity scores are a popular method for matching,
as are the likelihoods assigned by statistical or machine-learned models.
Synonyms for ``outcome'' include ``response'' and ``result,''
and the present paper will use all these synonyms interchangeably.
The responses are {\it random} variables,
whereas the scores are viewed as given, non-random.
In many practical settings, no score from among either subpopulation's members
is exactly equal to any score from among the two subpopulations' other members,
complicating the comparison and very concept of ``matching'';
the present paper addresses precisely these practical settings.
Some simpler settings are addressed already by~\cite{tygert} and others.

Prominent practical applications include the analysis of equity
for subpopulations (often the subpopulations considered are sensitive groups,
perhaps based on protected classes such as race, color, religion, gender,
national origin, age, disability, veteran status, or genetic information),
as by~\cite{corbett-davies-pierson-feller-goel-huq} and others,
as well as the comparison of control to treated subpopulations
in medical trials, as by~\cite{xu-kalbfleisch} and the references
in their introduction. Observational studies are another popular application,
especially when investigating differences between healthy, diseased, infected,
or treated subpopulations in biomedicine,
as reviewed by~\cite{luo-gardiner-bradley}.

Statistical questions arise when the responses are discrete,
taking values at random according to probability distributions
whose parameter values can only be estimated from the observed data.
Perhaps the most common scenario is when each response is either a success
or a failure, typically encoded as taking the values 1 or 0, respectively.
If the underlying probability of success is 0.5, for example,
then the actual observation will be 1 half the time and 0 half the time.
Thus, some averaging is necessary to obtain reliable estimates
when the responses are discrete.

The traditional ``reliability diagram'' plots binned responses
against binned scores. Namely, the diagram partitions the real line
into disjoint intervals known as ``bins'' and takes the (arithmetic) average
of the scores in each bin paired with the average
of the responses corresponding to the scores in that bin.
The reliability diagram then graphs the average responses
against the average scores.
Typically, each subpopulation under consideration gets its own graph,
superimposed on the same diagram.
Copious examples are available in the figures below,
as detailed in Section~\ref{results} below.
Another name for ``reliability diagram''
(popularized by~\cite{corbett-davies-pierson-feller-goel-huq})
is ``calibration plot,'' especially when the responses are Bernoulli variates.
A comprehensive, textbook review of reliability diagrams
for plotting calibration is available in Chapter~8 of~\cite{wilks}.

There are two canonical choices for the bins that partition the real line
in the reliability diagram: \{1\} make the width of every bin be the same
or \{2\} set the widths of the bins such that each bin contains
roughly the same number of scores from the observed data set.
Naturally, the second choice can adapt to each subpopulation
under consideration. In both cases, increasing the number of bins
trades off statistical confidence in the estimates
for enhanced resolution in detecting deviations as a function of score;
after all, narrower bins perform less averaging, averaging away less
of the randomness in the observations.
The trade-off between resolution and statistical confidence
is inherent in methods based on binning or kernel density estimation
such as that of~\cite{srihera-stute}.
The methods proposed in the present paper avoid making
such an explicit trade-off and also avoid the rather arbitrary decisions
about which bins or kernels to use.
The present paper extensively compares its methods against
both standard choices of bins for the classical methods.

The present paper follows the cumulative approach introduced into statistics
by~\cite{kolmogorov} and~\cite{smirnov}.
The methodology of Kolmogorov and Smirnov,
as well as the refinement (``Kuiper's statistic'') introduced by~\cite{kuiper},
yields scalar summary statistics useful for screening large numbers
of data sets and subpopulations. After identification via the scalar statistics
of potentially statistically significant deviations in a data set
for two subpopulations, graphical methods allow for in-depth investigation
into the variation of the deviations as a function of score.
The graphical methods (and hence an intuitive interpretation
of the associated scalar summary statistics) rely on the weighting
used by~\cite{delgado}, \cite{diebolt}, and~\cite{stute},
which is different from the weighting used by the otherwise
closely related approach of~\cite{scheike} and the others
cited by~\cite{gonzalez-manteiga-crujeiras}.
The scalar summary statistics of~\cite{delgado} are almost the same
as those in the present paper, but for the simpler setting in which each score
comes with precisely one observation from one subpopulation
and one observation from the other subpopulation.
The scalar summary statistics of~\cite{diebolt} and~\cite{stute} are analogues
of those from the appendix of~\cite{tygert}
in the special case that the parametric regression function they consider
is nothing but the identity function on the unit interval $[0, 1]$.

\begin{figure}
\begin{centering}

(a)
\parbox{\imsize}{\includegraphics[width=\imsize]
{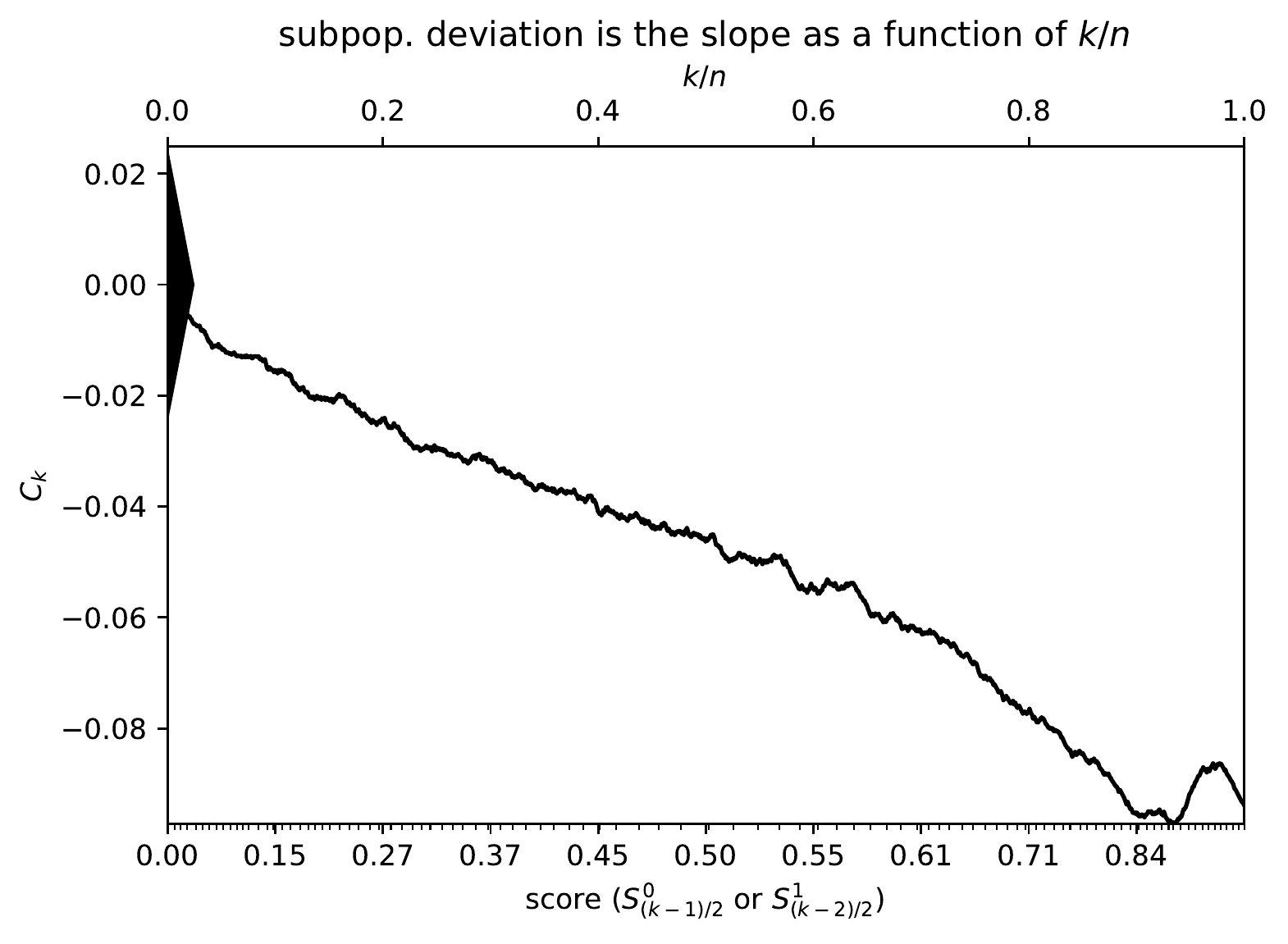}}
\quad\quad
(b)
\parbox{\imsize}{\includegraphics[width=\imsize]
{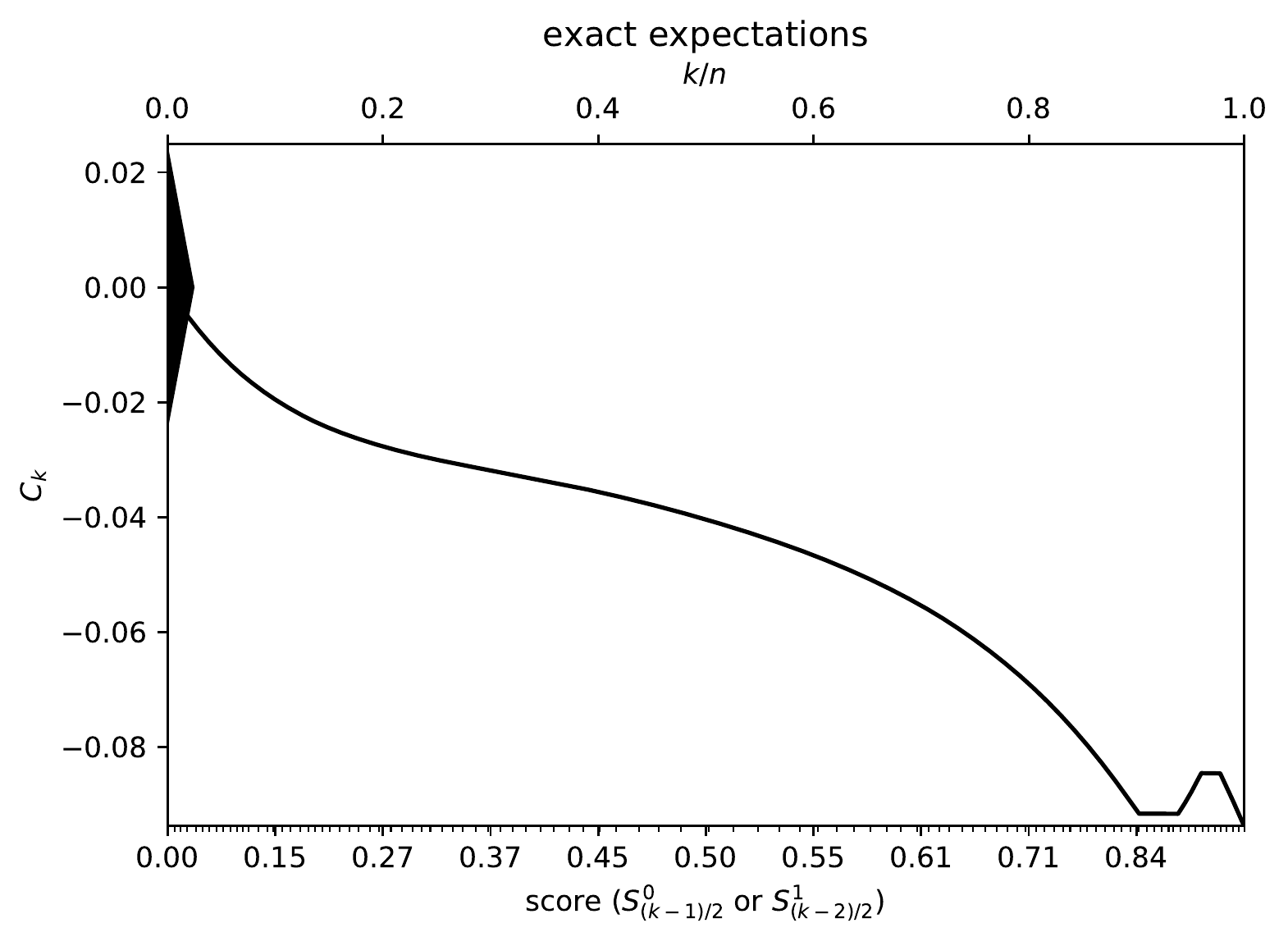}}

\vspace{\vertsep}

(c)
\parbox{\imsize}{\includegraphics[width=\imsize]
{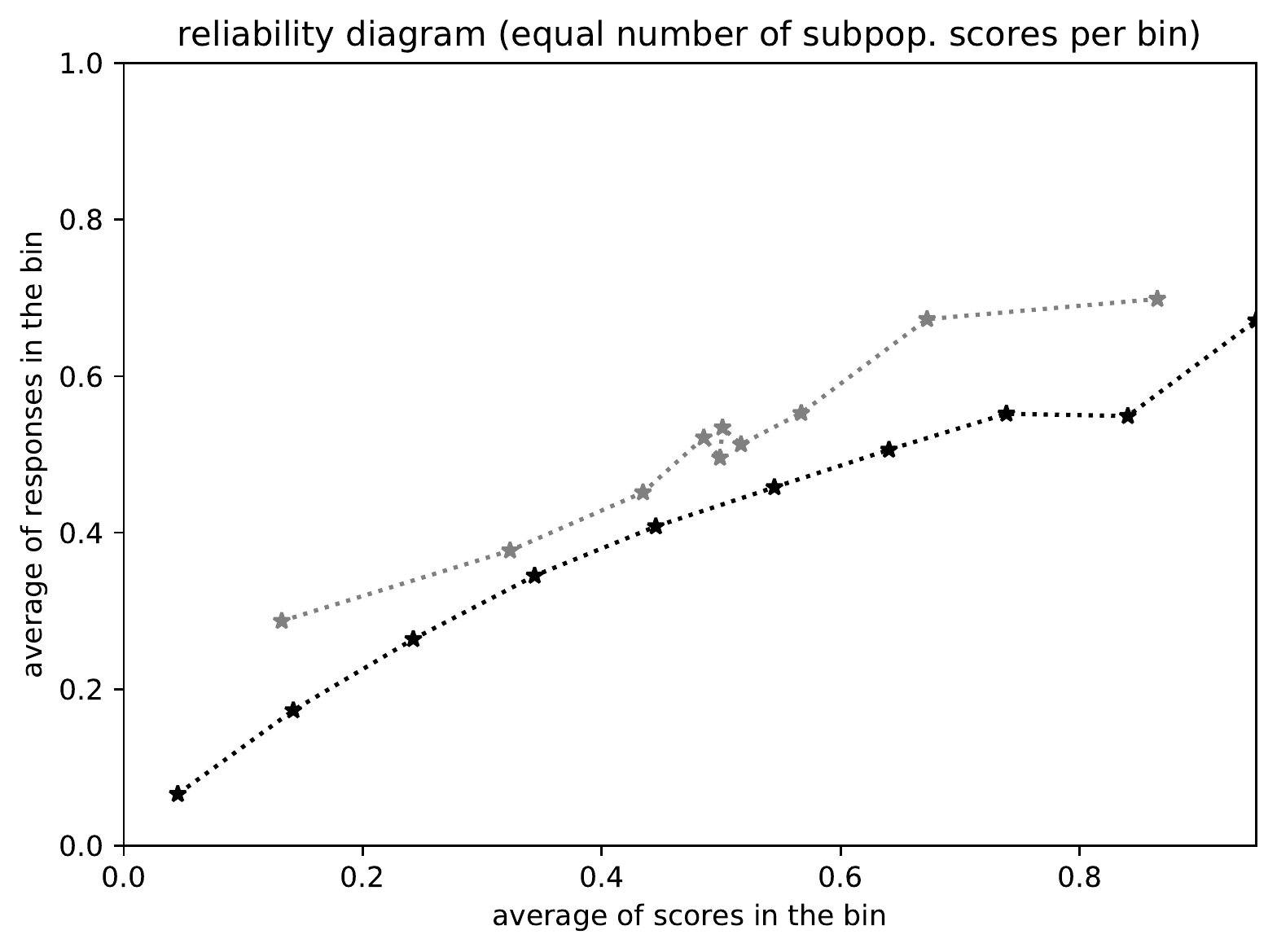}}
\quad\quad
(d)
\parbox{\imsize}{\includegraphics[width=\imsize]
{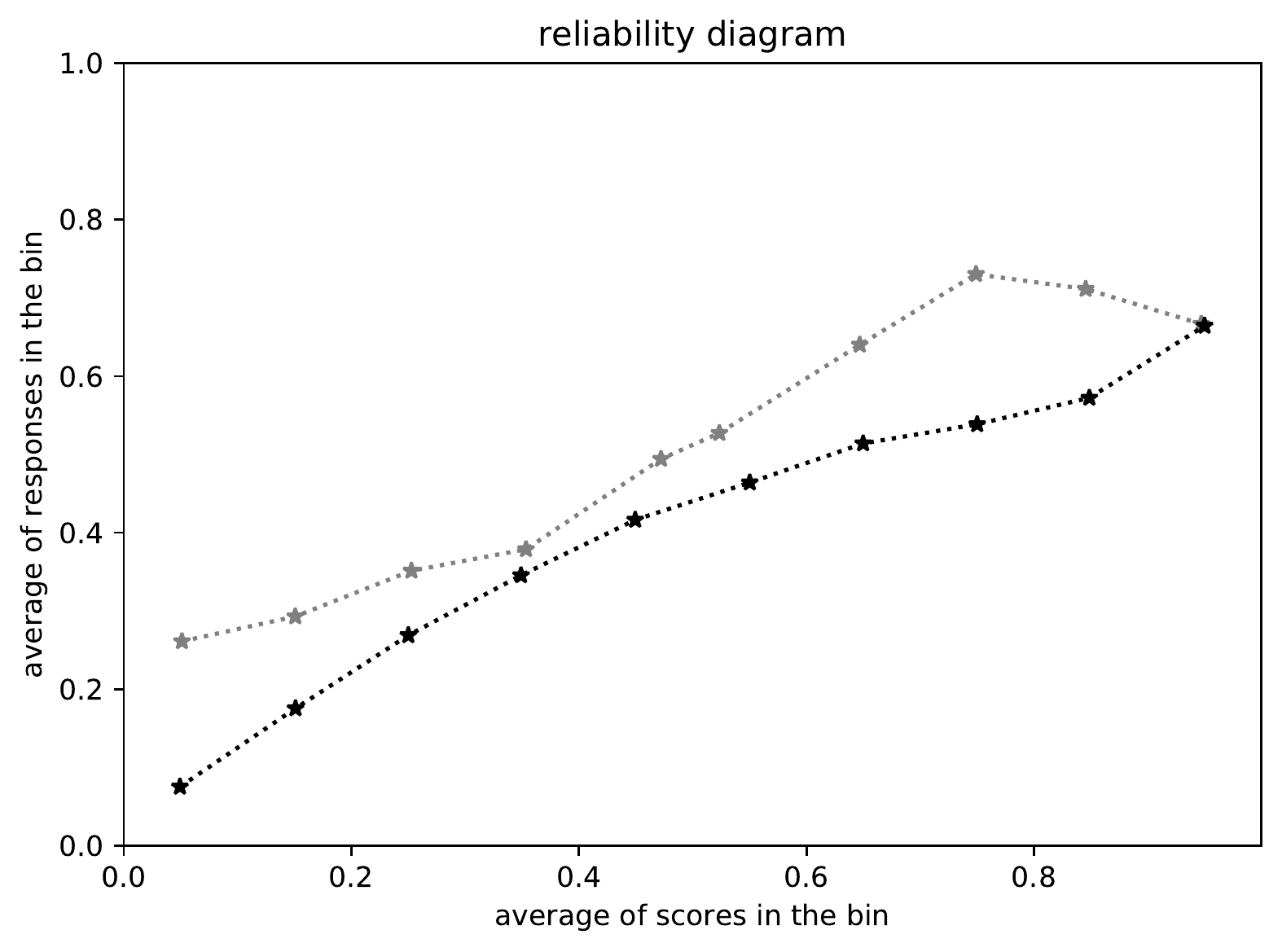}}

\vspace{\vertsep}

(e)
\parbox{\imsize}{\includegraphics[width=\imsize]
{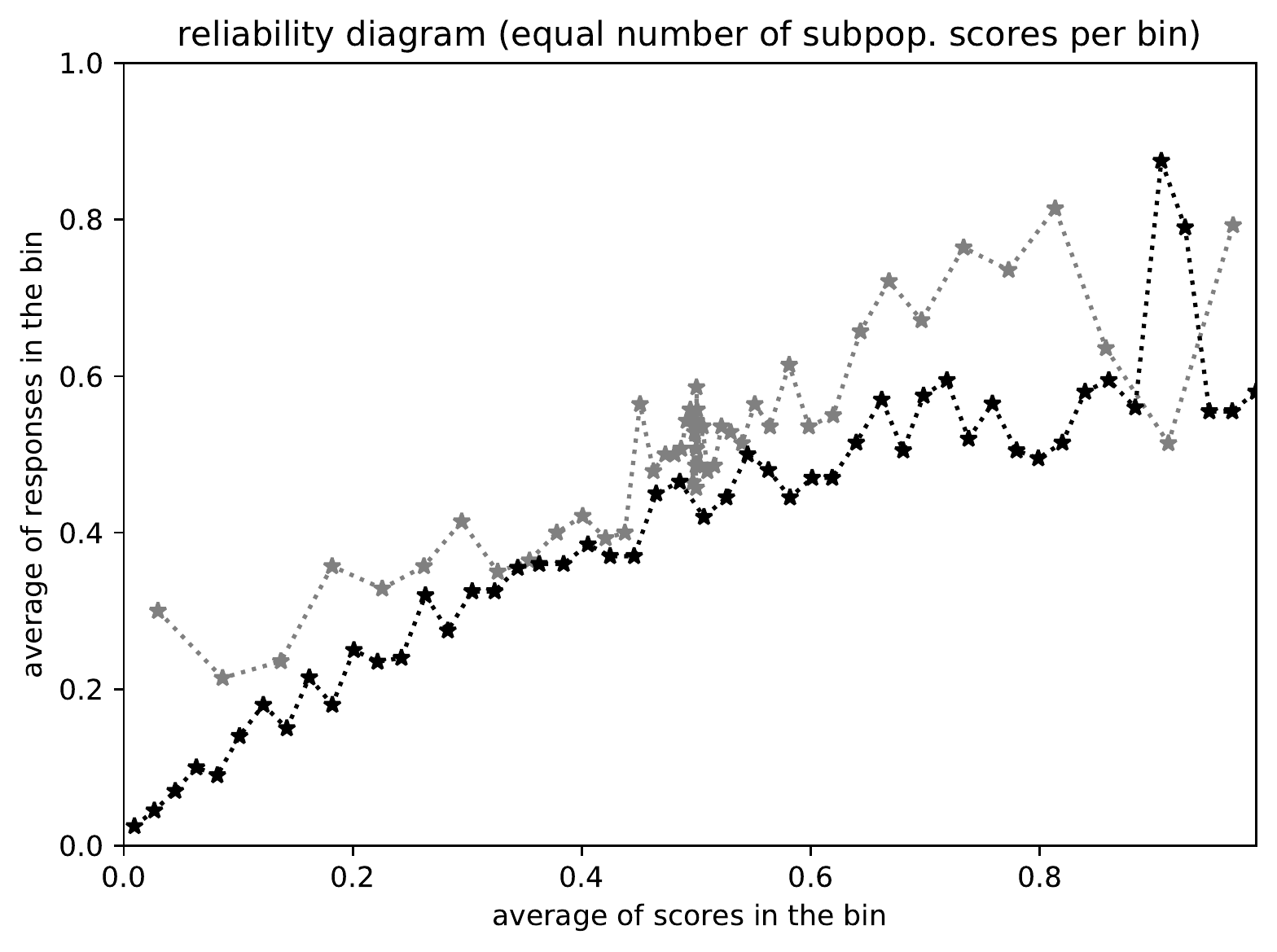}}
\quad\quad
(f)
\parbox{\imsize}{\includegraphics[width=\imsize]
{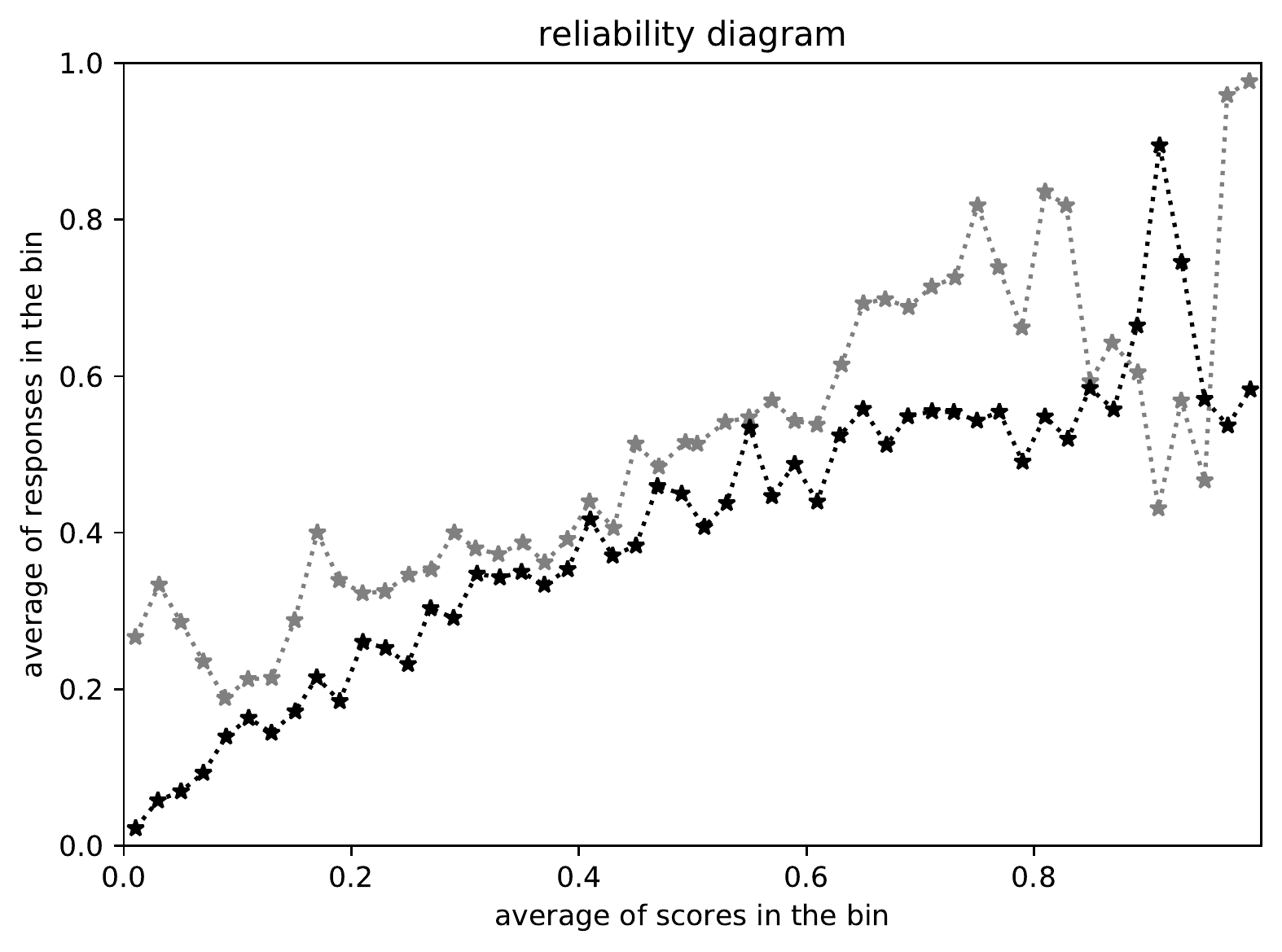}}

\vspace{\vertsep}

(g)
\parbox{\imsize}{\includegraphics[width=\imsize]
{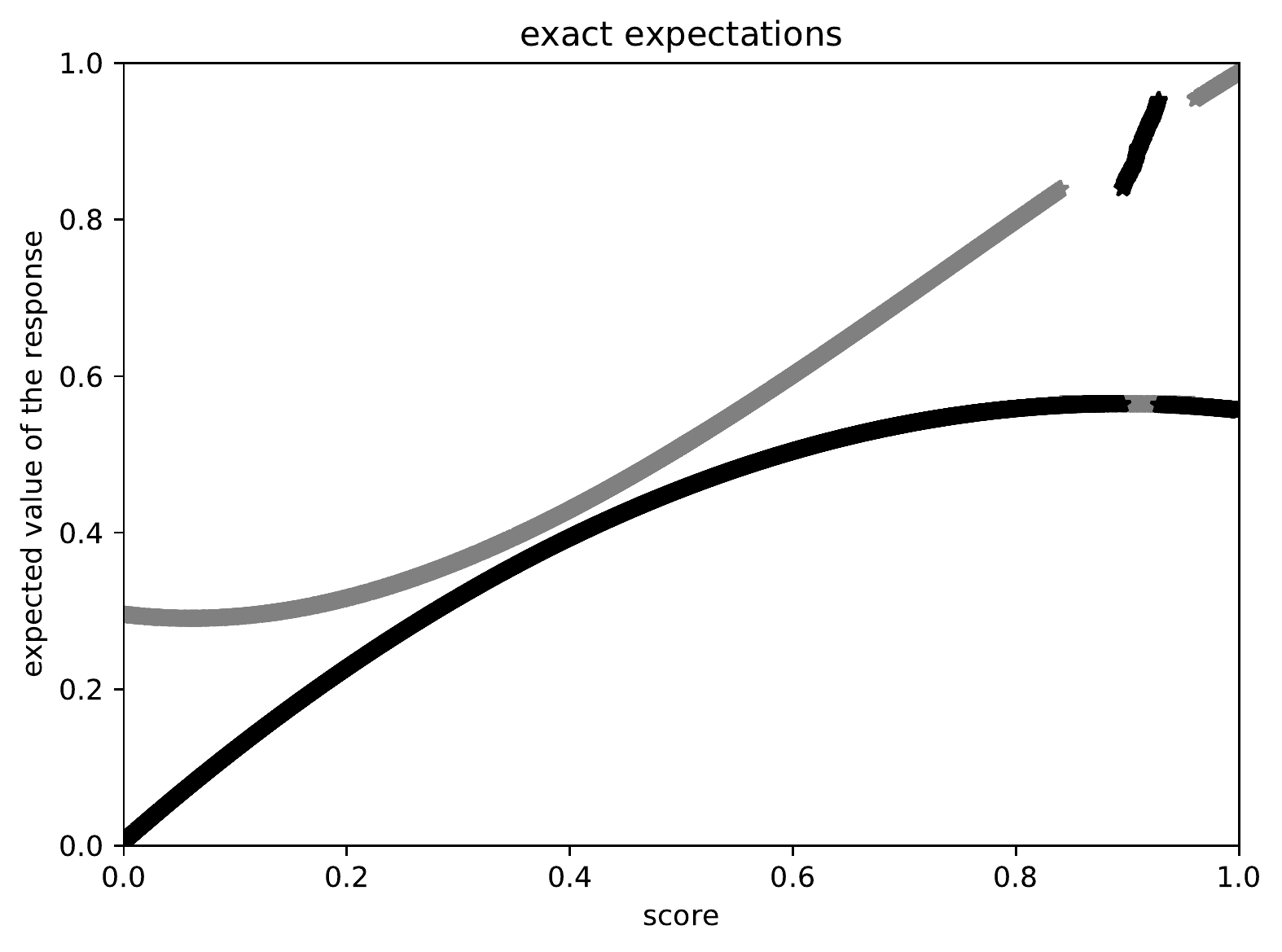}}

\end{centering}
\caption{$n =$ 6,451; Kuiper's statistic is $0.09740 / \sigma = 7.823$,
         Kolmogorov's and Smirnov's is $0.09724 / \sigma = 7.810$;
         the reliability diagrams with only 10 bins each (c and d) smooth out
         the jumps at high scores, and while the reliability diagrams
         with 50 bins each (e and f) give some indication of the jumps,
         the jumps still get smoothed over, while the bins for lower scores are
         too narrow to average away noise well. The cumulative graph (a)
         clearly displays the jumps, while remaining easily interpretable
         at lower scores.
         The statistics of Kuiper and of Kolmogorov and Smirnov are both
         several times greater than $\sigma$,
         so both reflect that the deviation displayed in the graphs
         is highly statistically significant.
}
\label{ex0}
\end{figure}

The graphs introduced in the present paper are easy to interpret.
For instance, in the topmost plots (a and b) of Figure~\ref{ex0},
the deviation between the two subpopulations over a range of scores
is simply the expected slope of the secant line for the graph
over that range of scores, as a function of the index $k/n$
(positive slope indicates that the responses
for one subpopulation are greater on average than those
for the other subpopulation, while negative slope indicates
that the responses for the former subpopulation are less than the latter's
on average).
Long ranges of steep slopes correspond to ranges of scores for which
the average responses are significantly different
between the two subpopulations;
the triangle along the vertical axis on the left of each plot
indicates the magnitude of the deviation across the full range of scores
that would be statistically significant at around the 95\% confidence level.
The connection with statistical significance also motivated related works,
including that of~\cite{gupta-rahimi-ajanthan-mensink-sminchisescu-hartley}
and~\cite{roelofs-cain-shlens-mozer}, which offer Kolmogorov-Smirnov metrics
to help gauge calibration of probabilistic predictions,
much like in the appendix of~\cite{tygert}.
Similarly, Section~3.2 of~\cite{gneiting-balabdaoui-raftery} and
Chapter~8 of~\cite{wilks} propose cumulative reliability diagrams,
albeit without leveraging the key to the approach of the present paper,
namely that slope is easy to assess visually even when the constant offset
of the part of a graph under consideration is arbitrary and uninformative.
Detailed explanation of statistical significance and Figure~\ref{ex0}
is available in Sections~\ref{methods} and~\ref{results} below.

Section~\ref{methods} introduces the methodology of cumulative differences,
both for graphs of the differences and for the scalar metrics
of Kuiper and of Kolmogorov and Smirnov that summarize the graphs' deviation
away from being perfectly flat.
Section~\ref{results} presents several illustrative examples,
via both simple synthetic and complicated real data sets.\footnote{Permissively
licensed open-source software that can automatically reproduce all figures
and statistics reported below is available at
\url{https://github.com/facebookresearch/fbcddisgraph}}
Section~\ref{conclusion} concludes the paper with a brief discussion.
Table~\ref{notation} summarizes the notation used throughout the present paper.
Readers interested mainly in seeing results and comparisons
of the proposed methods to the old standbys may wish to start
with Section~\ref{results}.

\begin{table}
\caption{Notational conventions
(The symbols in the table are in alphabetical order.)}
\label{notation}
\vspace{-.5em}
\begin{center}
\resizebox{\textwidth}{!}{%
\begin{tabular}{llll}
\hline
& & equation for the & equation for the \\
symbol & meaning & unweighted case & case with weights \\\hline
$A_k$ & abscissa for the cumulative graph in the case with weights &
(Not applicable) & (\ref{abscissae}) \\
$C_k$ & cumulative average difference between the subpopulations &
(\ref{cumulative}) & (\ref{cumulativew}) \\
$D_k$ & average difference between the subpopulations &
(\ref{diff_even}) and~(\ref{diff_odd}) &
(\ref{diff_even}) and~(\ref{diff_odd}) \\
$\Delta_k$ & expected slope of $C_j$ from $j = k$ to $j = k+1$ &
(\ref{delta}) & (\ref{deltaw}) \\
$G$ & Kolmogorov-Smirnov statistic & (\ref{Kolmogorov-Smirnov}) &
(\ref{Kolmogorov-Smirnov}) \\
$H$ & Kuiper statistic & (\ref{Kuiper}) & (\ref{Kuiper}) \\
$R^j_k$ & (average) response for subpopulation $j$'s $k$th block &
(Step~\ref{defining} within & (Readjusted in \\
& --- random dependent variable, outcome, or result &
Subsection~\ref{unweighted}) & Subsection~\ref{weighted}) \\
$S^j_k$ & (average) score for subpopulation $j$'s $k$th block &
(Step~\ref{defining} within & (Readjusted in \\
& --- non-random independent variable & Subsection~\ref{unweighted}) &
Subsection~\ref{weighted}) \\
$\sigma$ & scale of random fluctuations over the full range of scores &
(\ref{stddev}) & (\ref{stddevw}) \\
$T_k$ & total weight for $R^0_{k/2}$ or $R^1_{(k-1)/2}$ &
(Not applicable) & (\ref{aggregatew}) \\
$W_k$ & aggregated weight & (Not applicable) & (\ref{aggregatew}) \\
\hline
\end{tabular}}
\end{center}
\end{table}

\section{Methods}
\label{methods}

This section details the methodology proposed in the present paper.
Subsection~\ref{high-level} breaks data analysis into two stages:
a first, broad-brush stage of screening for potentially significant deviations
across many data sets and pairs of subpopulations,
and a second, finely detailed investigation of the variations
in the deviations as a function of score.
Subsection~\ref{unweighted} develops the graphical method
for the second stage, in the simplest case of unweighted sampling.
Subsection~\ref{scalarstats} then collapses the graphs
of Subsection~\ref{unweighted} into scalar statistics
useful for the first, broad-brush stage.
Subsection~\ref{significance} explains how to gauge statistical significance.
Finally, Subsection~\ref{weighted} treats the case of weighted sampling,
generalizing the previous subsections to the more complicated case
of data with weights.

\subsection{Approach to big data}
\label{high-level}

This subsection proposes a two-step approach to analyzing
multiple data sets and subpopulations (the same approach taken by~\cite{tygert}
in a related setting):
\begin{enumerate}
\item Calculate a single scalar summary statistic
for each data set for each pair of subpopulations of interest,
such that the size of the statistic
measures the deviation between the subpopulations.
\item Analyze in graphic detail each data set and pair of subpopulations
whose scalar summary statistic is large, graphing how the deviation
between the subpopulations varies as a function of score.
\end{enumerate}

The scalar statistic for the first step simply summarizes
the overall deviation across all scores,
as either the maximum absolute deviation of the second step's graph
or the size of the range of deviations in the graph.
Thus, both steps rely on a graph, with the first stage collapsing
the graphical display into a single scalar summary statistic.
The following subsection details the construction of this graph,
for the case of unweighted sampling (later, Subsection~\ref{weighted} treats
the weighted case).

\subsection{Unweighted sampling}
\label{unweighted}

This subsection presents the special case in which the observations
are unweighted (or, equivalently, uniformly or equally weighted).
Subsection~\ref{weighted} treats the more general case
of weighted observations, which is more complicated.

The present and all following subsections focus on a single data set
together with a single pair of subpopulations;
the previous subsection outlines a strategy for handling multiple data sets
and pairs of subpopulations, based on the processing of individual cases.
The data being considered should be observations of independent responses,
with each response taking one of finitely many real-valued possibilities,
and with each (random) response being paired with a real-valued score
viewed as given not random
(the responses across the different scores should be independent).
Hence, the scores can take on any real values,
whereas the responses should be drawn from discrete distributions.
{\it In the present paper, the scores from the observations
in both subpopulations put together must be distinct
--- the score for every observation from either subpopulation must be unique
or else slightly perturbed to become different from all the other scores
(perturbing as little as possible while accounting for roundoff,
for instance).}

Under this assumption of uniqueness, a graphical method for analyzing
deviation between the outcomes of the two subpopulations as a function
of score comprises the following procedure:
\begin{enumerate}
\item Merge all scores into a single sequence.
\item Sort the merged sequence into ascending order and
let ``subpopulation 0'' denote the subpopulation associated
with the first (the least) score in the sorted sequence.
\item Partition the sorted sequence into blocks such that
the scores in every other block all come from subpopulation 0,
interleaved with blocks in which all scores come from subpopulation 1;
that is to say:
\begin{enumerate}
\item the scores in the first (lowest) block all come from subpopulation 0,
\item the scores in the second lowest block all come from subpopulation 1,
\item the scores in the third lowest block all come from subpopulation 0,
\item the scores in the fourth lowest block all come from subpopulation 1,
\item and so on, alternating between the two subpopulations,
with all scores in each block coming from only one of the subpopulations.
\end{enumerate}
\item\label{defining} Denote by $S^0_k$ the (arithmetic) average of the scores
in the $(2k+1)$th block
and denote by $S^1_k$ the average of the scores in the $(2k+2)$th block;
denote by $R^0_k$ the average of the responses (the random outcomes)
corresponding to the scores in the $(2k+1)$th block and denote by $R^1_k$
the average of the responses (the random outcomes) corresponding to the scores
in the $(2k+2)$th block.
\item Form the sequence of average differences with even-indexed entries
\begin{equation}
\label{diff_even}
D_{2k} = \frac{(R^0_k - R^1_k) + (R^0_{k+1} - R^1_k)}{2}
       = \frac{R^0_k + R^0_{k+1} - 2R^1_k}{2}
\end{equation}
and odd-indexed entries
\begin{equation}
\label{diff_odd}
D_{2k+1} = \frac{(R^0_{k+1} - R^1_k) + (R^0_{k+1} - R^1_{k+1})}{2}
         = \frac{2R^0_{k+1} - R^1_k - R^1_{k+1}}{2}.
\end{equation}
\item Graph as a function of $j/n$ the sequence
of cumulative average differences
\begin{equation}
\label{cumulative}
C_j = \frac{1}{n} \sum_{k=0}^{j-1} D_k
\end{equation}
for $j = 1$, $2$, \dots, $n$,
where $n$ is the length of the sequence $D_0$,~$D_1$, \dots, $D_{n-1}$
from the previous step. Supplement $C_1$, $C_2$, \dots, $C_n$ with
\begin{equation}
\label{cumulative0}
C_0 = 0.
\end{equation}
\end{enumerate}

\begin{figure}
\begin{centering}
\hfil\parbox{0.65\textwidth}
{\includegraphics[width=0.65\textwidth]{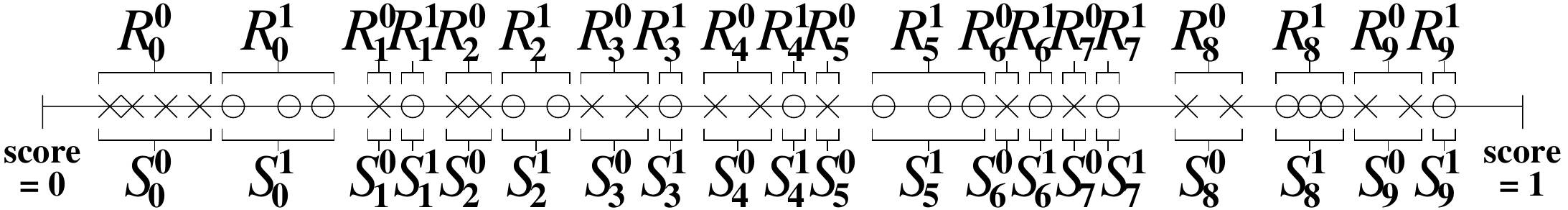}}
\end{centering}
\caption{The crosses (``x'') indicate the scores for subpopulation 0
while the circles (``o'') indicate the scores for subpopulation 1.
The averages of the scores for subpopulation 0 for the indicated blocks
of observed scores are $S^0_0$, $S^0_1$, \dots, $S^0_9$,
while the averages of the scores for subpopulation 1 are
$S^1_0$, $S^1_1$, \dots, $S^1_9$.
The averages of the responses for subpopulation 0 corresponding
to the indicated blocks of observed scores are
$R^0_0$, $R^0_1$, \dots, $R^0_9$, while the averages of the responses
for subpopulation 1 are $R^1_0$, $R^1_1$, \dots, $R^1_9$.
The scores need not range from 0 to 1 as in the present figure,
but that is a common case.
}
\label{partition}
\end{figure}

\begin{figure}
\vspace{.2in}
\begin{centering}
\hfil
(a) \parbox{0.111\textwidth}
{\includegraphics[width=0.111\textwidth]{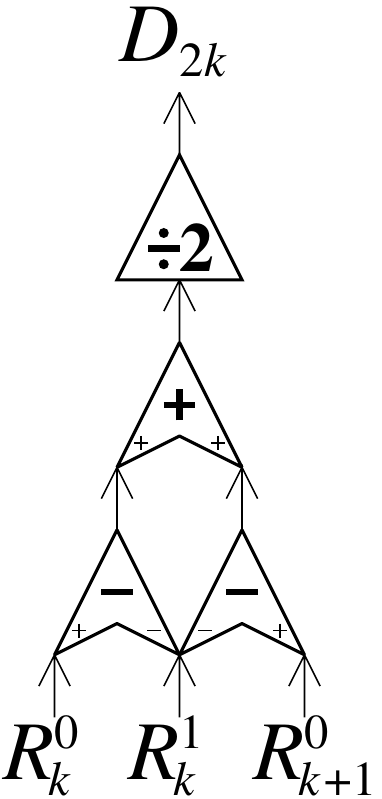}}
\hfil
(b) \parbox{0.111\textwidth}
{\includegraphics[width=0.111\textwidth]{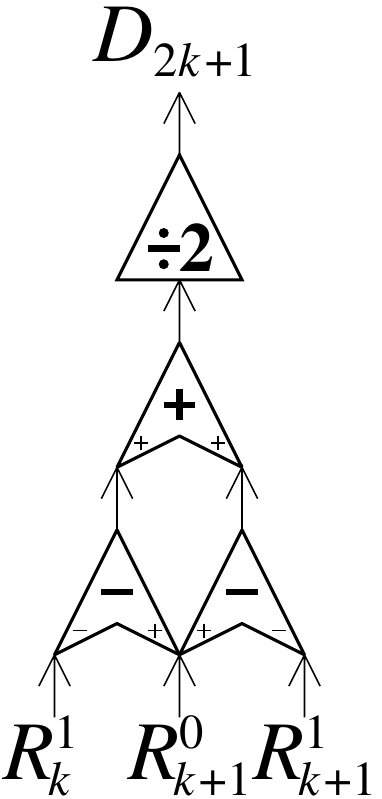}}
\end{centering}
\caption{In each of these subfigures, the operation indicated by ``$+$'' sums
its two inputs and the operations indicated by ``$-$'' subtract their inputs,
with one of these ``$-$'' operations subtracting its rightmost input
from its leftmost input, while the other subtracts its leftmost input
from its rightmost input.
In all cases, the operations indicated by ``$-$'' subtract
subpopulation 1 from subpopulation 0, in that order.
The operation indicated by ``$\div 2$'' divides its input by 2.
These subfigures depict visually formulae~(\ref{diff_even})
and~(\ref{diff_odd}), respectively.
}
\label{diffs}
\end{figure}

Figure~\ref{partition} illustrates Steps~1--4,
while Figure~\ref{diffs} illustrates Step~5.
The increment in the expected cumulative average difference
from $j = k$ to $j = k+1$ is
\begin{equation}
\label{fundamental}
\E[ C_{k+1} - C_k ] = \frac{\E[D_k]}{n},
\end{equation}
so that the expected slope of a graph of $C_k$ versus $k/n$ is
\begin{equation}
\label{delta}
\Delta_k = \E[D_k],
\end{equation}
which is simply the expected value of the difference
between the two subpopulations.
Thus, {\it the slope of a secant line over a long range of $k/n$
for the graph of $C_k$ versus $k/n$ becomes the average difference
in responses between the subpopulations}.

Figure~\ref{ex0} presents a synthetic example
from Subsection~\ref{synthetic} below for which the ground-truth
is known explicitly.
In accord with~(\ref{fundamental}),
the topmost plots (a and b) of Figure~\ref{ex0}
display deviation between the two subpopulations over a range of scores
as the expected slope of the secant line for the graph
over that range of scores, as a function of the index $k/n$
given along the horizontal axis.
As mentioned in the introduction,
long ranges of steep slopes correspond to ranges of scores for which
the average responses are significantly different
between the two subpopulations,
with the triangle along the vertical axis on the left of each plot
indicating the magnitude of the deviation across the full range of scores
that would be statistically significant at around the 95\% confidence level.
Subsection~\ref{significance} below provides details
on statistical significance and the computation of the triangle's height.

\begin{remark}
The blocked sequence of responses is
$R^0_0$, $R^1_0$, $R^0_1$, $R^1_1$, $R^0_2$, $R^1_2$, \dots.
The backward differences are
\begin{equation}
\label{back0}
R^0_k - R^1_k
\end{equation}
and
\begin{equation}
\label{back1}
R^1_k - R^0_{k+1},
\end{equation}
while the forward differences are
\begin{equation}
\label{forward0}
R^0_{k+1} - R^1_k
\end{equation}
and
\begin{equation}
\label{forward1}
R^1_{k+1} - R^0_{k+1},
\end{equation}
so that $D_{2k}$ from~(\ref{diff_even}) is the average of~(\ref{back0})
and~(\ref{forward0}) while $D_{2k+1}$ from~(\ref{diff_odd}) is the negative
of the average of~(\ref{back1}) and~(\ref{forward1}).
The reason for $D_{2k+1}$ to be the negative
is to align with $D_{2k}$ when summing them in~(\ref{cumulative})
--- the differences need to be in the same direction for the sum to make sense,
and the negative synchronizes the directions of the differences
(which would otherwise be alternating or staggered in the sequence);
with the negative, the differences always compare
subpopulation 0 to subpopulation 1, in that order.
\end{remark}

\begin{remark}
In the absence of any reason to prefer backward differences
to forward differences (or vice versa),
we opt to average the two possibilities together.
In the absence of any reason to prefer entries in the sequence
with even indices ($D_0$, $D_2$, $D_4$, \dots) to entries with odd indices
($D_1$, $D_3$, $D_5$, \dots), we include both.
\end{remark}

\subsection{Scalar summary statistics}
\label{scalarstats}

This subsection constructs standardized statistics
which summarize in single scalars the plots of the previous subsection.

Two standard metrics for the overall deviation between the two subpopulations
over the full range of scores
and that take into account expected random fluctuations are that due
to Kolmogorov and Smirnov, the maximum absolute deviation
\begin{equation}
\label{Kolmogorov-Smirnov}
G = \max_{1 \le k \le n} |C_k|,
\end{equation}
and that due to Kuiper, the size of the range of the deviations
\begin{equation}
\label{Kuiper}
H = \max_{0 \le k \le n} C_k - \min_{0 \le k \le n} C_k,
\end{equation}
where $C_0$ is defined in~(\ref{cumulative0})
and $C_1$, $C_2$, \dots, $C_n$ are defined in~(\ref{cumulative}).
Under appropriate statistical models,
$G$ and $H$ can form the basis for tests of statistical significance,
the context in which they originally appeared;
see, for example, Section~14.3.4 of~\cite{press-teukolsky-vetterling-flannery}.
To assess statistical significance (rather than absolute effect size),
$G$ and $H$ should be rescaled larger by a factor proportional to $\sqrt{n}$;
further discussion of the rescaling is available in the next subsection.
Needless to say, if the graph constructed in the previous subsection
is fairly flat for all scores (which indicates a lack of deviation
between the subpopulations for all scores),
then both the maximum absolute deviation of the graph and the size of the range
of deviations ($G$ and $H$, respectively) will be close to 0.
The captions of the figures report the values of these scalar statistics
for numerical examples.

\begin{remark}
Remark~1 of~\cite{tygert} explains the reason for including $C_0$
in the definition of Kuiper's statistic $H$ in~(\ref{Kuiper}),
as well as why $H$ is often slightly preferable to $G$.
\end{remark}

\subsection{Significance of stochastic fluctuations}
\label{significance}

This subsection discusses statistical significance
both for the graphical methods of Subsection~\ref{unweighted}
and for the summary statistics of Subsection~\ref{scalarstats}.

The graph of $C_k$ as a function of $k/n$ generally displays
some ``confidence bands'' due to $C_k$ fluctuating randomly
as the index $k$ increments; the ``thickness'' of the plot
arising from the random fluctuations gives some sense of ``error bars.''
To indicate the rough size of the fluctuations
of the maximum deviation expected under the hypothesis that
the actual underlying response distributions of the two subpopulations
are the same, the plots should include a triangle centered at the origin
whose height above the origin is proportional to $1/\sqrt{n}$.
The triangle is similar to the conventional confidence bands
around an empirical cumulative distribution function
introduced by Kolmogorov and Smirnov, as reviewed by~\cite{doksum}
--- a driftless, purely random walk deviates from zero
by roughly $\sqrt{n}$ after $n$ steps, so a random walk scaled by $1/n$
deviates from zero by roughly $1/\sqrt{n}$.
Identification of deviation between the two subpopulations
is reliable when focusing on long ranges of steep slopes
(as a function of $k/n$) for $C_k$; the triangle gives a sense
of the length scale for the largest stochastic variations that
are likely to happen even when there is no underlying deviation
between the subpopulations. The remainder of the present subsection
derives this conservative upper bound on the length scale
in cases for which the value of every observed response is either 0 or 1.

The long-range deviations of $C_0$, $C_1$, $C_2$, \dots, $C_n$ from zero
can be biased even when the two subpopulations are drawn
from the same underlying distribution as a function of score;
however, the use of centered, second-order differences in~(\ref{diff_even})
and~(\ref{diff_odd}) makes this a second-order effect.
In the sequel, we make two assumptions about bias:
\{1\} the bias arising from averaging together multiple responses
at slightly different scores into a single $R^0_k$ or $R^1_k$ is offset
by the reduction in variance due to the averaging, and
\{2\} the bias arising from taking differences of responses
from the different subpopulations at slightly different scores
is negligible in comparison with the square root of the accumulated variance.
The first assumption can be especially reasonable when the scores
considered for a single $R^0_k$ or $R^1_k$
are in reality drawn at random from some probability distribution,
such that the variance in the probabilities of success
for the associated Bernoulli responses is comparable
to the variance of a Bernoulli variate with a given probability of success.
In such cases, the first assumption permits us to regard each $R^0_k$
or $R^1_k$ as contributing no more to the long-range deviation
than a single Bernoulli variate would.
The second assumption means that we will neglect
the second-order effect of accumulated bias,
which is often reasonable due to the use of second-order differences
in~(\ref{diff_even}) and~(\ref{diff_odd}).

In cases for which the value of every observed response is either 0 or 1,
the tip-to-tip height of the triangle centered at the origin should be $8/n$
times the standard deviation of the sum of $n$ independent Bernoulli variates.
This is simply $8/n$ times the square root of the sum of the variances
of $n$ Bernoulli variates, which could be at most
$(8/n)(\sqrt{n/4}) = 4\sigma$, where
\begin{equation}
\label{stddev}
\sigma = \frac{1}{\sqrt{n}},
\end{equation}
since the variance of a Bernoulli variate is $p(1-p) \le 1/4$,
where $p$ is the unknown probability of success.
Note that the factor 8 incorporates a factor of 2 for the triangle
extending both above and below the origin, a factor of 2 to extend
for 2 standard deviations rather than just 1
(setting the confidence level at approximately 95\%), a factor of $\sqrt{2}$
due to the dependency between the even- and odd-indexed entries
in the sequence of second-order differences from~(\ref{diff_even})
and~(\ref{diff_odd}), and a factor of $\sqrt{2}$ to account
for having 2 independently drawn subpopulations.
Needless to say, the upper bound of $4\sigma$ is often somewhat loose
in practice, as the two assumptions discussed in the previous paragraph
yield rather conservative guarantees.
Tighter bounds may exist in settings for which the scores are drawn
from a specified probability distribution
(unlike in the setting of the present paper).

\subsection{Weighted sampling}
\label{weighted}

This subsection presents the general case in which the observations
come with weights, where each weight is a positive real number
associated with the corresponding observation.
Subsection~\ref{unweighted} treats the special case of unweighted
(or, equivalently, uniformly or equally weighted) observations,
which is simpler.

The weighted case uses the same procedure as in Subsection~\ref{unweighted},
but with $S^0_k$, $S^1_k$, $R^0_k$, and $R^1_k$ being weighted averages rather
than unweighted averages (the weighted average for each $S^0_k$, $S^1_k$,
$R^0_k$, and $R^1_k$ should be normalized separately).
Then, we define $T_{2k}$ to be the average of the weights associated
with the scores whose weighted average is $S^0_k$,
and define $T_{2k+1}$ to be the average of the weights associated
with the scores whose weighted average is $S^1_k$.
Setting $W_k$ to be the sum of the weights associated
with $D_k$ defined in~(\ref{diff_even}) and~(\ref{diff_odd}), that is,
\begin{equation}
\label{aggregatew}
W_k = T_k + 2T_{k+1} + T_{k+2},
\end{equation}
the formula~(\ref{cumulative}) generalizes to
\begin{equation}
\label{cumulativew}
C_j = \frac{\sum_{k=0}^{j-1} W_k D_k}{\sum_{k=0}^{n-1} W_k}
\end{equation}
for $j = 1$, $2$, \dots, $n$,
while $C_0 = 0$ exactly as before in formula~(\ref{cumulative0}).
In the weighted case, the abscissae (that is, the horizontal coordinates)
for the graph consist of the normalized aggregated weights
\begin{equation}
\label{abscissae}
A_j = \frac{\sum_{k=0}^{j-1} W_k}{\sum_{k=0}^{n-1} W_k}
\end{equation}
for $j = 1$, $2$, \dots, $n$, and
\begin{equation}
A_0 = 0.
\end{equation}
The original, unweighted procedure of Subsection~\ref{unweighted}
yields precisely the same results as the weighted procedure
of the present subsection in the special case that the weights
for the original observations are all the same.

The increment in the expected cumulative weighted average difference
from $j = k$ to $j = k+1$ is
\begin{equation}
\label{fundamentalw}
\E[ C_{k+1} - C_k ] = \frac{W_k \E[D_k]}{\sum_{j=0}^{n-1} W_j},
\end{equation}
while the increment in the normalized aggregated weights
from $j = k$ to $j = k+1$ is
\begin{equation}
\label{fundamentala}
A_{k+1} - A_k = \frac{W_k}{\sum_{j=0}^{n-1} W_j},
\end{equation}
so that the expected slope of a graph of $C_k$ versus $A_k$ is
the ratio of~(\ref{fundamentalw}) to~(\ref{fundamentala}), that is,
\begin{equation}
\label{deltaw}
\Delta_k = \E\left[\frac{C_{k+1} - C_k}{A_{k+1} - A_k}\right] = \E[D_k],
\end{equation}
which is none other than the expected value
of the difference between the two subpopulations.
Thus, {\it the slope of a secant line over a long range of $k$
for the graph of $C_k$ versus $A_k$ becomes the average difference
in responses between the subpopulations}.

The scalar summary statistics in the weighted case are given
by the same formulae from Subsection~\ref{scalarstats}
as for the unweighted case, just using $C_j$ from~(\ref{cumulativew})
in place of $C_j$ from~(\ref{cumulative}).
In cases for which the value of every observed response is either 0 or 1,
the tip-to-tip height of the triangle centered at the origin
analogous to that from Subsection~\ref{significance}
could be set conservatively at $4\sigma$, where
\begin{equation}
\label{stddevw}
\sigma = \frac{\sqrt{\sum_{k=0}^{n-1} (W_k)^2}}{\sum_{k=0}^{n-1} W_k},
\end{equation}
which is an upper bound on the worst case under the same two assumptions
as in Subsection~\ref{significance}.

\begin{remark}
\label{weightedremark}
The classical methods for reliability diagrams discussed in the introduction
easily adapt to the case of weighted sampling.
Rather than plotting the plain, unweighted average of responses
against the unweighted average of scores in each bin,
the weighted case involves plotting the weighted average of responses
against the weighted average of scores in each bin.
Two natural choices of bins in the weighted case are
\{1\} make the widths of the bins all be the same or
\{2\} use the binning of the following remark (Remark~\ref{equierrs}).
As in the unweighted case, the second choice can adapt to each subpopulation
under consideration, with each subpopulation having its own binning.
\end{remark}

\begin{remark}
\label{equierrs}
In the case of weighted sampling, the most useful reliability diagrams
are usually those entitled,
``reliability diagram ($\|W\|_2/\|W\|_1$ is similar for every bin).''
These diagrams construct bins such that, for every bin,
the ratio of the sum of the squares of the bin's weights
to the square of the sum of the bin's weights is similar for every bin.
Remark~5 of~\cite{tygert} details the specific procedure employed
for setting the bins.
\end{remark}

\section{Results and discussion}
\label{results}

This section illustrates via numerous examples
the previous section's methods, including comparisons
with the canonical plots --- the ``reliability diagrams'' ---
discussed in the introduction.\footnote{Permissively licensed open-source
software that can automatically reproduce all figures and statistics reported
in the present paper is available at
\url{https://github.com/facebookresearch/fbcddisgraph}}
Subsection~\ref{synthetic} presents several synthetic examples.
Subsection~\ref{imagenetex} gives examples
from a popular, unweighted data set of images, ImageNet.
Subsection~\ref{census} considers a weighted data set,
the year 2019 American Community Survey of the United States Census Bureau.
Finally, Subsection~\ref{caution} issues a warning
about possible overinterpretations of the plots (both for the cumulative graphs
and for the classical reliability diagrams)
and suggests following~\cite{tygert} by comparing a subpopulation
to the full population (when apposite).

The figures display the reliability diagrams
(that is, the classical calibration plots)
as well as both the graphs of cumulative differences and the exact expectations
in the absence of the random sampling's noise (the figures include
the exact expectations only when they are known, as for the synthetic data).
The captions of the figures discuss the numerical results depicted.

The title, ``subpopulation deviation is the slope as a function of $k/n$,''
labels a plot of $C_k$ from~(\ref{cumulative}) as a function of $k/n$.
In each such plot, the upper axis specifies $k/n$,
while the lower axis specifies the score for the corresponding value of $k$.
The title, ``subpopulation deviation is the slope as a function of $A_k$,''
labels a plot of $C_k$ from~(\ref{cumulativew}) versus 
the cumulative weight $A_k$ from~(\ref{abscissae}).
In each such plot, the major ticks on the upper axis specify $k/n$,
while the major ticks on the lower axis specify the score
for the corresponding value of $k$; the points in the plot
are the ordered pairs $(A_k, C_k)$ for $k = 1$,~$2$, \dots, $n$,
with $A_k$ being the abscissa and $C_k$ being the ordinate.
(The abscissa is the horizontal coordinate;
the ordinate is the vertical coordinate.)

In all cases, if the second subpopulation ends up being subpopulation 0
in the notation of Section~\ref{methods}, then the cumulative graph technically
actually plots $-C_k$ rather than $C_k$
(in the same notation of Section~\ref{methods}).

The titles, ``reliability diagram,''
``reliability diagram (equal number of subpopulation scores per bin),''
and ``reliability diagram ($\|W\|_2/\|W\|_1$ is similar for every bin),''
label plots of the pairs from the introduction (in the unweighted case)
or from Remark~\ref{weightedremark} (in the case of weighted sampling),
with the pairs from the first subpopulation in black
and the pairs from the second subpopulation in gray.

In the traditional, binned plots,
we vary the number of bins to see how the plotted values vary.
Displaying the bin frequencies is another way to indicate uncertainties,
as suggested, for example, by~\cite{murphy-winkler}.
Still other possibilities for uncertainty quantification could use
kernel density estimation, as suggested, for example,
by~\cite{brocker}, \cite{srihera-stute}, and~\cite{wilks}.
Such uncertainty estimates involve setting widths for the bins
or kernel smoothing; such settings are fairly arbitrary
and actually unnecessary when varying the widths as in the plots
of the present paper.
A comprehensive review of the various possibilities is available
in Chapter~8 of~\cite{wilks}.

As the introduction discusses, there are two standard choices for the bins
when the sampling is unweighted (or uniformly weighted):
\{1\} make the average of the scores in each bin
be roughly equidistant from the average of the scores
in each neighboring bin or
\{2\} make the number of scores in every bin
(except perhaps for the last) be the same.
The figures label the first, more conventional possibility
with the short title, ``reliability diagram,'' and the second possibility
with the longer title,
``reliability diagram (equal number of subpopulation scores per bin).''
As noted in Remark~\ref{weightedremark}, there are two typical choices
for the bins when the sampling is weighted:
\{1\} make the weighted average of the scores in each bin
be roughly equidistant from the weighted average of the scores
in each neighboring bin or
\{2\} follow Remark~\ref{equierrs} above.
The figures label the first possibility with the short title,
``reliability diagram,'' and the second possibility
with the longer title,
``reliability diagram ($\|W\|_2/\|W\|_1$ is similar for every bin).''

Needless to say, reliability diagrams with fewer bins provide estimates
that are less noisy, at the cost of restricting the resolution
for detecting deviations and for resolving variations
as a function of the score.

\subsection{Synthetic}
\label{synthetic}

This subsection presents several toy examples
that consider instructive ``ground-truth'' statistical models
and generate observations at random from them.
The examples set values for the scores and expected values of the responses,
and then independently draw the observed responses
from the Bernoulli distributions whose probabilities of success
are those expected values.

\begin{figure}
\begin{centering}

(a)
\parbox{\imsize}{\includegraphics[width=\imsize]
{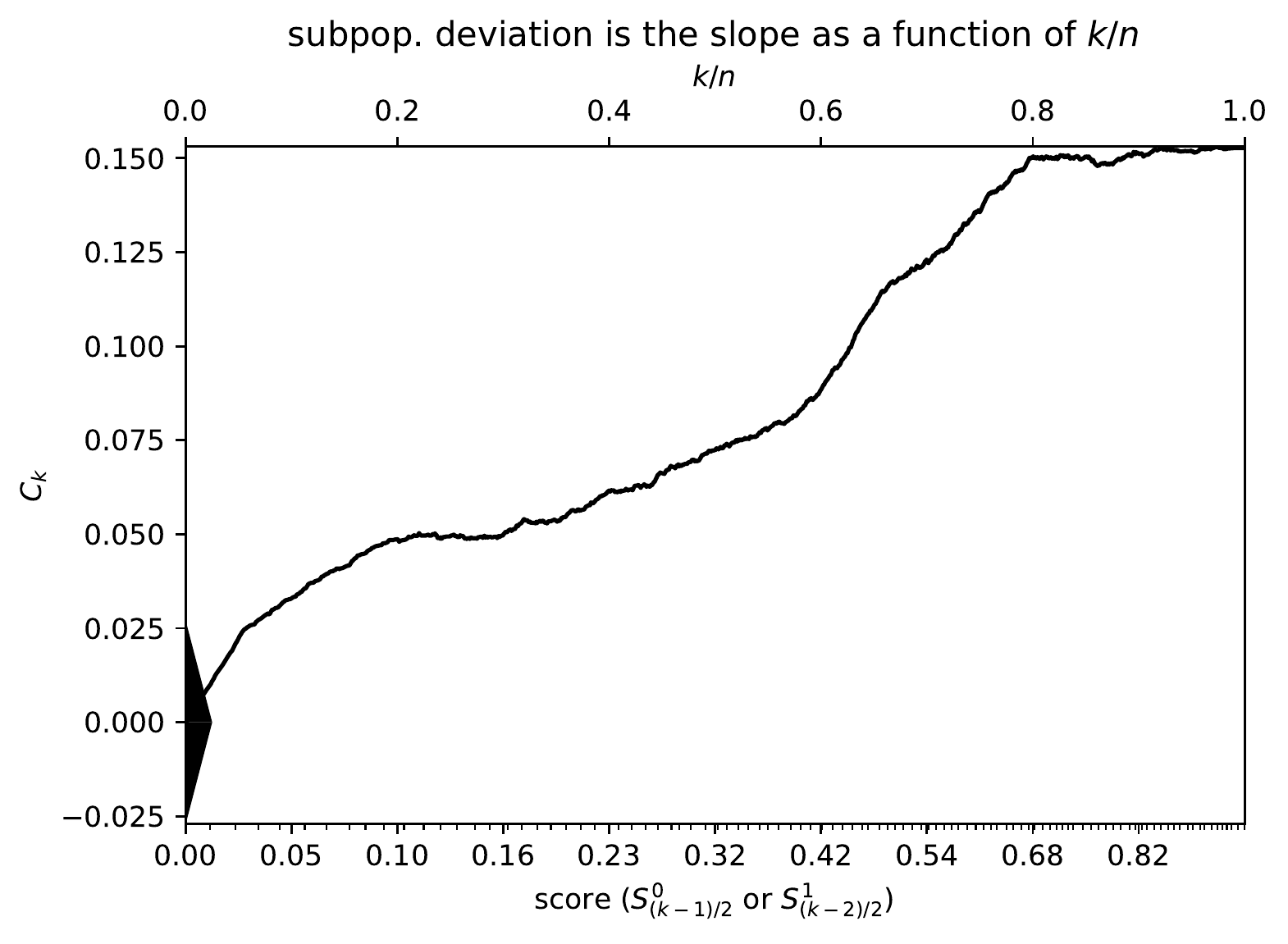}}
\quad\quad
(b)
\parbox{\imsize}{\includegraphics[width=\imsize]
{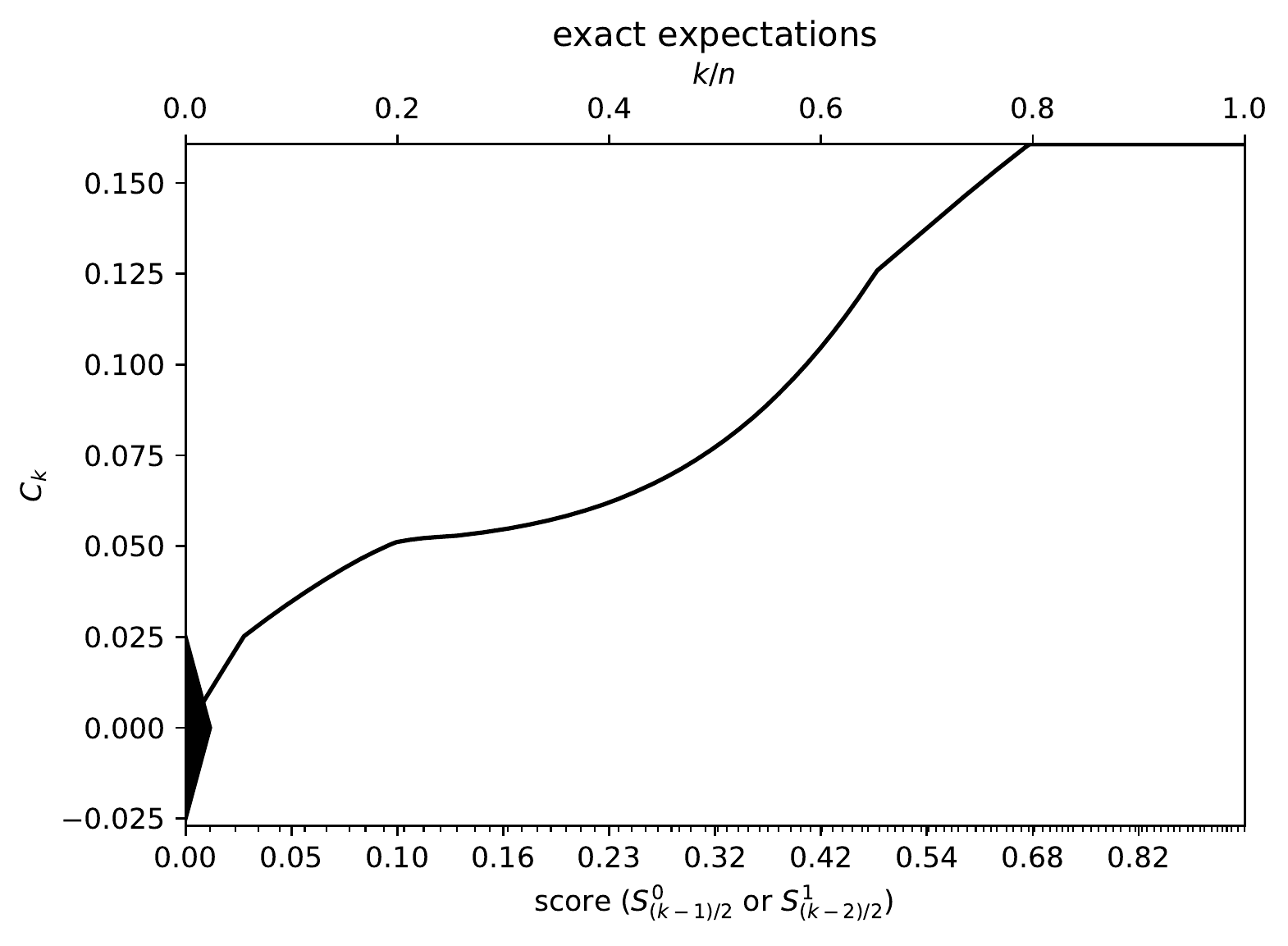}}

\vspace{\vertsep}

(c)
\parbox{\imsize}{\includegraphics[width=\imsize]
{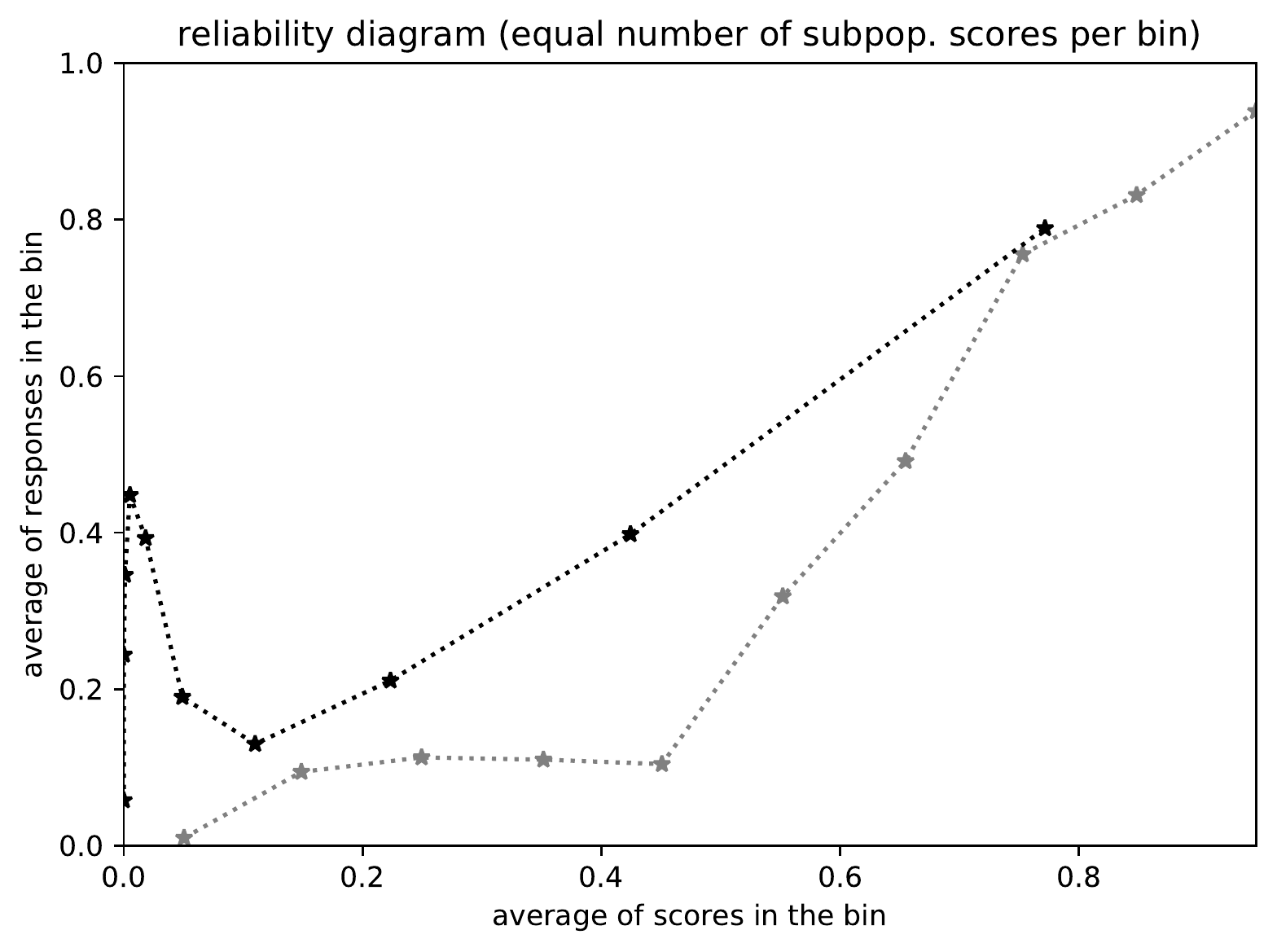}}
\quad\quad
(d)
\parbox{\imsize}{\includegraphics[width=\imsize]
{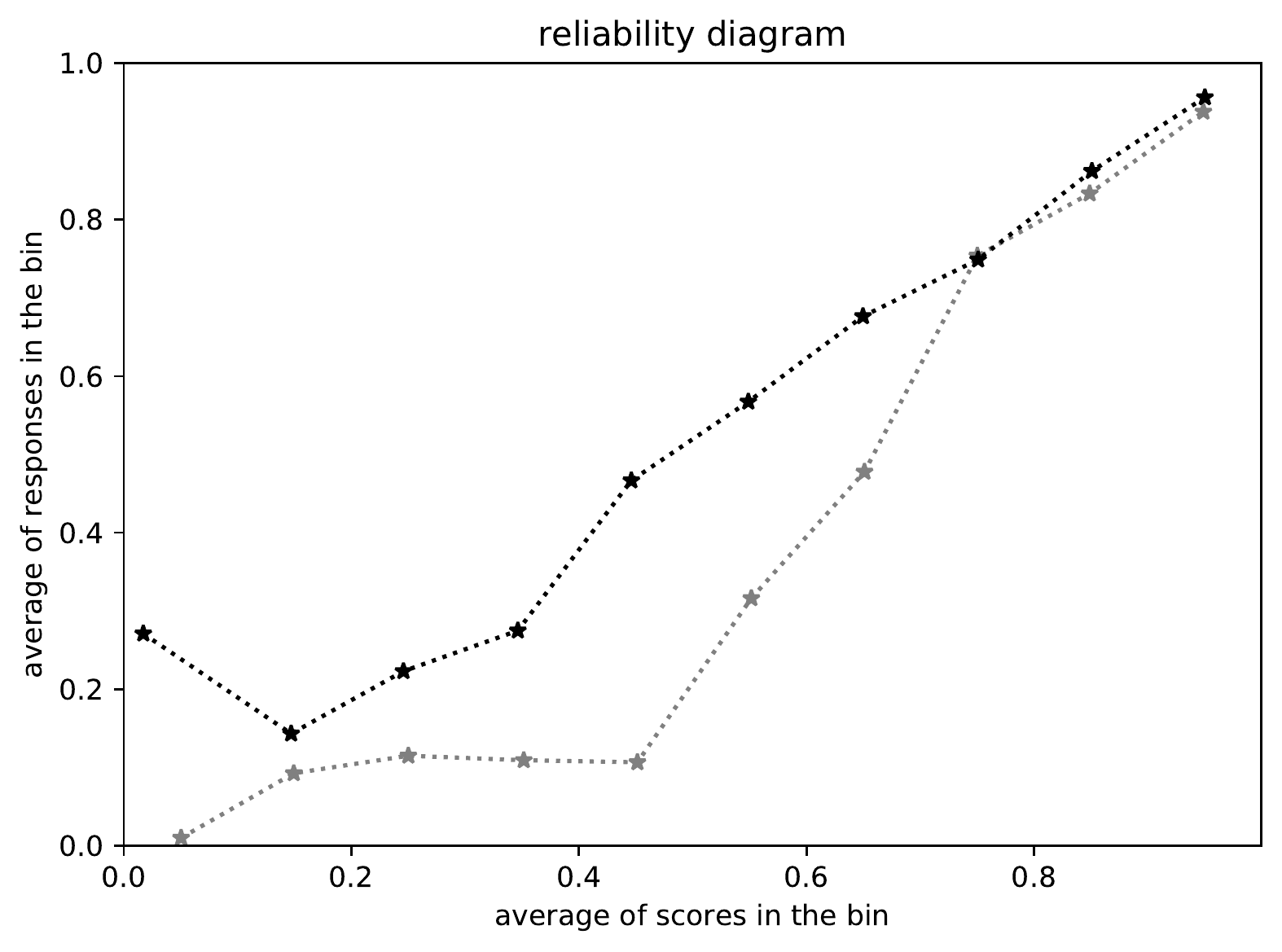}}

\vspace{\vertsep}

(e)
\parbox{\imsize}{\includegraphics[width=\imsize]
{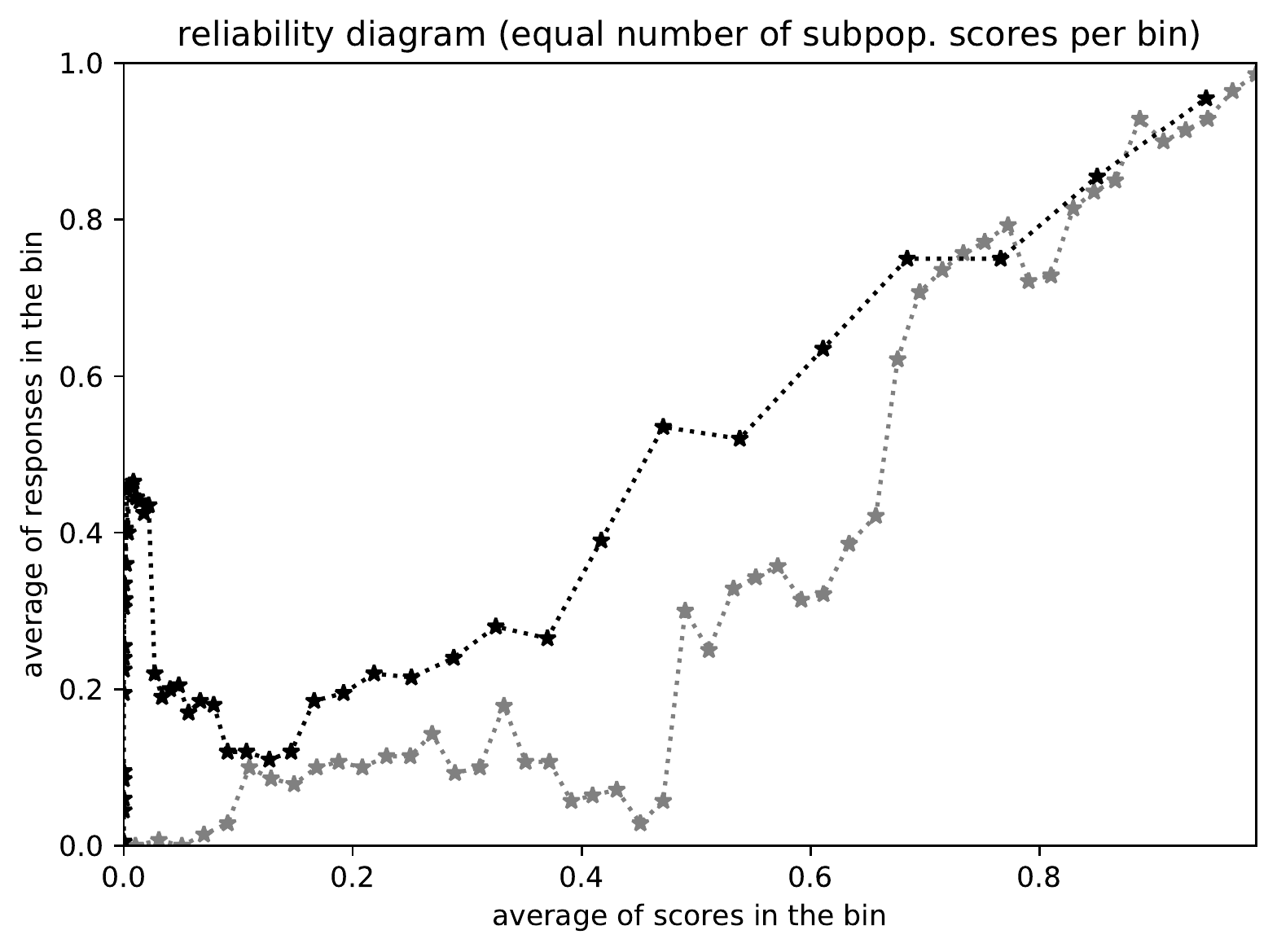}}
\quad\quad
(f)
\parbox{\imsize}{\includegraphics[width=\imsize]
{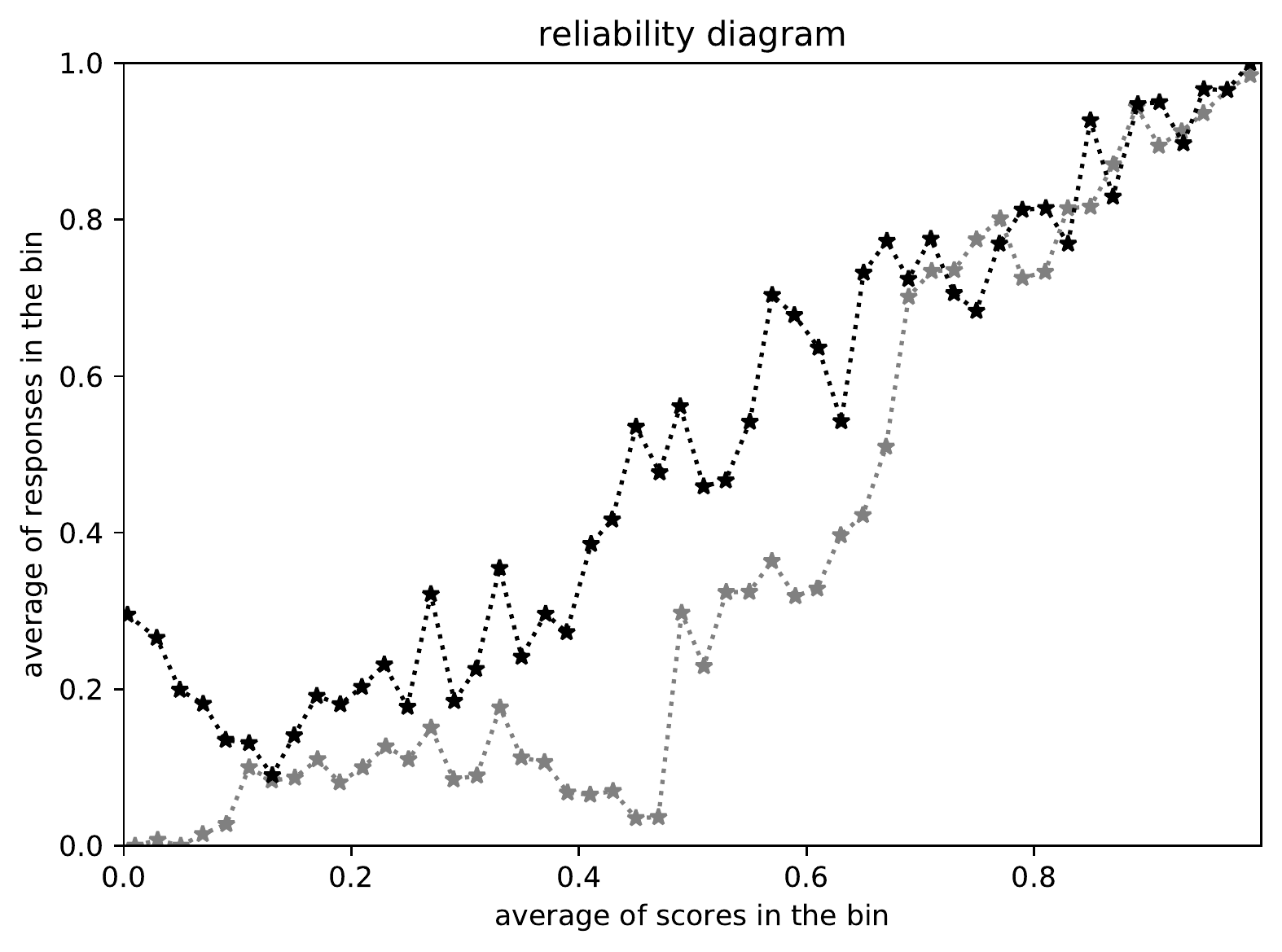}}

\vspace{\vertsep}

(g)
\parbox{\imsize}{\includegraphics[width=\imsize]
{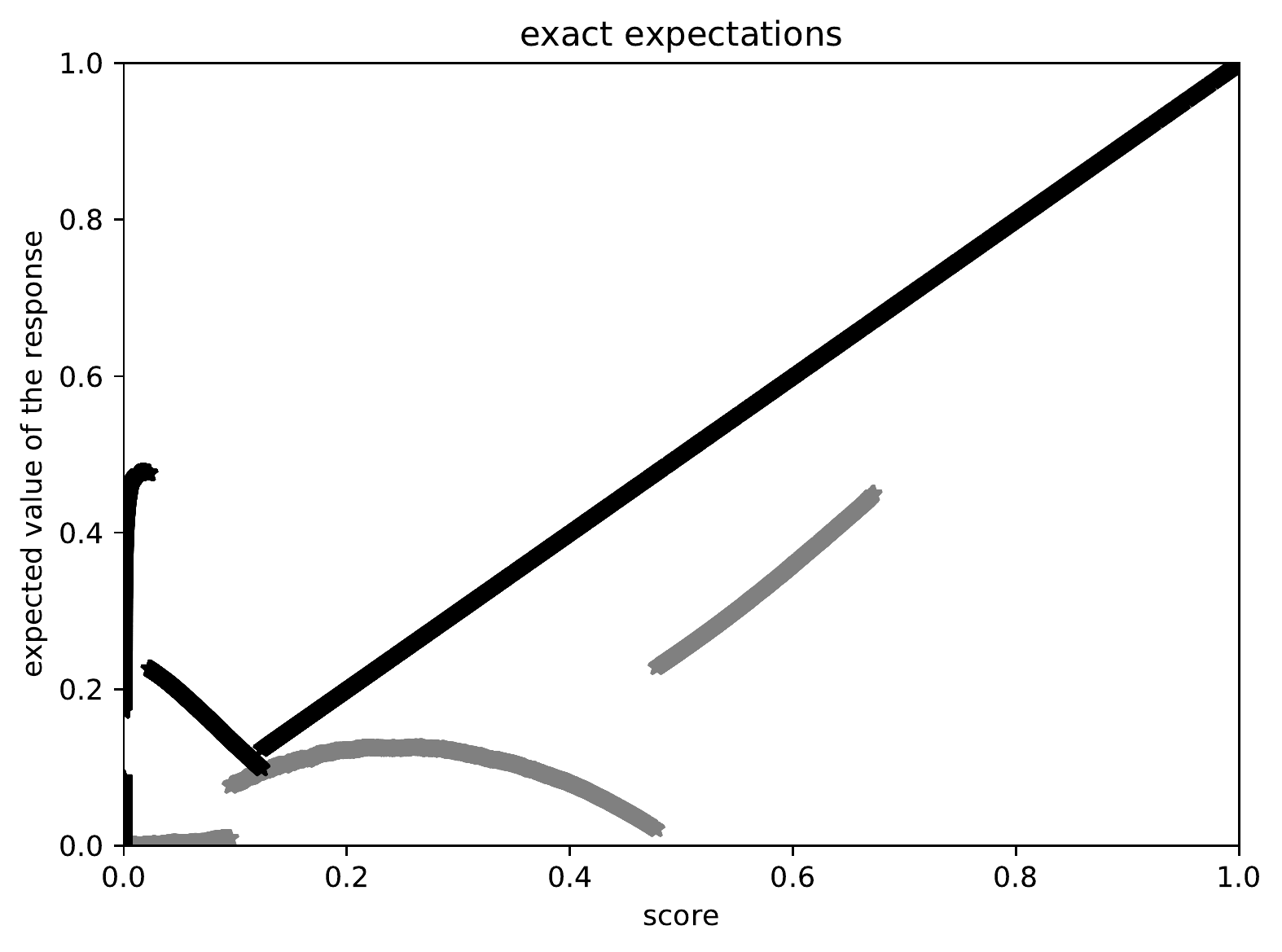}}

\end{centering}
\caption{$n =$ 5,472; Kuiper's statistic is $0.1531 / \sigma = 11.32$,
         Kolmogorov's and Smirnov's is $0.1531 / \sigma = 11.32$;
         the reliability diagrams all have trouble resolving the sharp behavior
         corresponding to the relatively sharp corners in the cumulative graphs
         (a and b), though the reliability diagram with 50 bins that
         has an equal number of subpopulation scores per bin (e) is decent.
         The metrics of Kuiper and of Kolmogorov and Smirnov
         report extremely statistically significant deviation,
         taking values of many times $\sigma$.
}
\label{ex1}
\end{figure}

\begin{figure}
\begin{centering}

(a)
\parbox{\imsize}{\includegraphics[width=\imsize]
{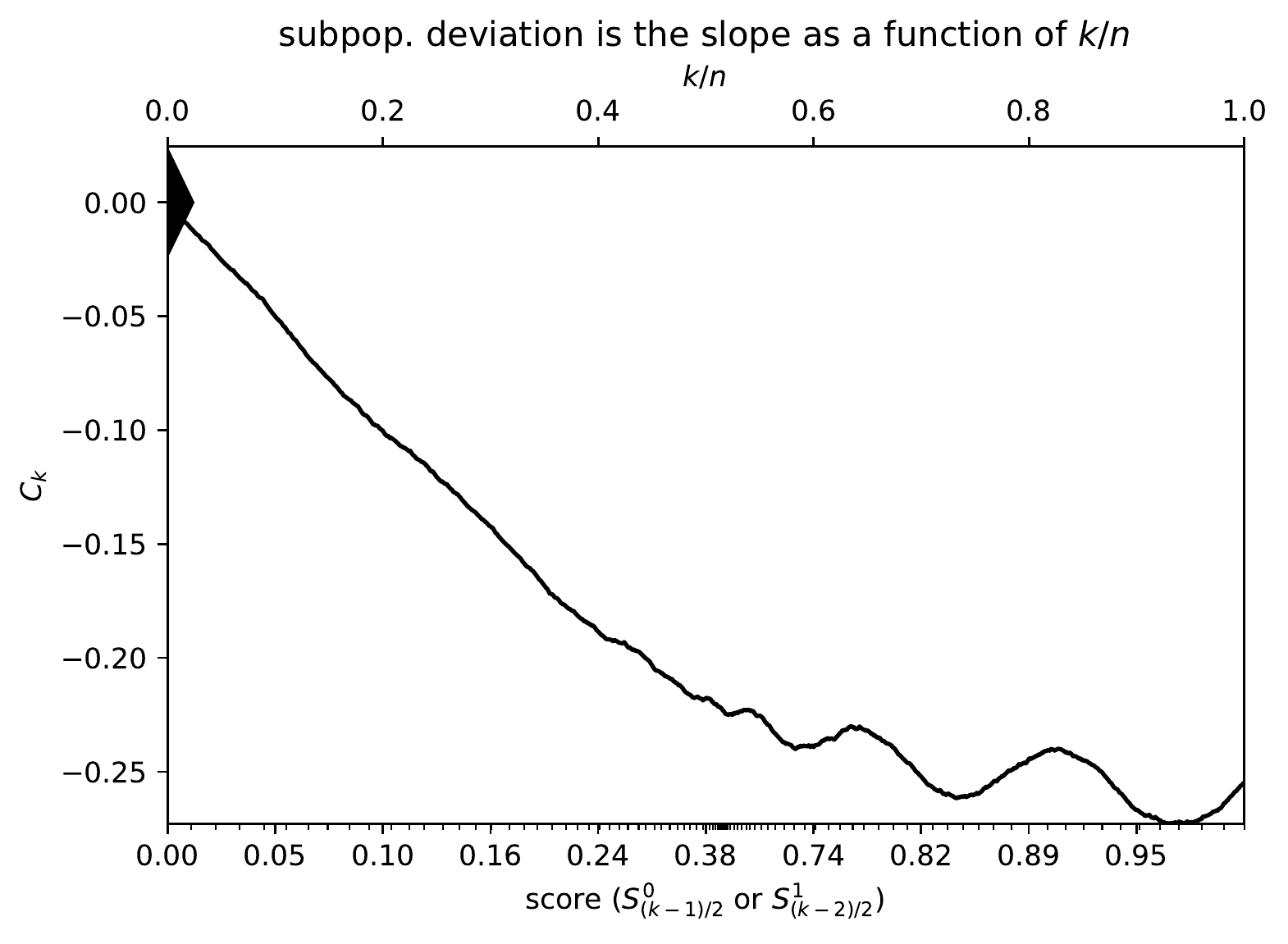}}
\quad\quad
(b)
\parbox{\imsize}{\includegraphics[width=\imsize]
{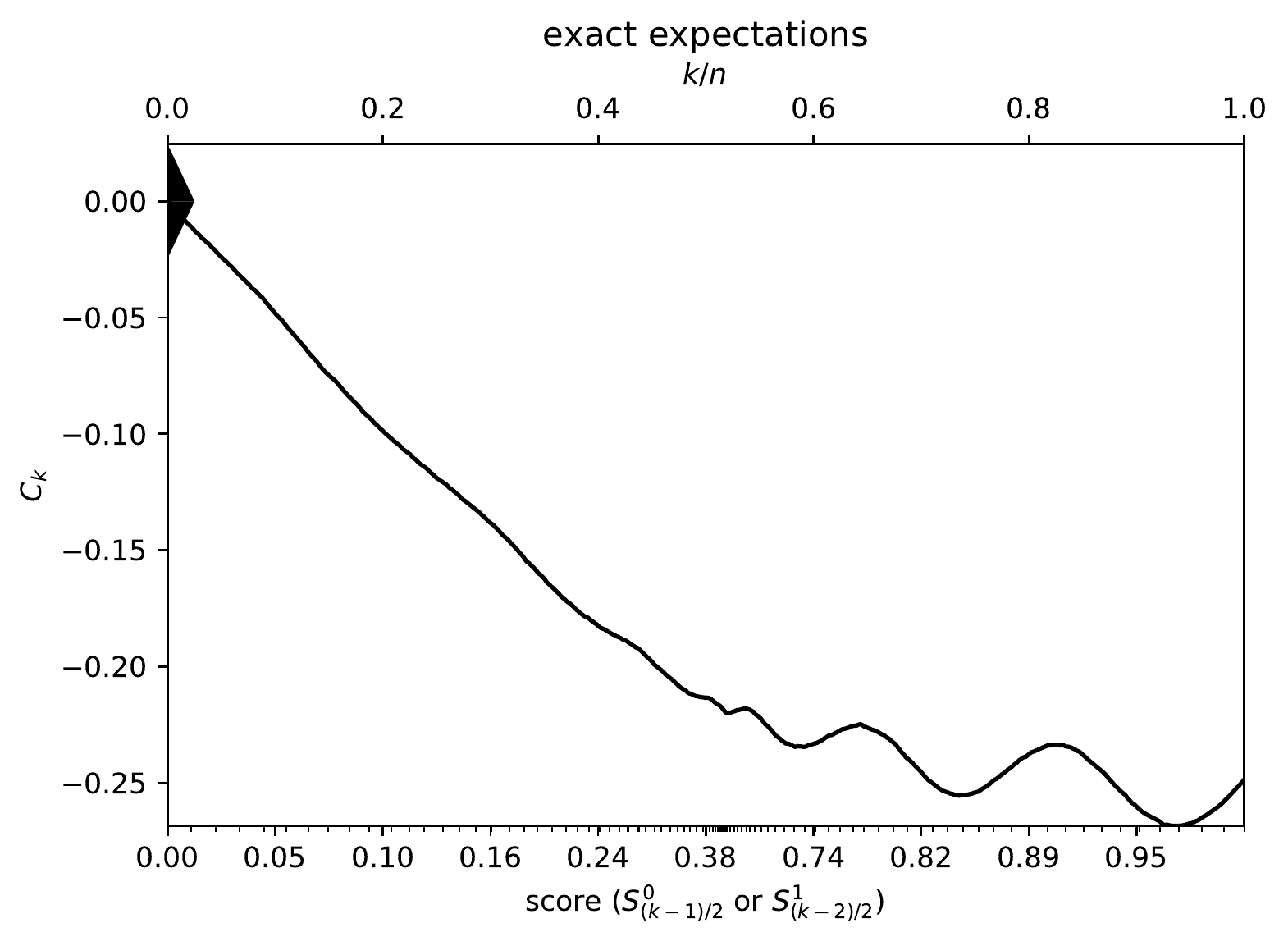}}

\vspace{\vertsep}

(c)
\parbox{\imsize}{\includegraphics[width=\imsize]
{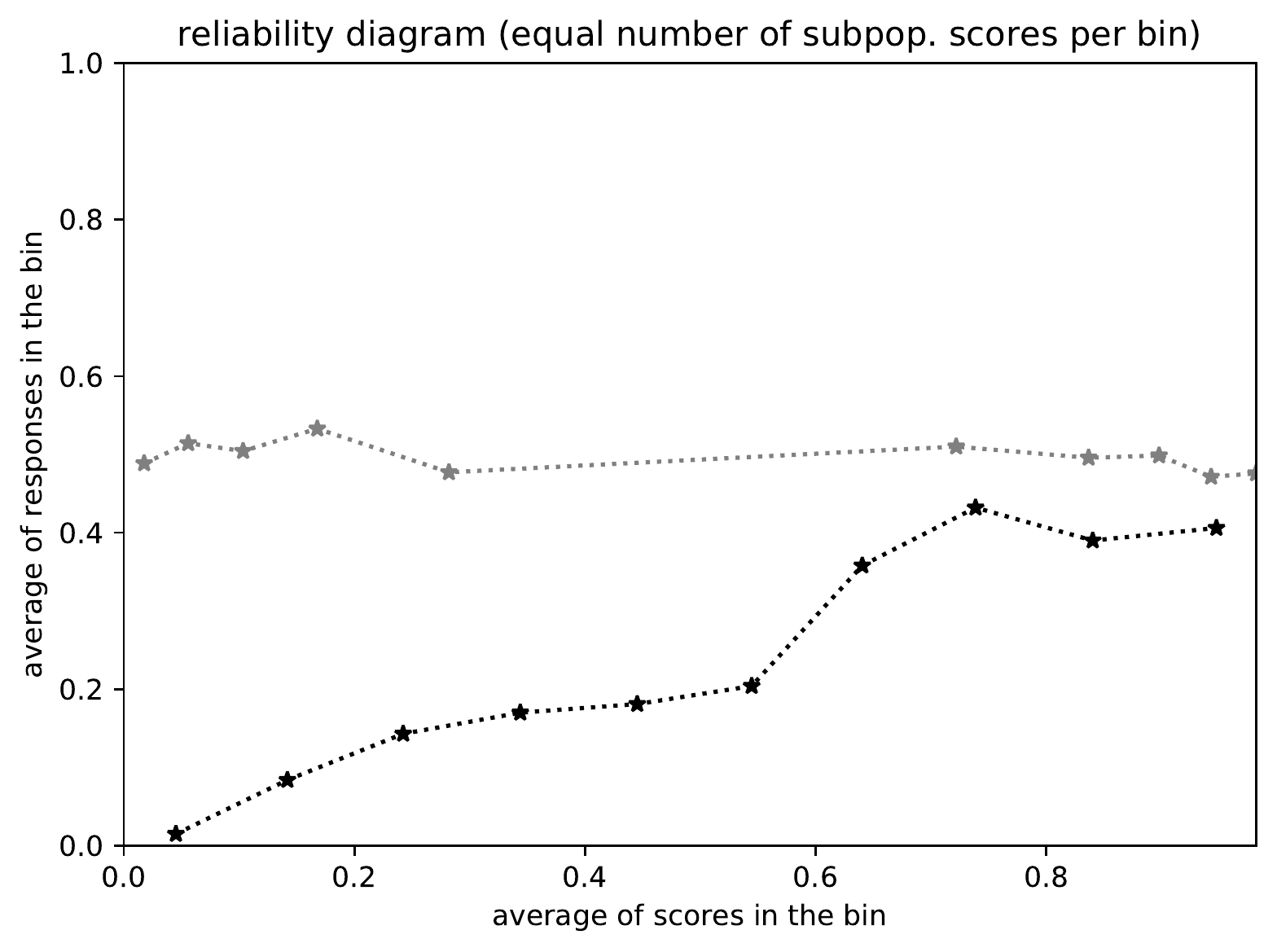}}
\quad\quad
(d)
\parbox{\imsize}{\includegraphics[width=\imsize]
{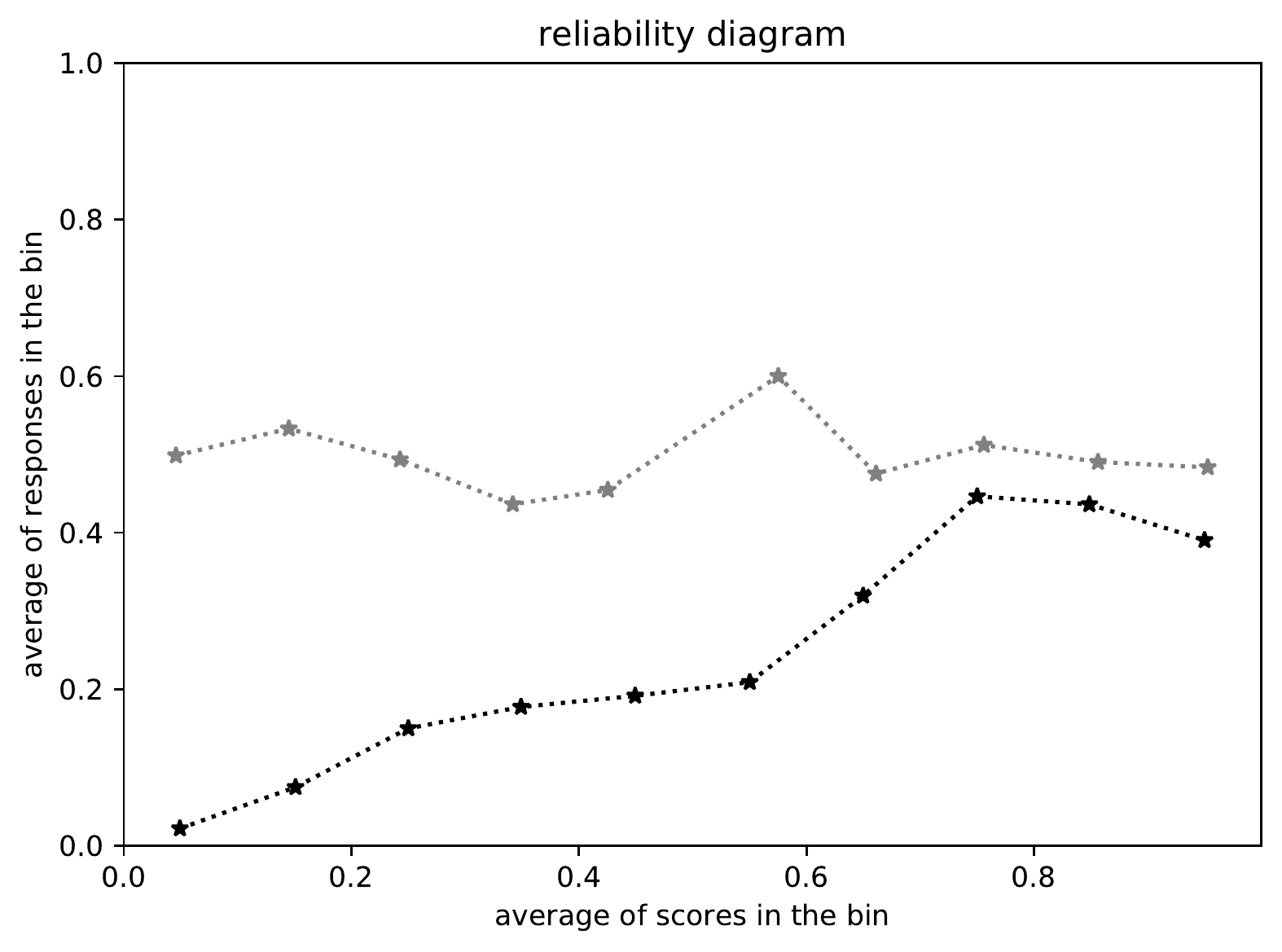}}

\vspace{\vertsep}

(e)
\parbox{\imsize}{\includegraphics[width=\imsize]
{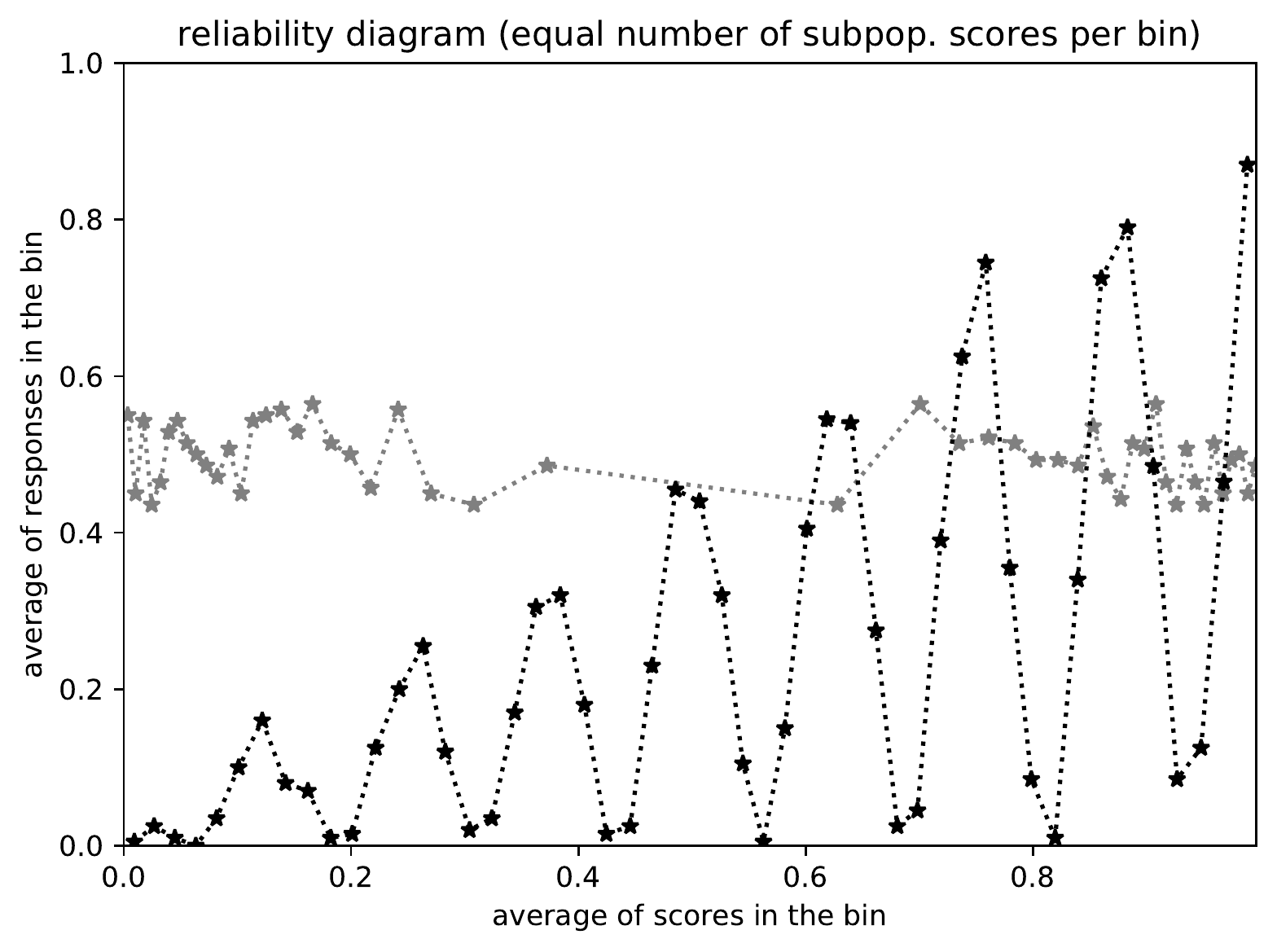}}
\quad\quad
(f)
\parbox{\imsize}{\includegraphics[width=\imsize]
{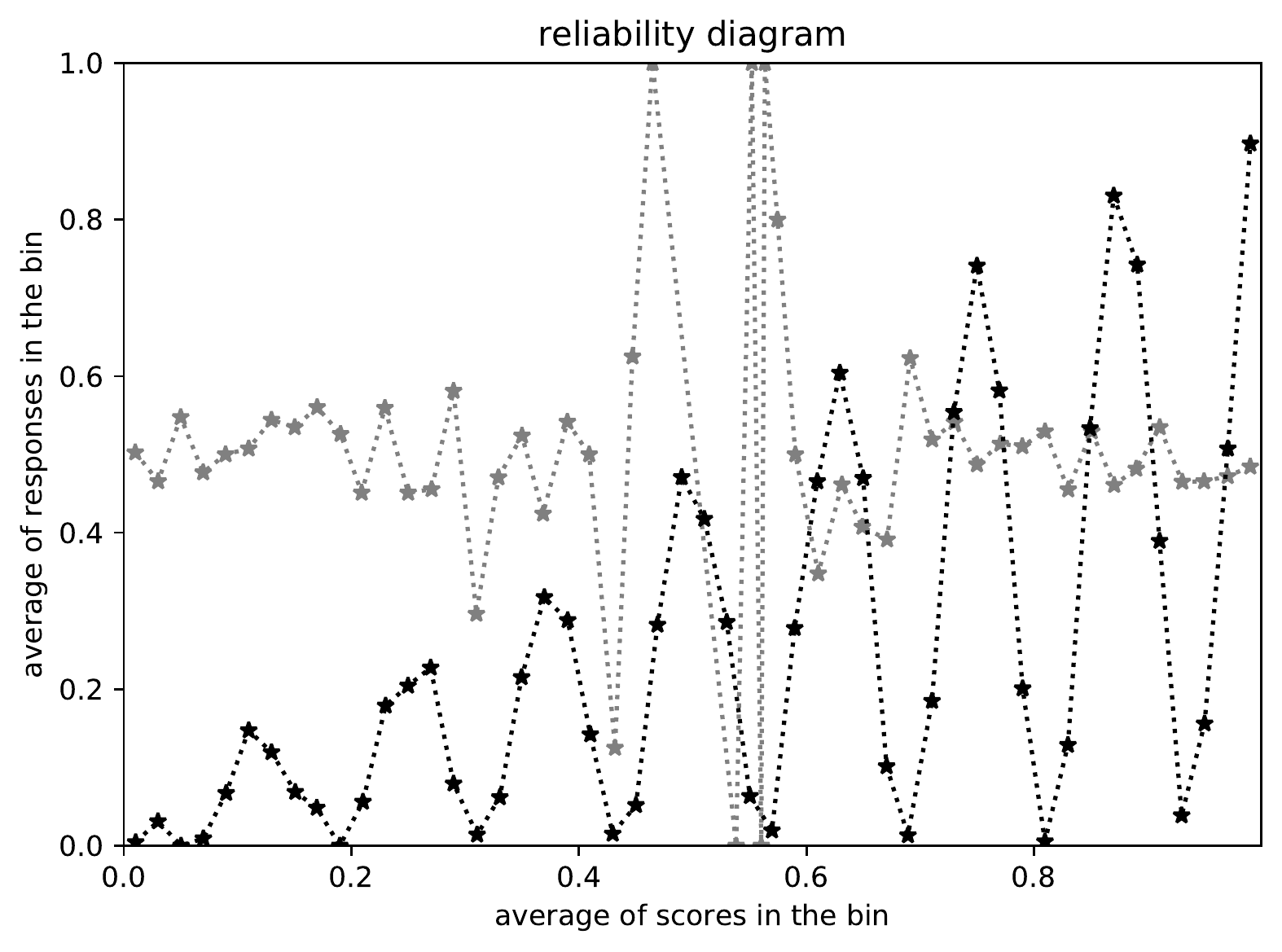}}

\vspace{\vertsep}

(g)
\parbox{\imsize}{\includegraphics[width=\imsize]
{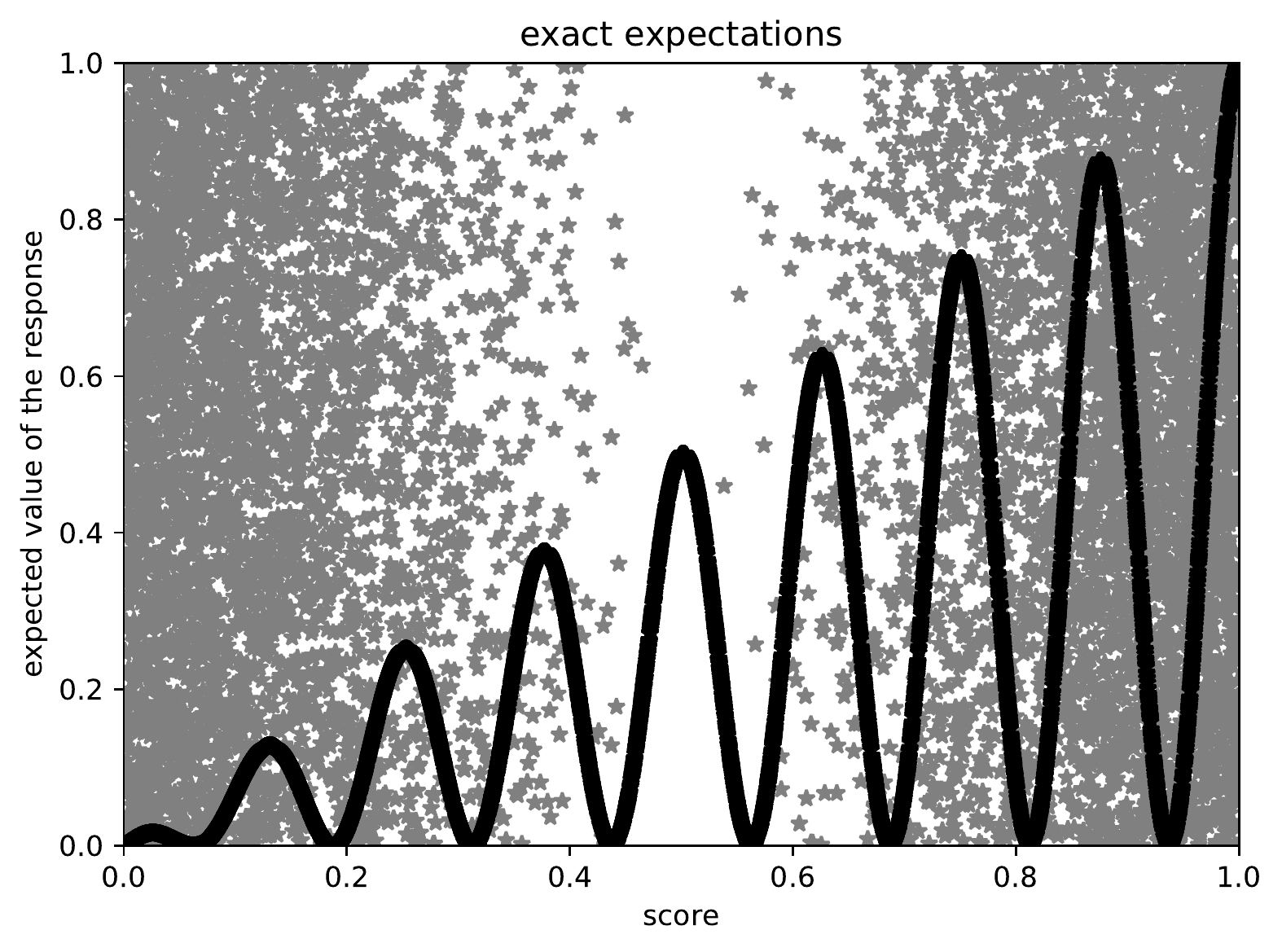}}

\end{centering}
\caption{$n =$ 6,637; Kuiper's statistic is $0.2730 / \sigma = 22.24$,
         Kolmogorov's and Smirnov's is $0.2730 / \sigma = 22.24$;
         the reliability diagrams with 10 bins each (c and d) smooth
         the black curve too much, while the reliability diagrams
         with 50 bins each (e and f) display overly noisy variations
         in the gray curve. The empirical cumulative graph (a) matches
         its ground-truth expectations (b) well, though the oscillations
         at low scores are a bit hard to discern in the cumulative graphs.
         The metrics of Kuiper and of Kolmogorov and Smirnov
         report profoundly statistically significant deviation,
         taking values many times larger than $\sigma$.
}
\label{ex2}
\end{figure}

\begin{figure}
\begin{centering}

(a)
\parbox{\imsize}{\includegraphics[width=\imsize]
{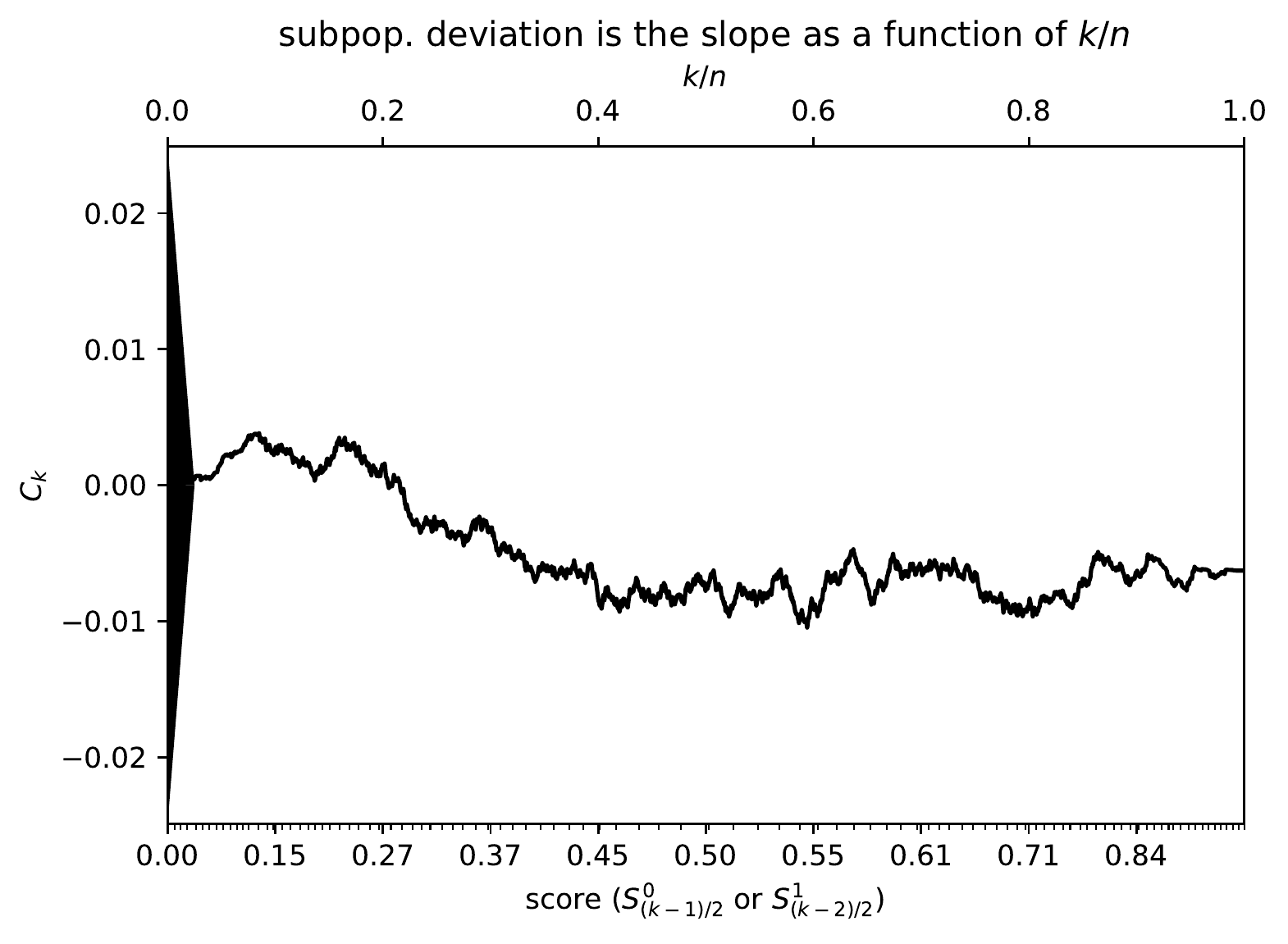}}
\quad\quad
(b)
\parbox{\imsize}{\includegraphics[width=\imsize]
{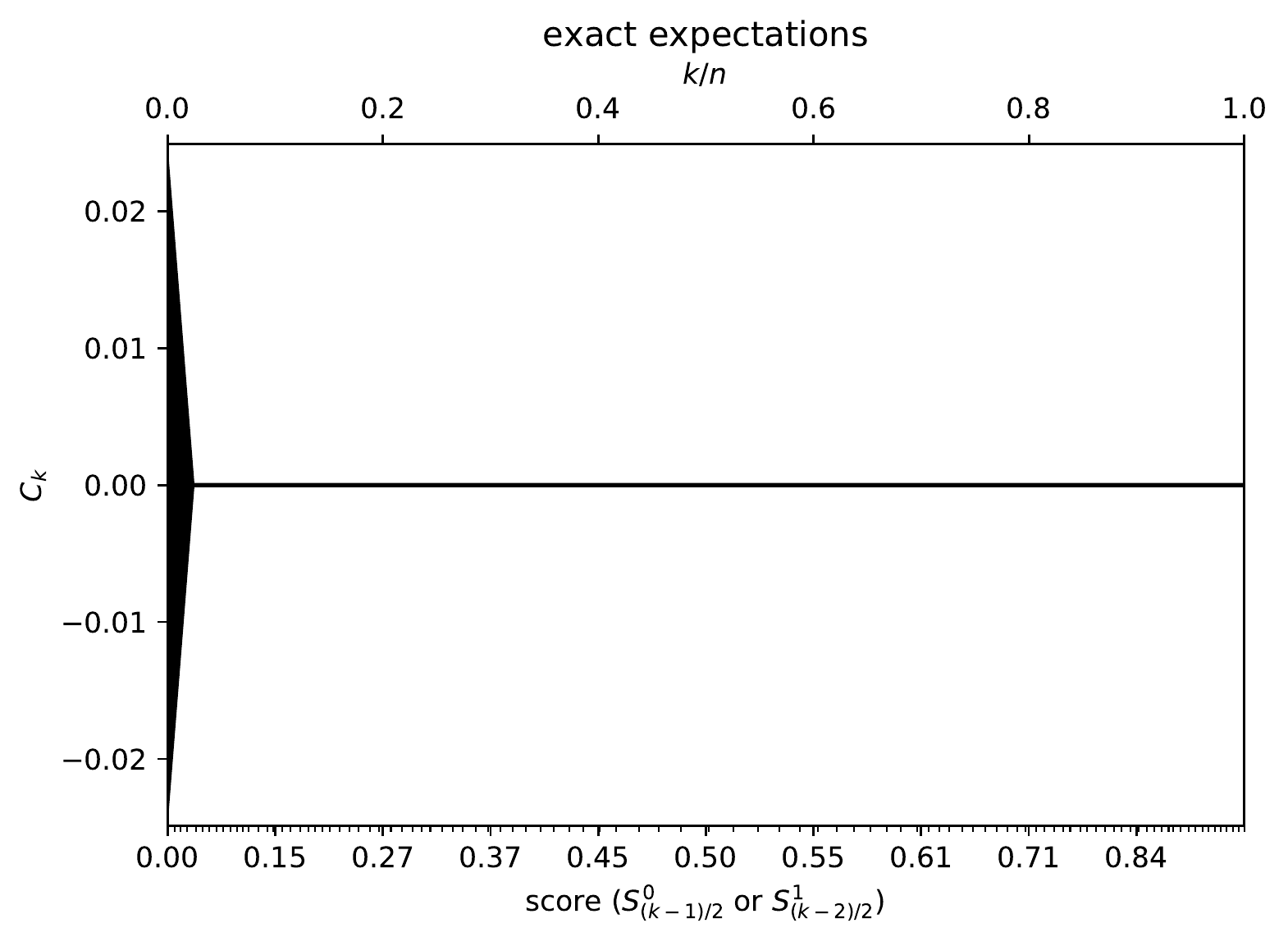}}

\vspace{\vertsep}

(c)
\parbox{\imsize}{\includegraphics[width=\imsize]
{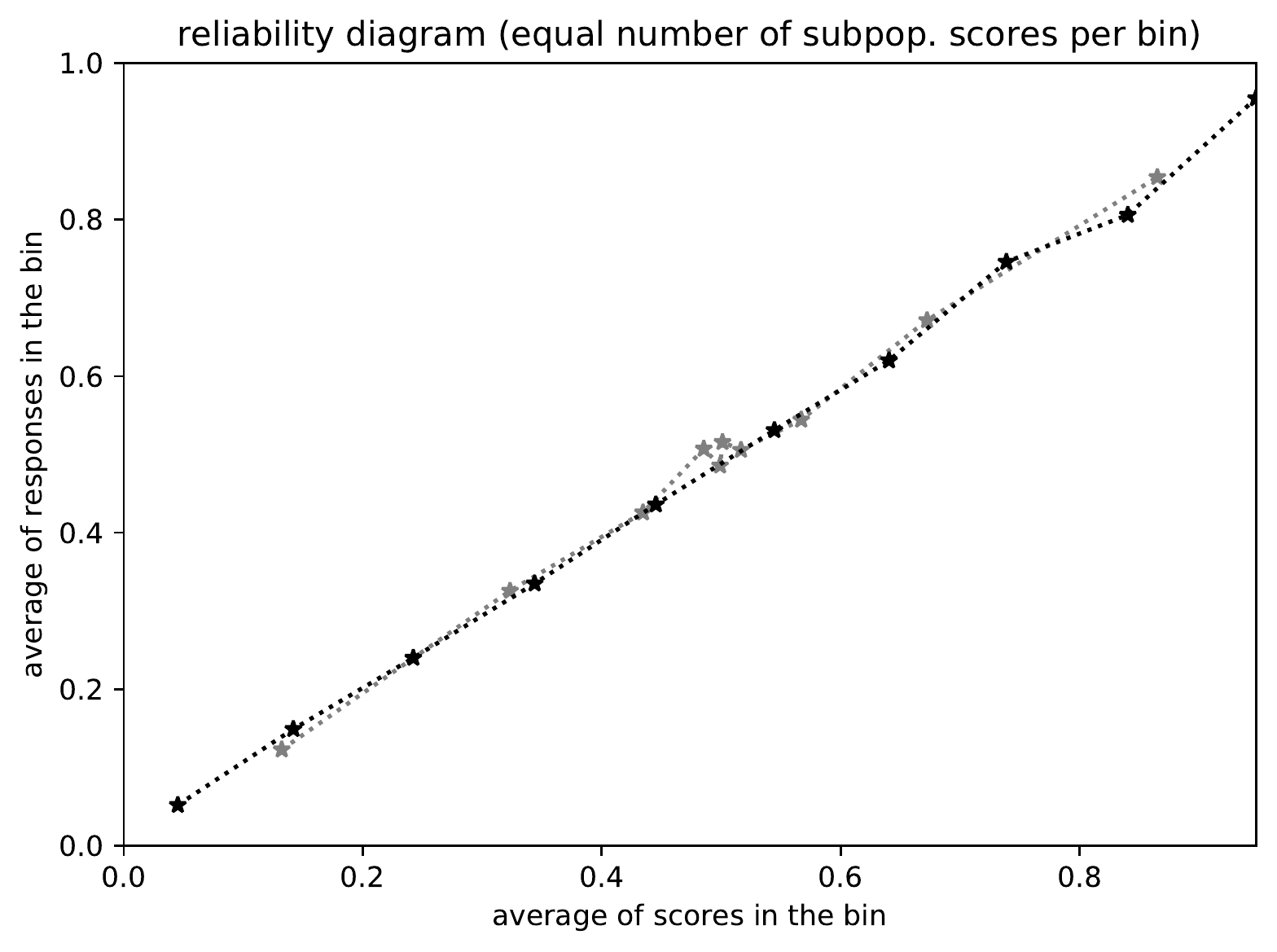}}
\quad\quad
(d)
\parbox{\imsize}{\includegraphics[width=\imsize]
{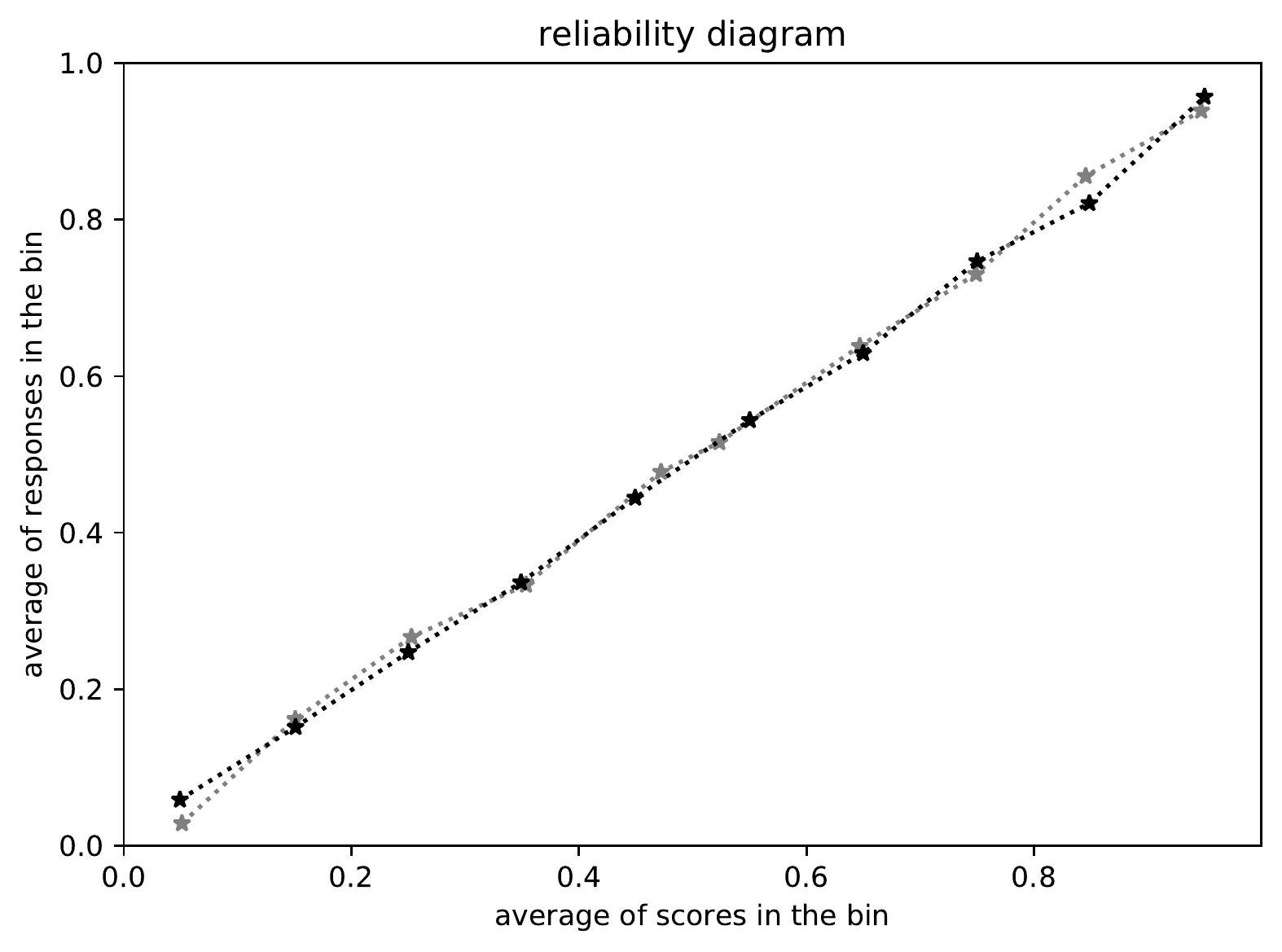}}

\vspace{\vertsep}

(e)
\parbox{\imsize}{\includegraphics[width=\imsize]
{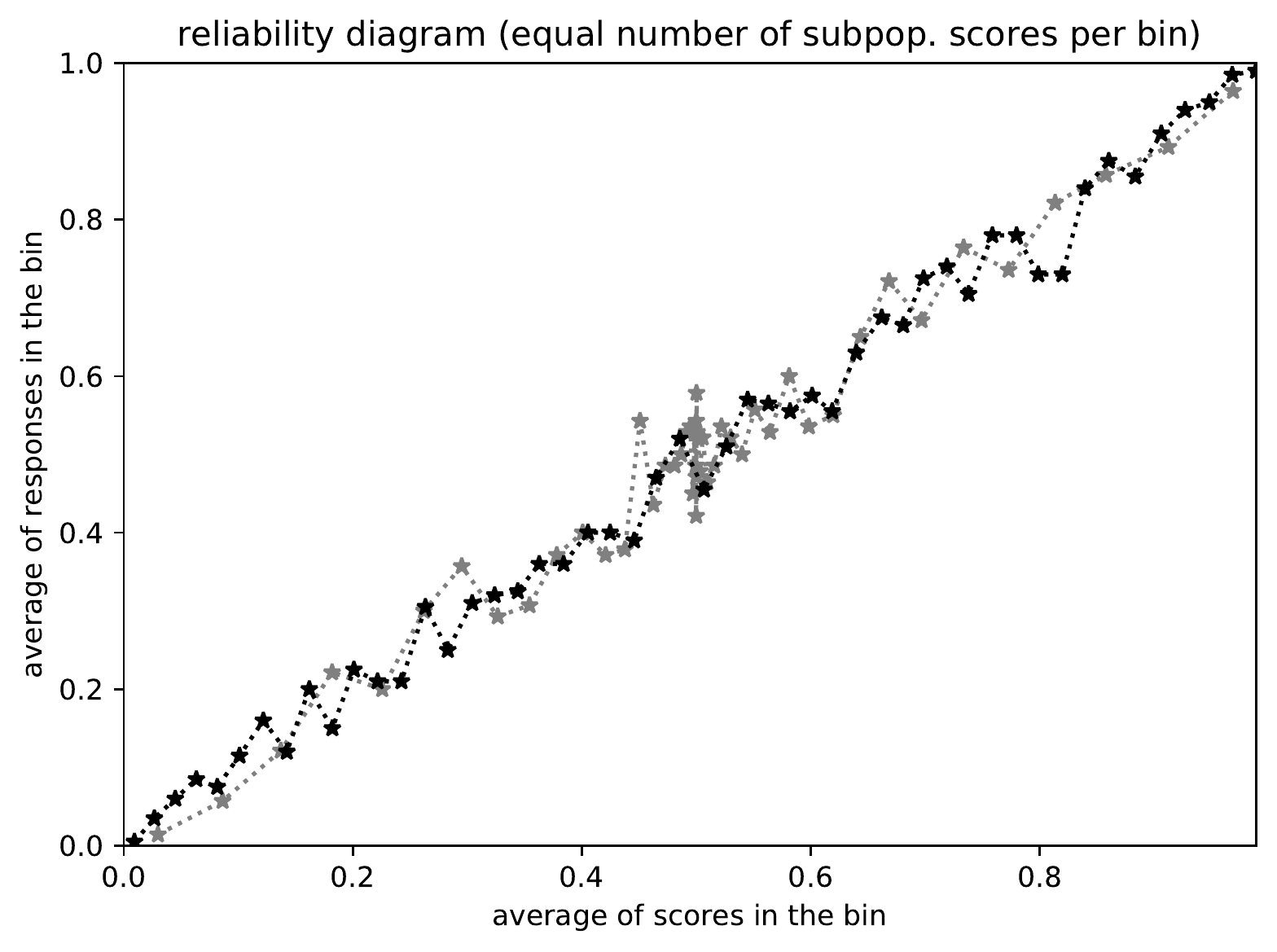}}
\quad\quad
(f)
\parbox{\imsize}{\includegraphics[width=\imsize]
{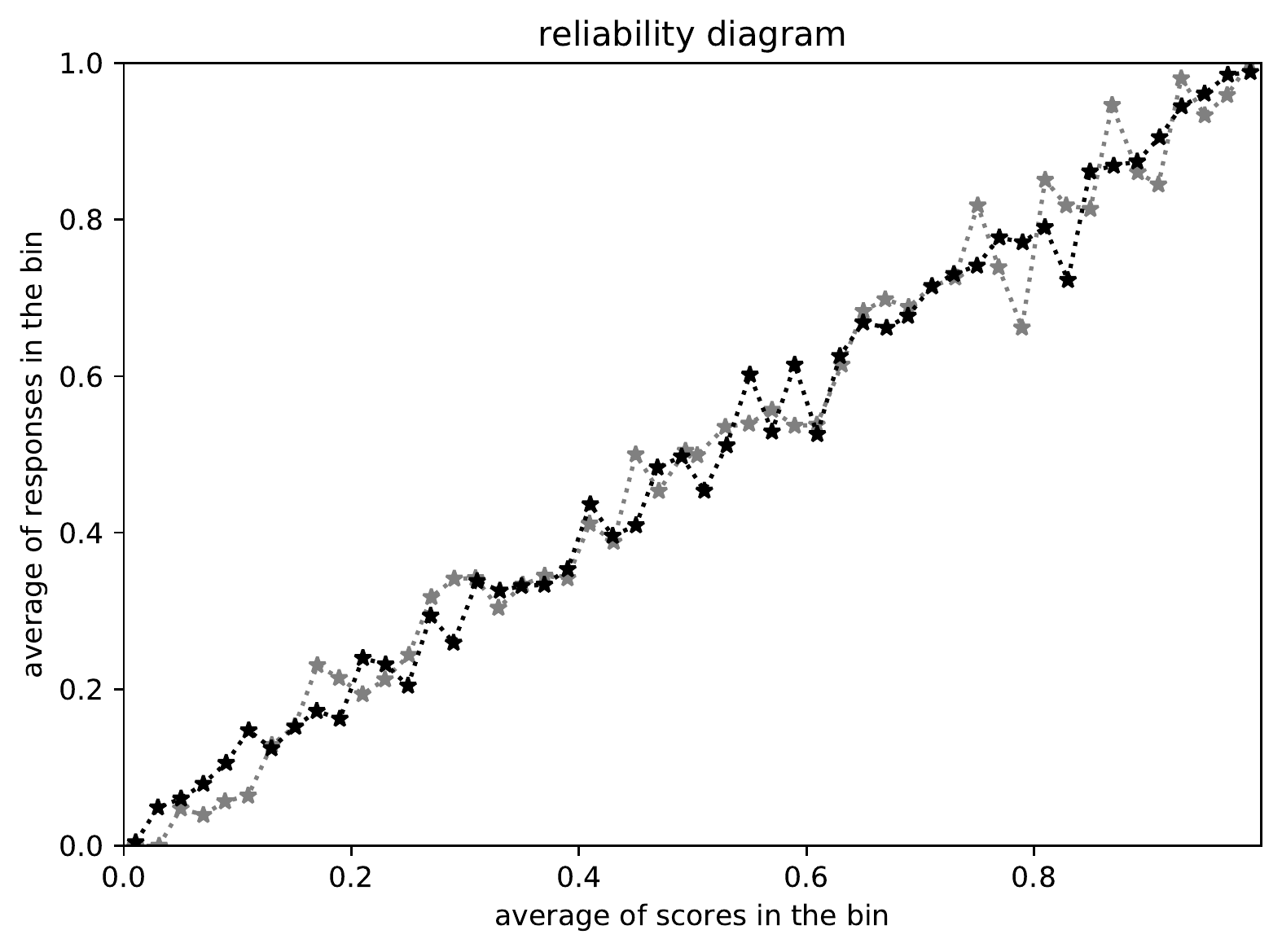}}

\vspace{\vertsep}

(g)
\parbox{\imsize}{\includegraphics[width=\imsize]
{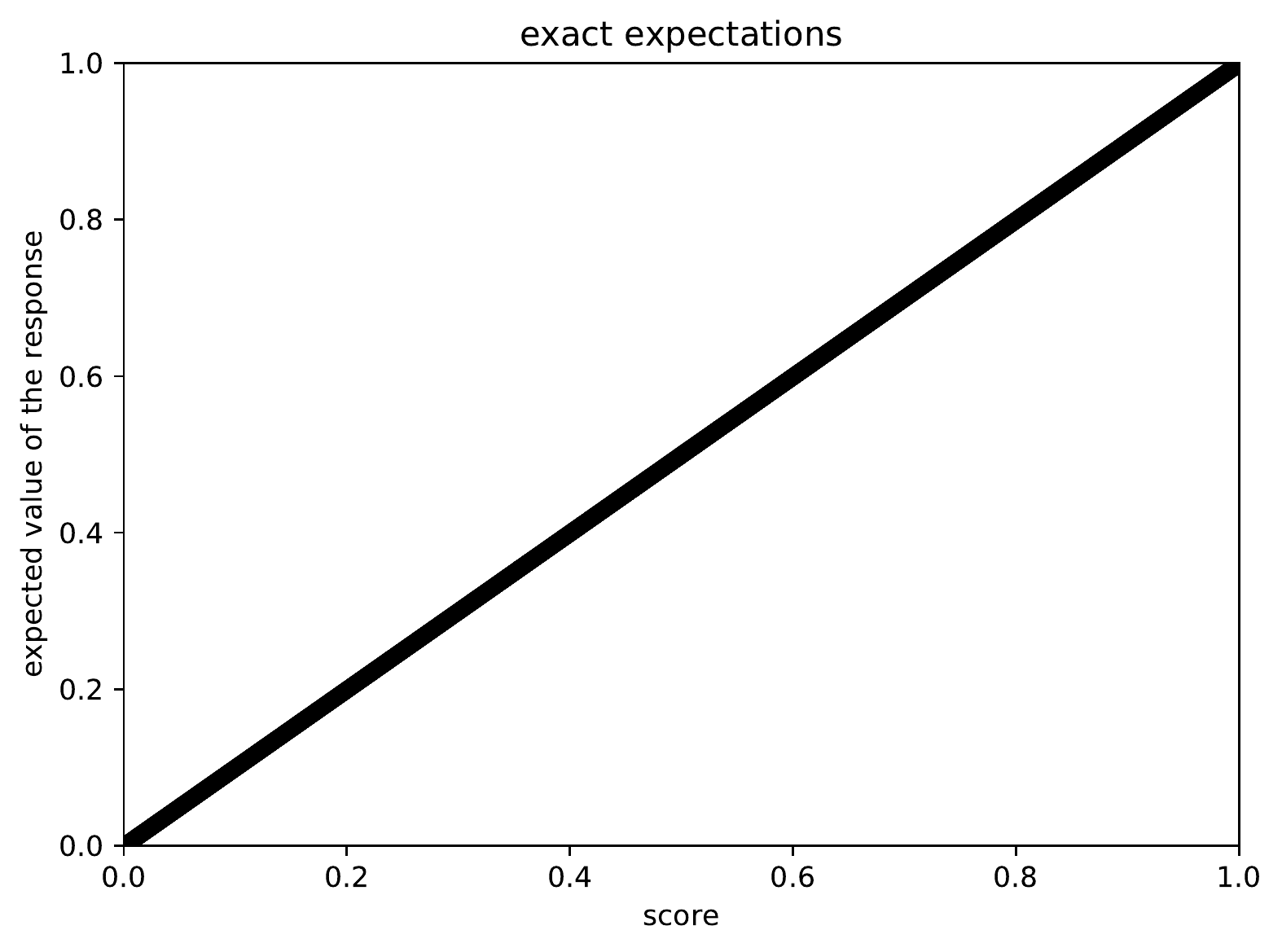}}

\end{centering}
\caption{$n =$ 6,451; Kuiper's statistic is $0.01429 / \sigma = 1.148$,
         Kolmogorov's and Smirnov's is $0.01046 / \sigma = 0.8402$;
         the stochastic variations in the empirical cumulative graph (a)
         are clearly within the expectations indicated by the triangle
         at the origin --- the graph looks like a perfectly random walk,
         and indeed really is a drift-free, perfectly random walk.
         The statistics of Kuiper and of Kolmogorov and Smirnov
         give no indication of any statistically significant deviation
         between the subpopulations, as both are less than $1.25 \sigma$ ---
         the expected value for the metric of Kolmogorov and Smirnov
         in the absence of any deviation between the subpopulations'
         expected responses, as detailed by Remark~2 of~\cite{tygert}.
}
\label{ex3}
\end{figure}

Each top row of Figures~\ref{ex0} and \ref{ex1}--\ref{ex3}
plots $C_1$, $C_2$, \dots, $C_n$
from~(\ref{cumulative}) as a function of $k/n$,
with the rightmost plot displaying its noiseless expected value
rather than using the random observations ($R^0_k$ and $R^1_k$).
(Technically speaking, the top row of Figure~\ref{ex2} actually plots
$-C_1$, $-C_2$, \dots, $-C_n$, since for Figure~\ref{ex2}
the second subpopulation ends up being subpopulation 0
in the notation of Section~\ref{methods}.)
Each bottom row of Figures~\ref{ex0} and \ref{ex1}--\ref{ex3} plots the pairs
of scores and expected values for the first subpopulation in black,
and plots the pairs for the second subpopulation in gray,
producing ground-truth diagrams that the middle two rows of plots
are trying to estimate using only the observations,
without access to the underlying probabilities.

The first three examples include substantial deviations
in the expected responses between the two subpopulations,
while the fourth example omits any deviation
in the expected responses between the two subpopulations.
The first three examples illustrate how well the various plots can detect
substantial deviations, while the fourth example illustrates how the plots look
in the absence of any deviation.

For the first example, corresponding to Figure~\ref{ex0},
the scores for the first subpopulation
are $0.5 (1 + 2^3 (x - 0.5)^3)$ for 10,000 values of $x$
drawn uniformly at random from the unit interval $[0, 1]$,
whereas the scores for the second subpopulation are 7,000 values
drawn uniformly at random from the unit interval $[0, 1]$
(the latter values are also equal to $0.5 (1 + 2 (x - 0.5))$
for 7,000 values of $x$ drawn uniformly at random from the unit interval
$[0, 1]$).
The expected values are as indicated in the lowermost plot of Figure~\ref{ex0},
with the expected values for each subpopulation varying smoothly
as a function of the score, aside from swapping the values between
the two subpopulations for scores in a short range near 0.9.
The deviation in the expected values between the subpopulations
is substantial for this example.

For the second example, corresponding to Figure~\ref{ex1},
the scores for the first subpopulation
are $x^5$ for 10,000 values of $x$
drawn uniformly at random from the unit interval $[0, 1]$,
whereas the scores for the second subpopulation are 7,000 values
drawn uniformly at random from the unit interval $[0, 1]$.
The expected values are as indicated in the lowermost plot of Figure~\ref{ex1},
with several discontinuities in the expected values.
The deviation in the expected values between the subpopulations
is substantial for this example, too.

For the third example, corresponding to Figure~\ref{ex2},
the scores for the first subpopulation
are $0.5 (1 + 2^{1/3} (x - 0.5)^{1/3})$ for 10,000 values of $x$
drawn uniformly at random from the unit interval $[0, 1]$,
whereas the scores for the second subpopulation are 7,000 values
drawn uniformly at random from the unit interval $[0, 1]$
(the latter values are also equal to $0.5 (1 + 2 (x - 0.5))$
for 7,000 values of $x$ drawn uniformly at random from the unit interval
$[0, 1]$).
The lowermost plot of Figure~\ref{ex2} displays the expected values,
with the expected values for the first subpopulation varying sinusoidally
within an envelope bounded below by 0 and bounded above by the diagonal line
on the plot extending from the origin $(0, 0)$ to the point $(1, 1)$,
and with the expected values for the second subpopulation drawn uniformly
at random from the unit interval $[0, 1]$.
The deviation in the expected values between the subpopulations
is substantial for this example, as well.

For the fourth example, corresponding to Figure~\ref{ex3},
the scores are the same as in the first example,
and the expected values are equal to the scores.
Since the expected values are equal to the scores, the expected values
are given by the same function of the score for both subpopulations,
and thus there is no deviation between the expected responses
for the subpopulations in this example.

The captions of the figures comment on the numerical results displayed.

\subsection{ImageNet}
\label{imagenetex}

\begin{figure}
\begin{centering}

(a)
\parbox{\imsize}{\includegraphics[width=\imsize]
{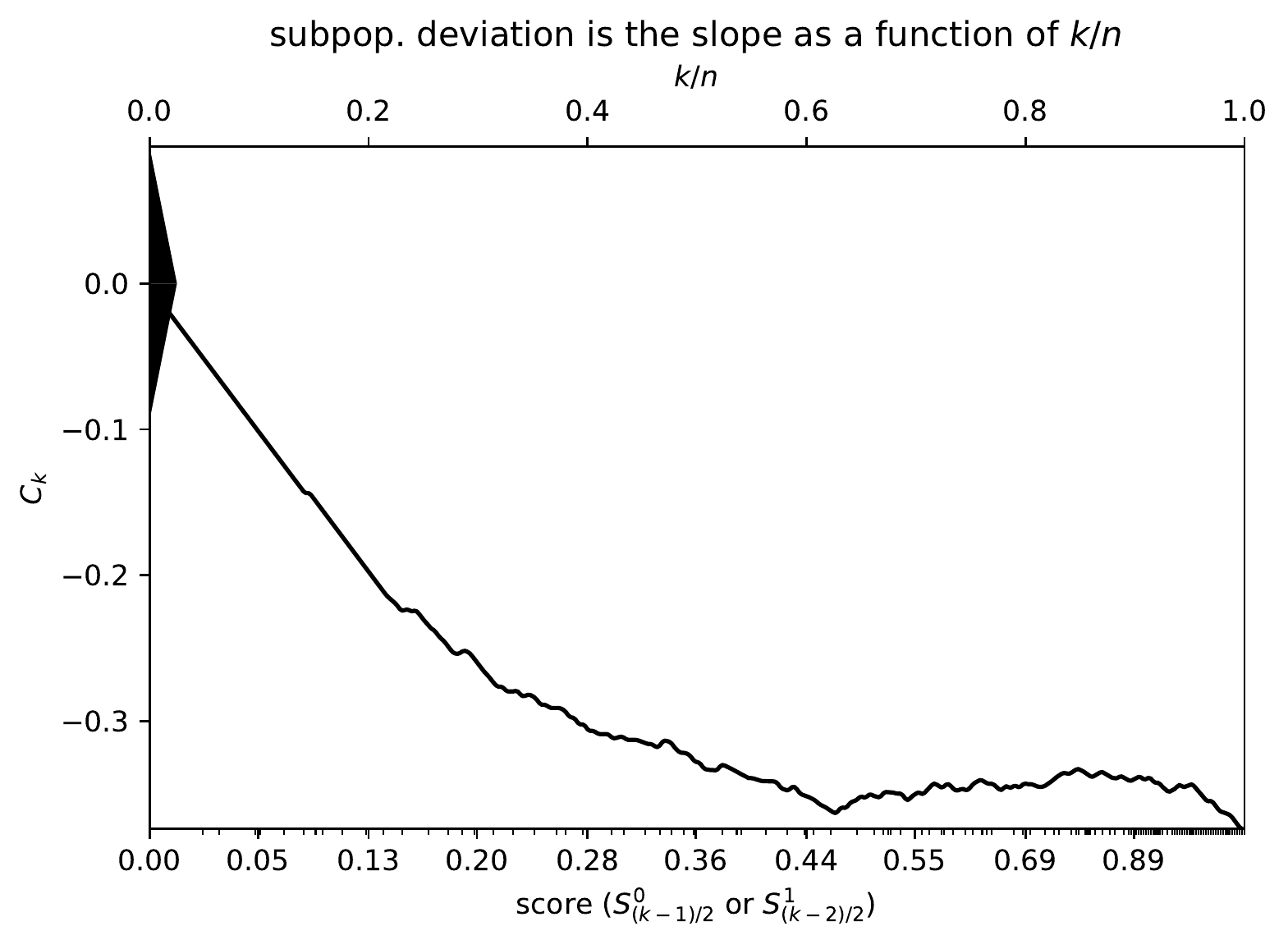}}

\vspace{\vertsep}

(b)
\parbox{\imsize}{\includegraphics[width=\imsize]
{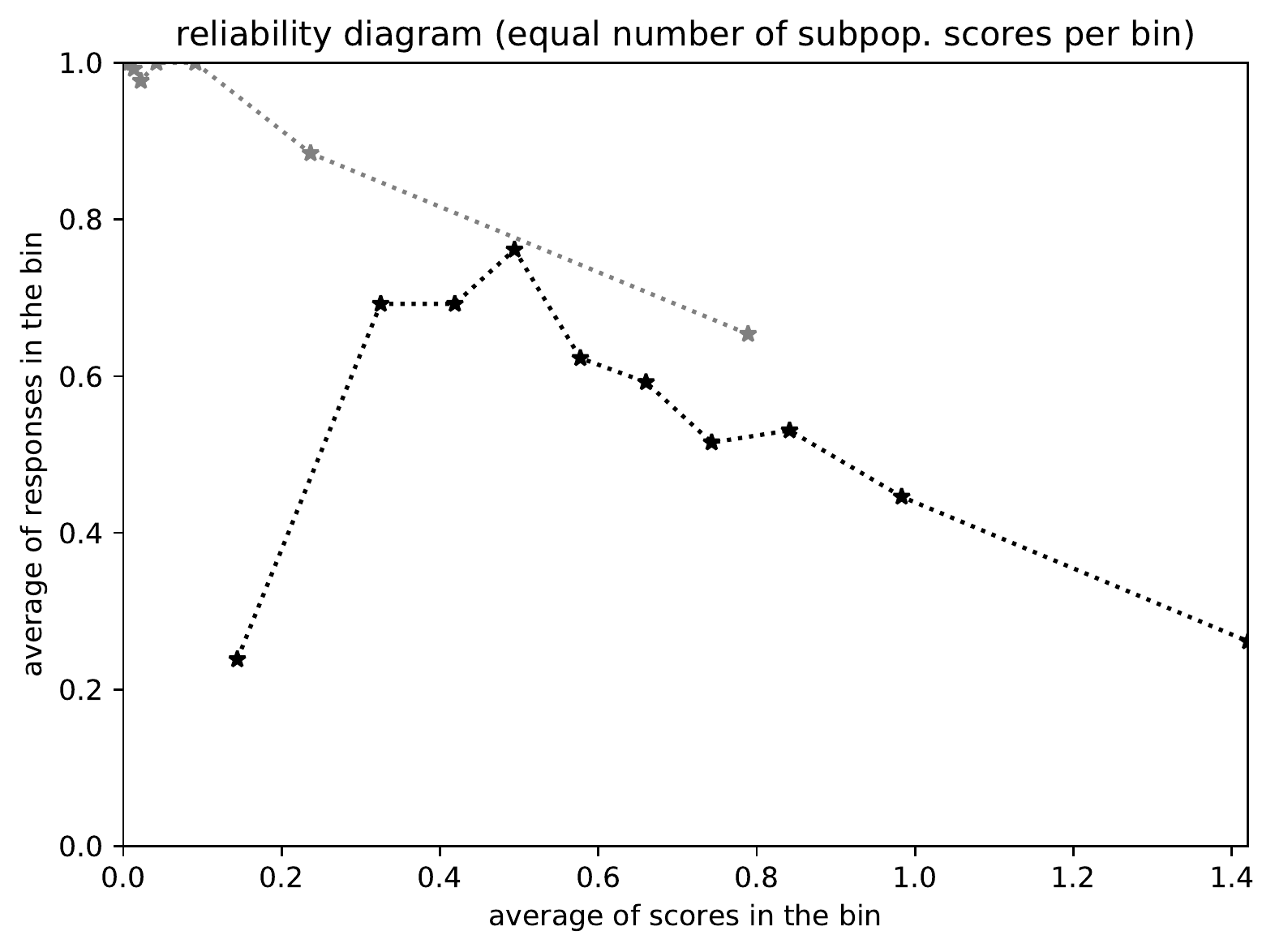}}
\quad\quad
(c)
\parbox{\imsize}{\includegraphics[width=\imsize]
{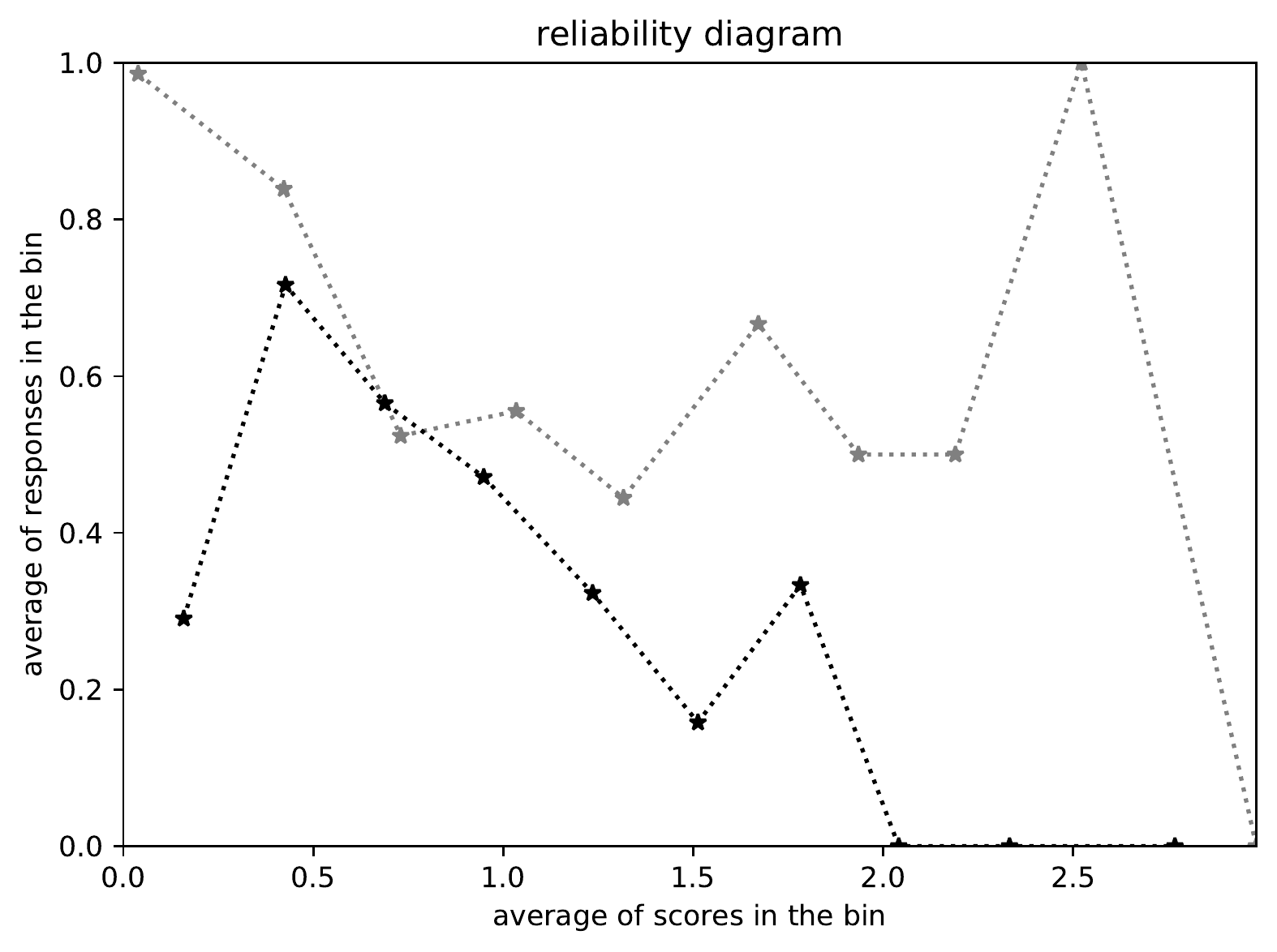}}

\vspace{\vertsep}

(d)
\parbox{\imsize}{\includegraphics[width=\imsize]
{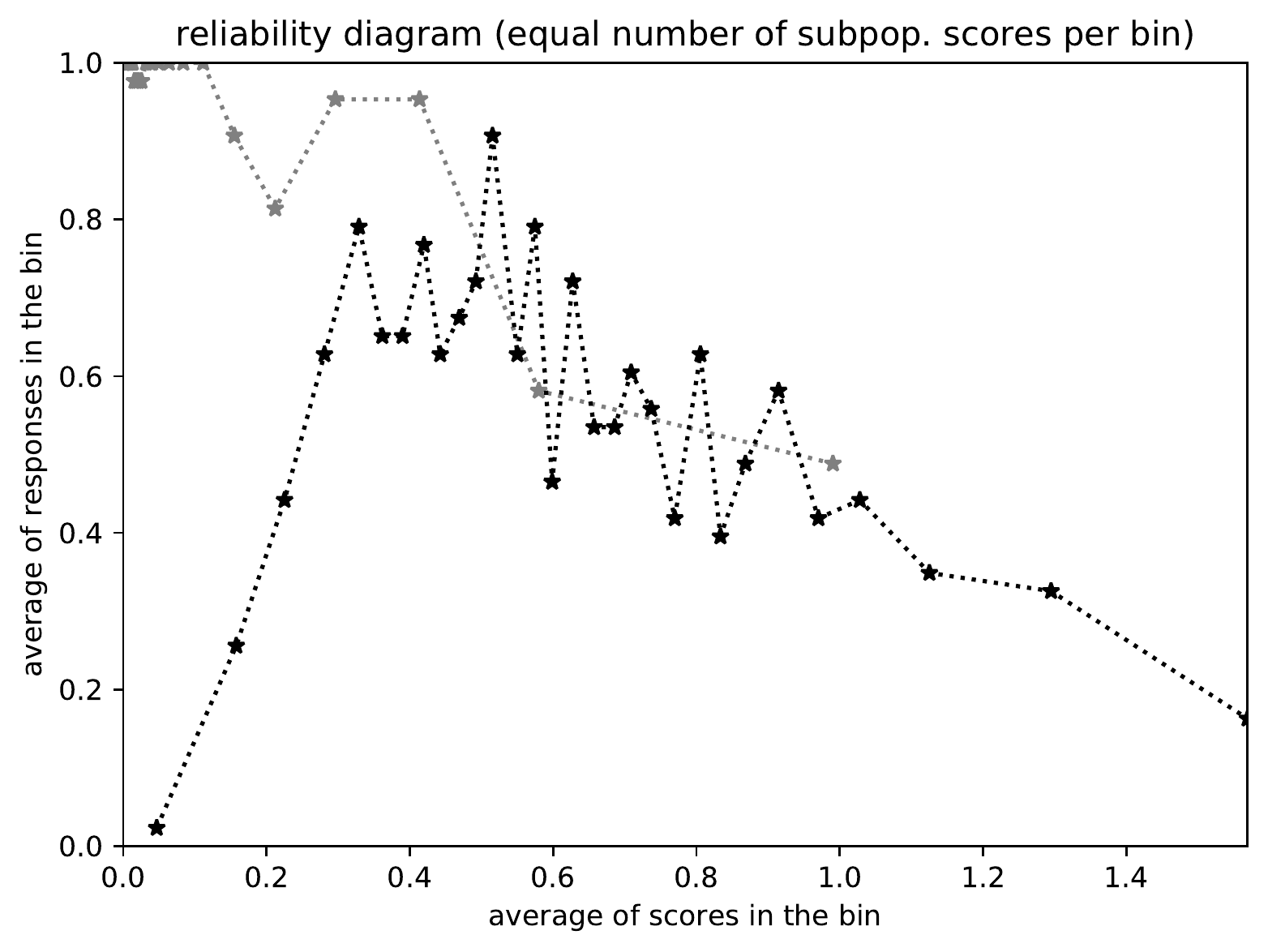}}
\quad\quad
(e)
\parbox{\imsize}{\includegraphics[width=\imsize]
{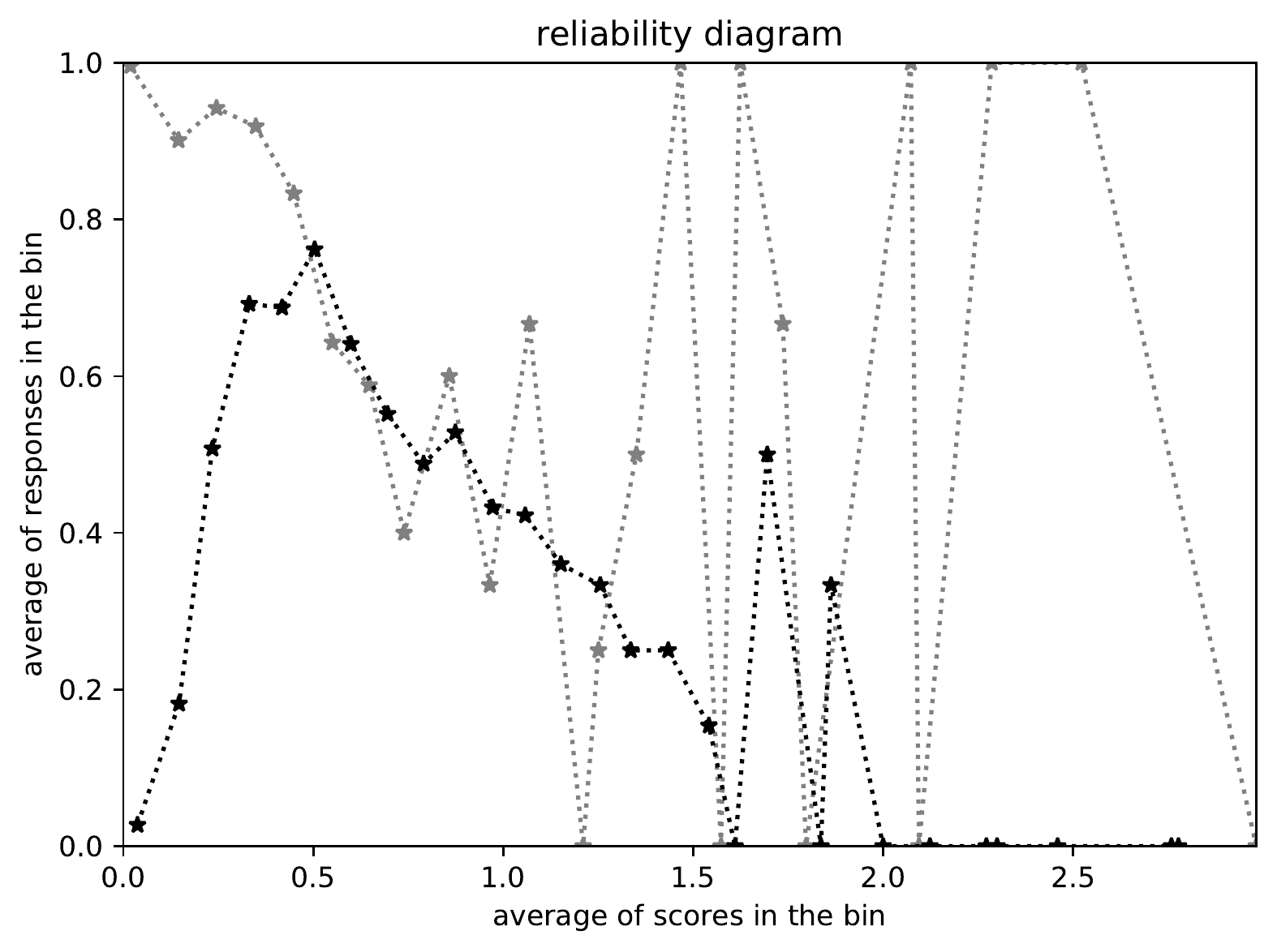}}

\vspace{\vertsep}

(f)
\parbox{\imsize}{\includegraphics[width=\imsize]
{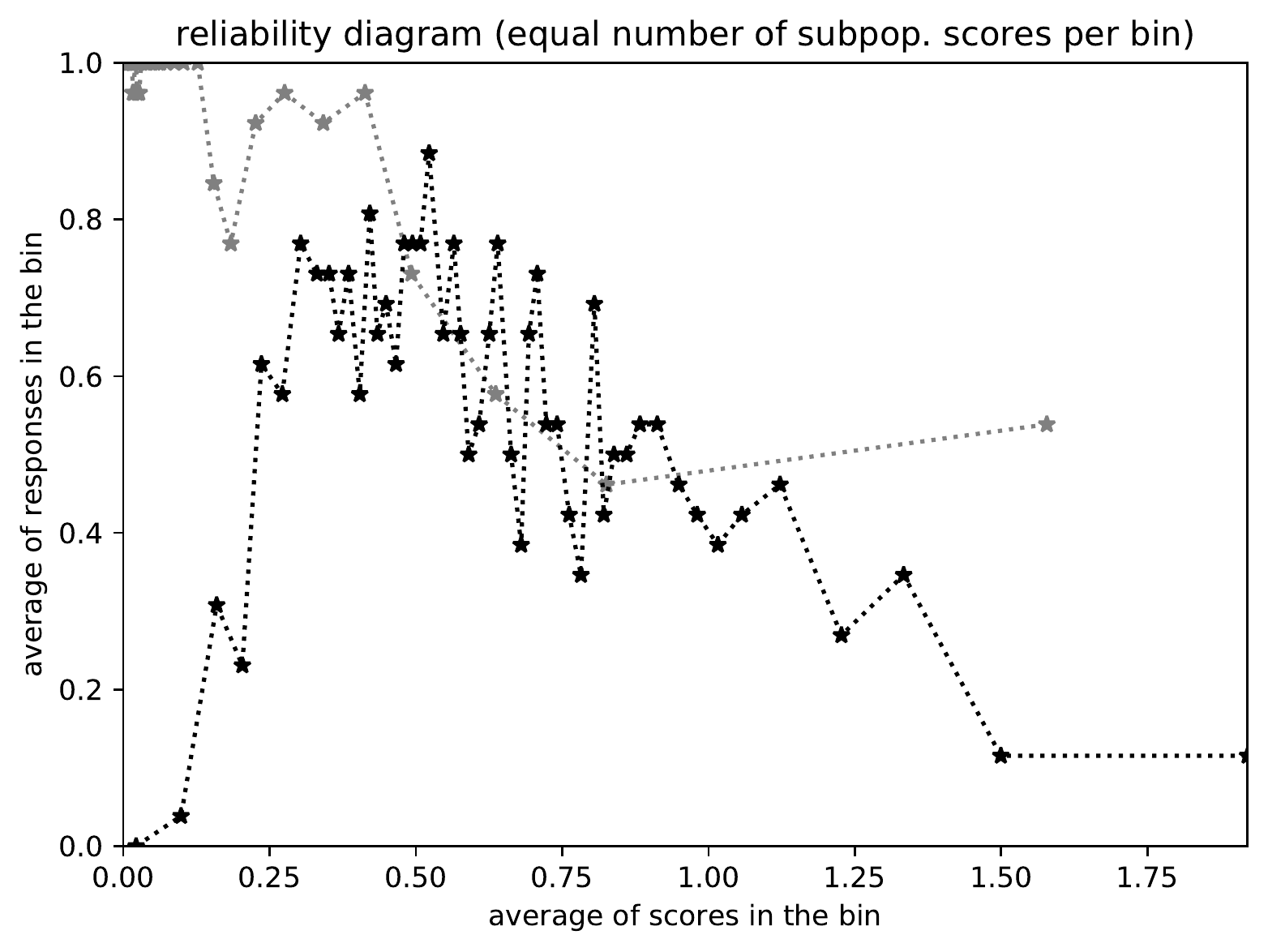}}
\quad\quad
(g)
\parbox{\imsize}{\includegraphics[width=\imsize]
{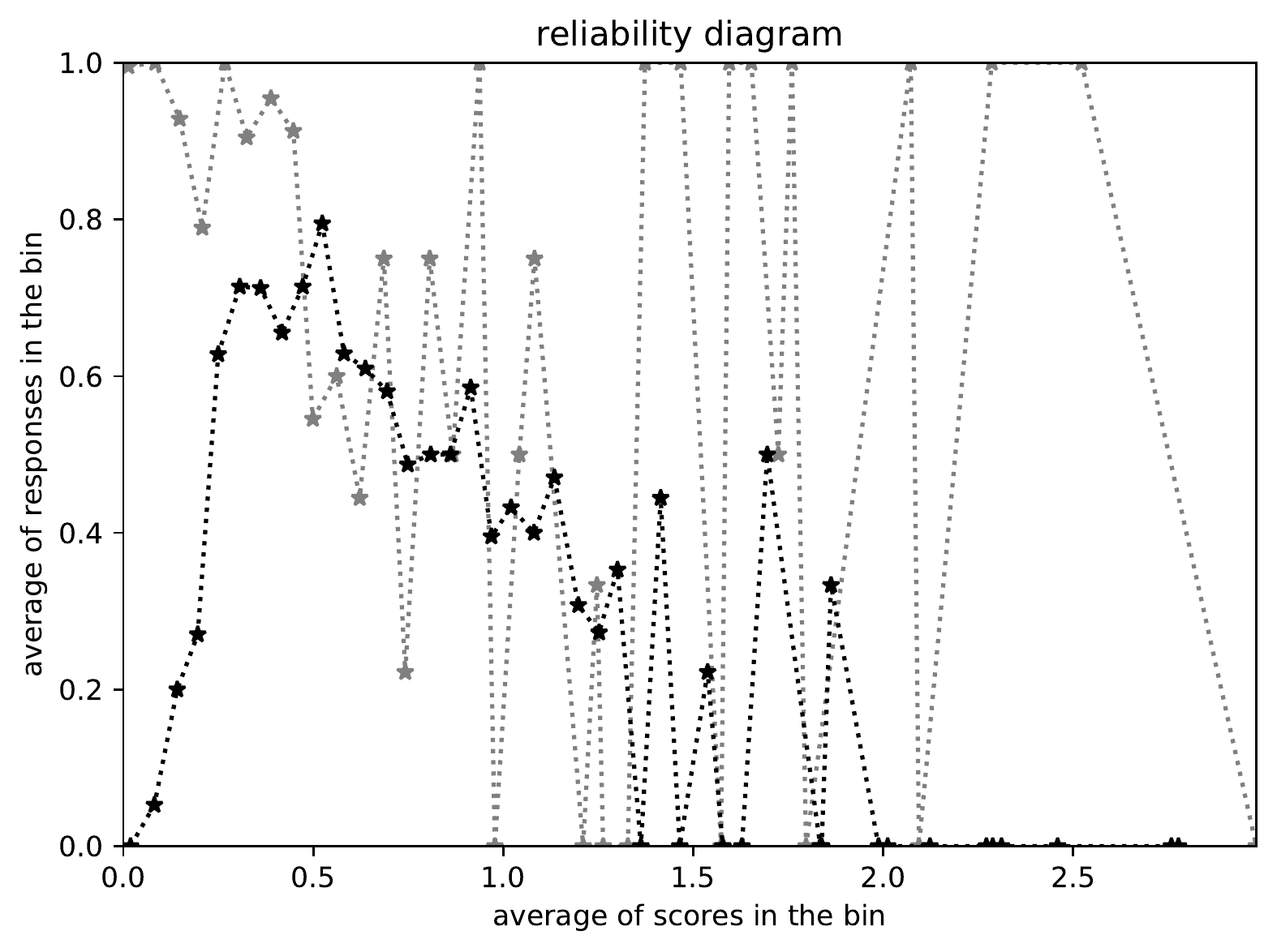}}

\end{centering}
\caption{Eskimo Dog (or Husky) vs.\ Cheetah (Acinonyx jubatus); $n =$ 455;
         Kuiper's statistic is $0.3738 / \sigma = 7.974$,
         Kolmogorov's and Smirnov's is $0.3738 / \sigma = 7.974$;
         in this case, the reliability diagrams with many bins can resolve
         the phenomena displayed in the graph of cumulative differences (a),
         but only by sacrificing confidence in their estimates,
         as they exhibit wild fluctuations.
         The metrics of Kuiper and of Kolmogorov and Smirnov both report
         extremely statistically significant deviations
         between the subpopulations.
}
\label{Eskimo-dog-husky_cheetah}
\end{figure}

\begin{figure}
\begin{centering}

(a)
\parbox{\imsize}{\includegraphics[width=\imsize]
{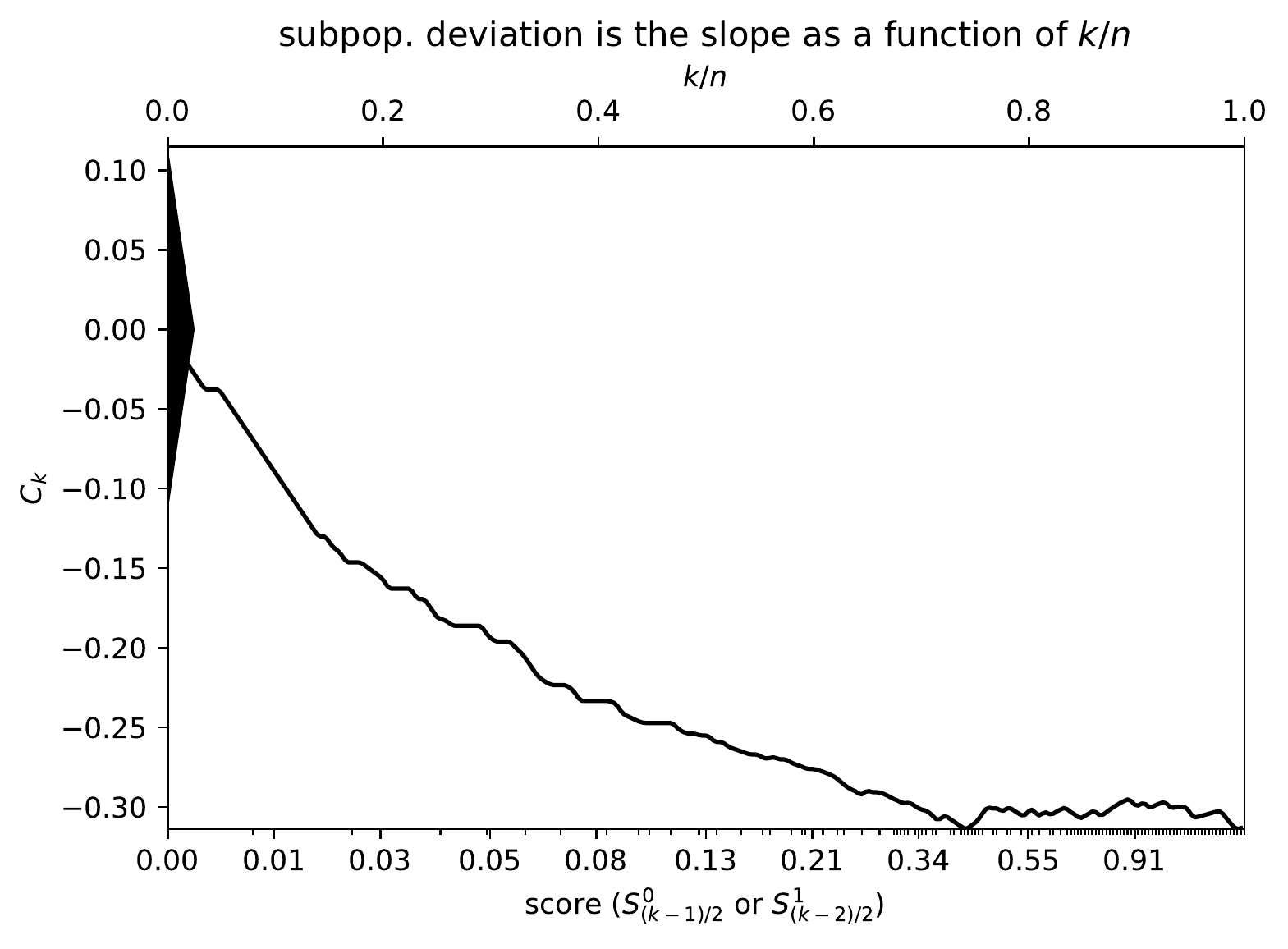}}

\vspace{\vertsep}

(b)
\parbox{\imsize}{\includegraphics[width=\imsize]
{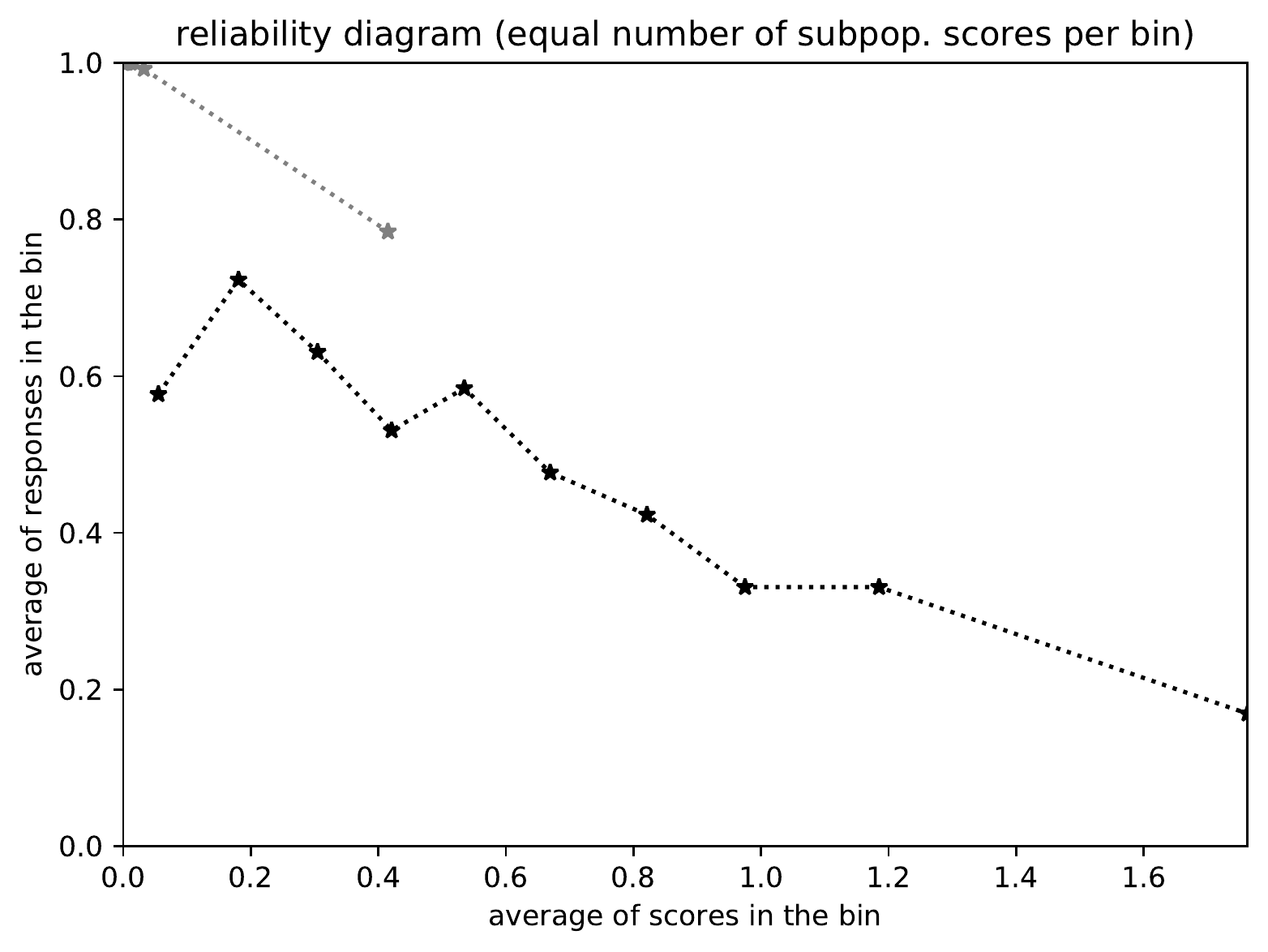}}
\quad\quad
(c)
\parbox{\imsize}{\includegraphics[width=\imsize]
{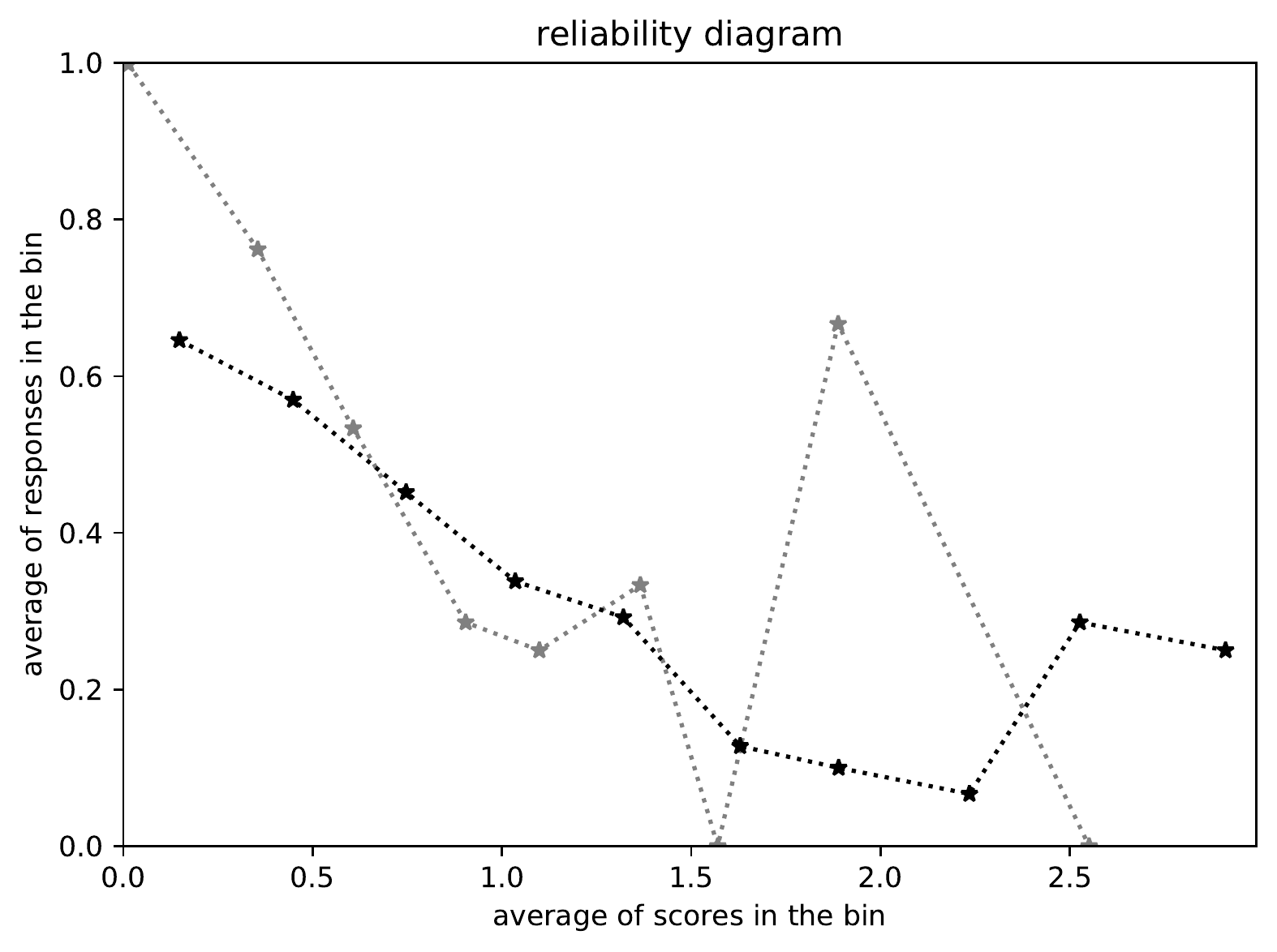}}

\vspace{\vertsep}

(d)
\parbox{\imsize}{\includegraphics[width=\imsize]
{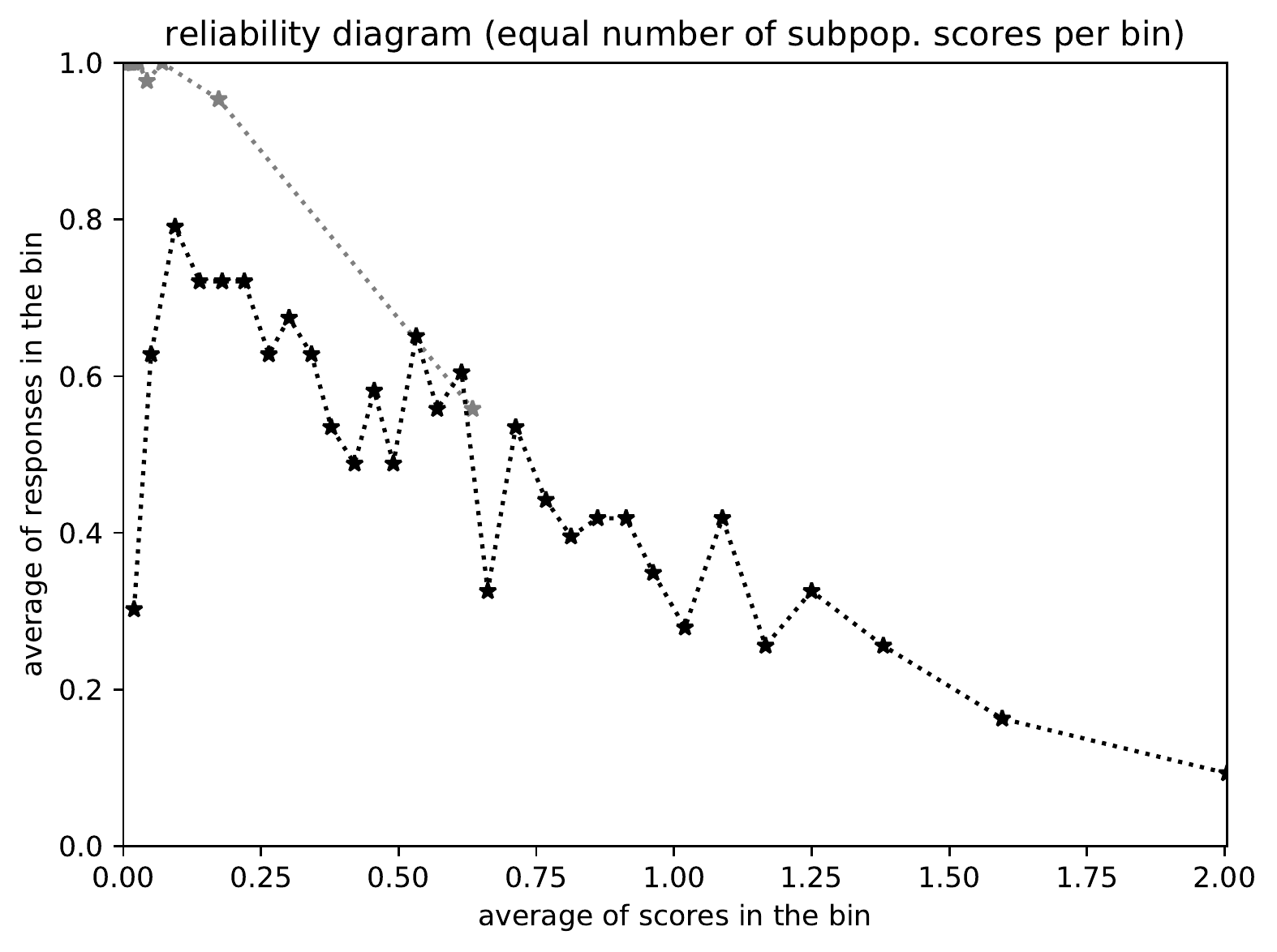}}
\quad\quad
(e)
\parbox{\imsize}{\includegraphics[width=\imsize]
{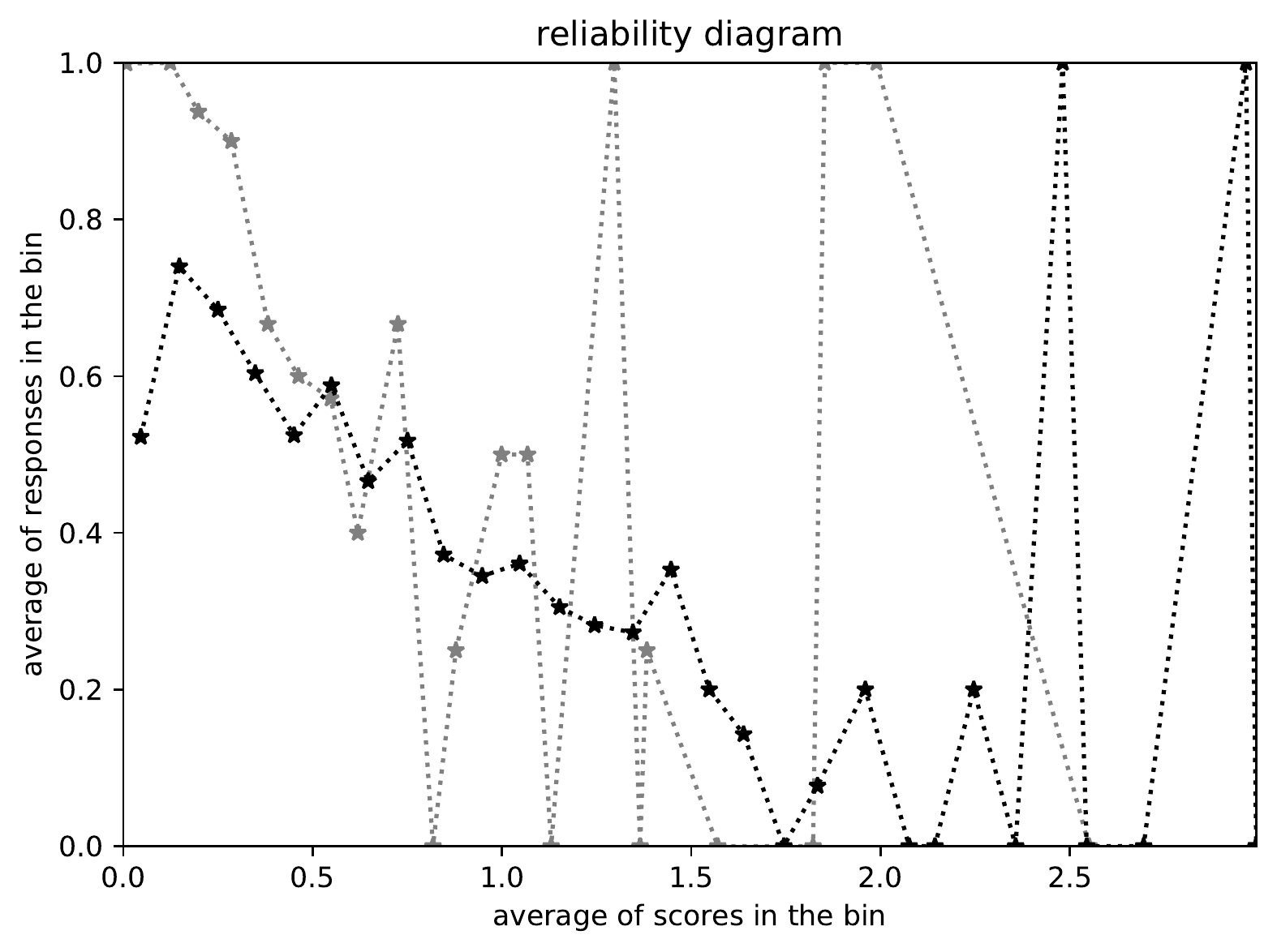}}

\vspace{\vertsep}

(f)
\parbox{\imsize}{\includegraphics[width=\imsize]
{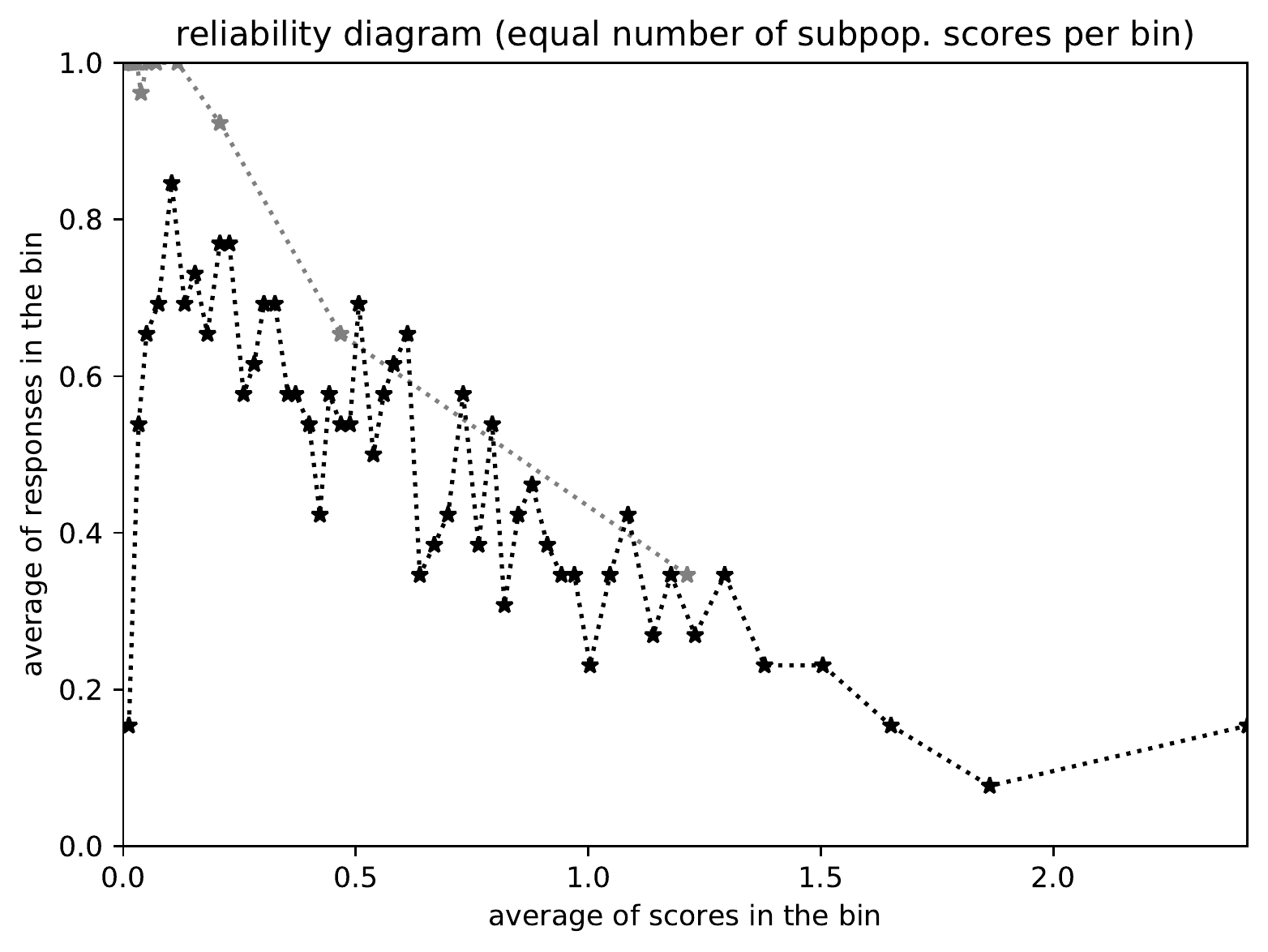}}
\quad\quad
(g)
\parbox{\imsize}{\includegraphics[width=\imsize]
{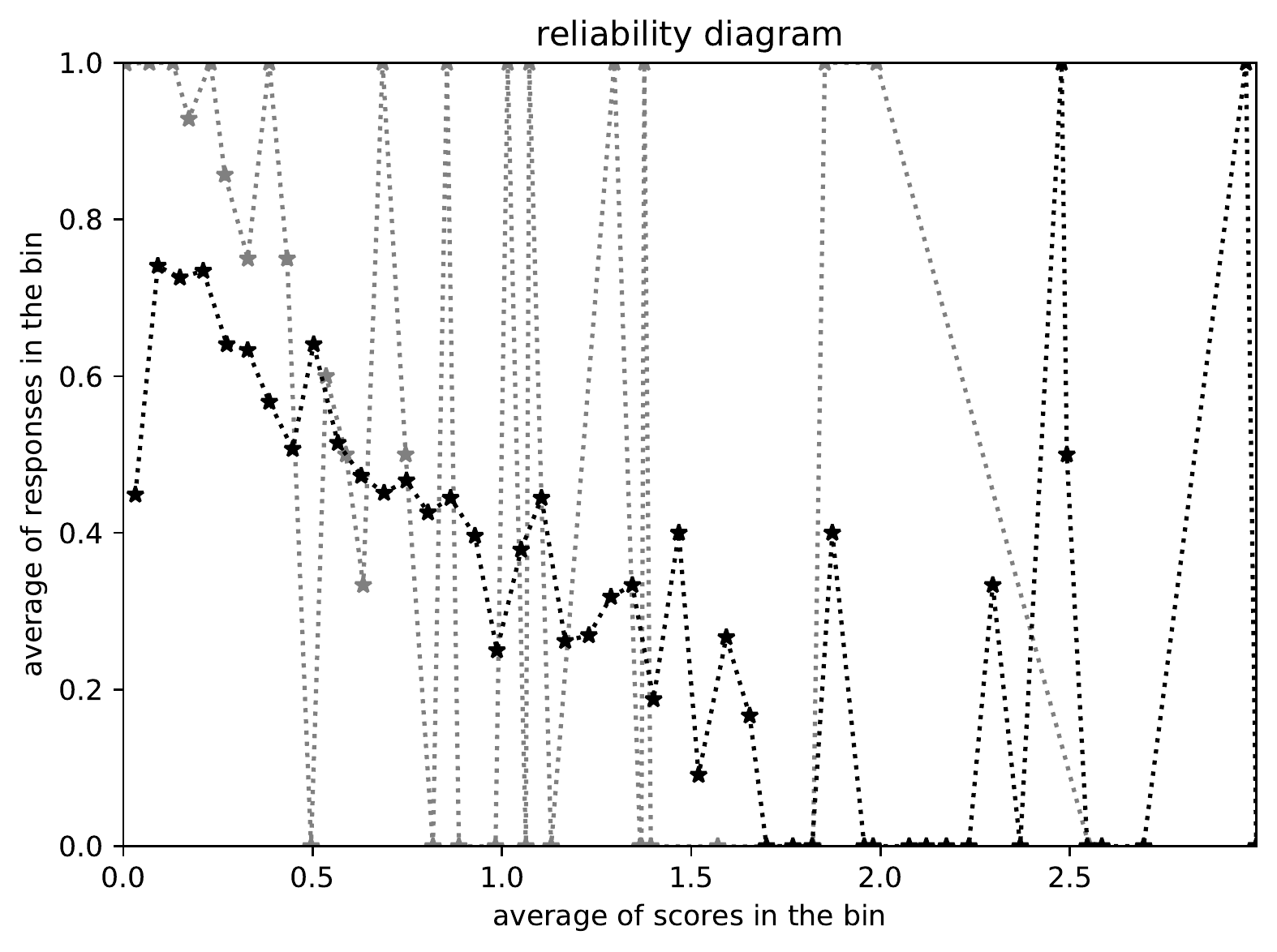}}

\end{centering}
\caption{Night snake (Hypsiglena torquata) vs.\ Monarch (or milkweed) butterfly
         (Danaus plexippus); $n =$ 304;
         Kuiper's statistic is $0.3138 / \sigma = 5.471$,
         Kolmogorov's and Smirnov's is $0.3138 / \sigma = 5.471$;
         the lack of deviation at large scores is hard to detect
         without 30 bins or more (d, e, f, and g),
         but then the reliability diagrams are too noisy for other scores.
         Moreover, the diagrams with only 10 or 30 bins (b, c, d, and e)
         smooth away the extreme deviation for the lowest scores.
         The graph of cumulative differences (a) captures all phenomena nicely
         simultaneously. The statistics of Kuiper and of Kolmogorov and Smirnov
         both report very highly statistically significant deviations
         between the subpopulations.
}
\label{night-snake_monarch-butterfly}
\end{figure}

\begin{figure}
\begin{centering}

(a)
\parbox{\imsize}{\includegraphics[width=\imsize]
{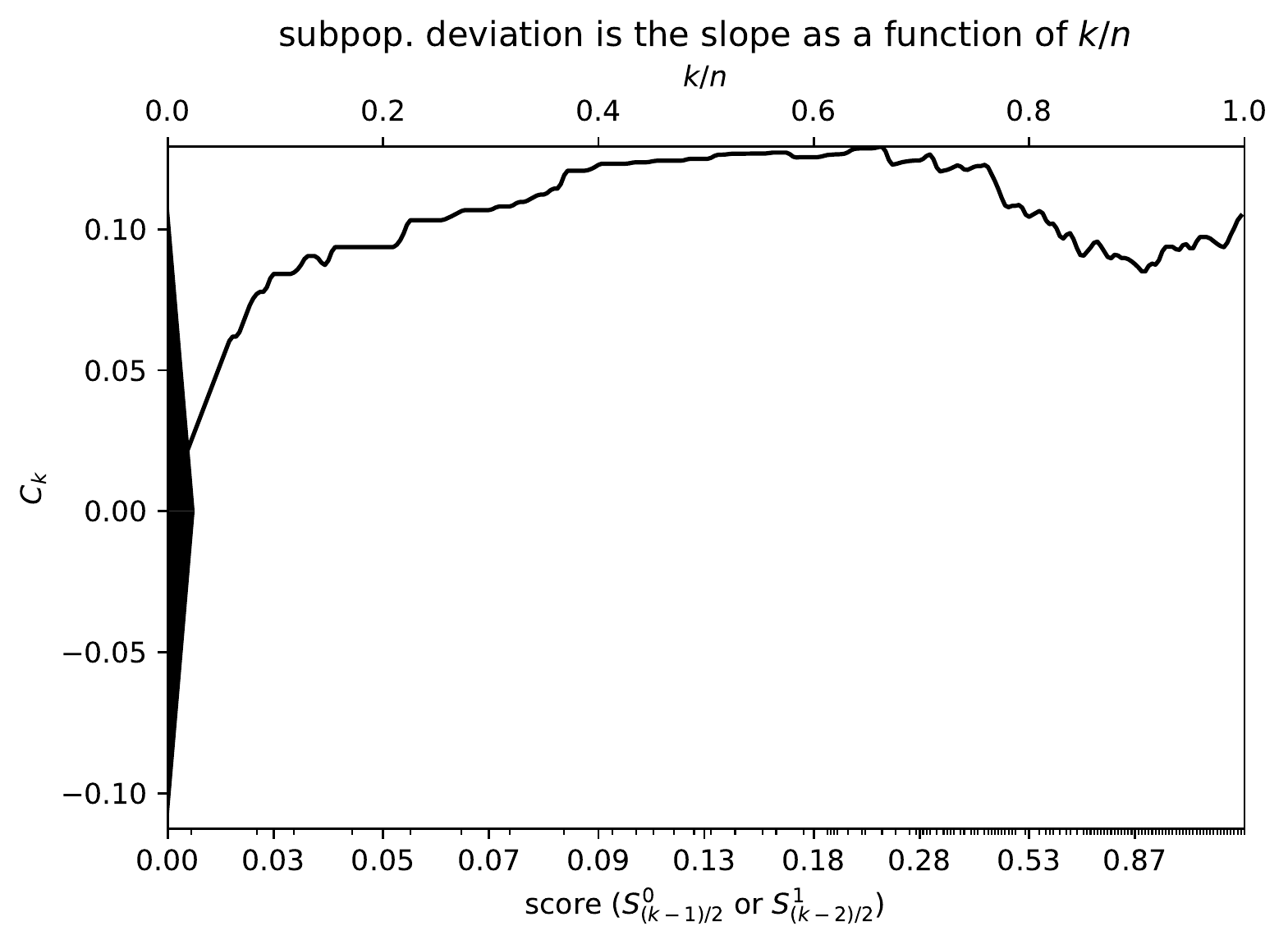}}

\vspace{\vertsep}

(b)
\parbox{\imsize}{\includegraphics[width=\imsize]
{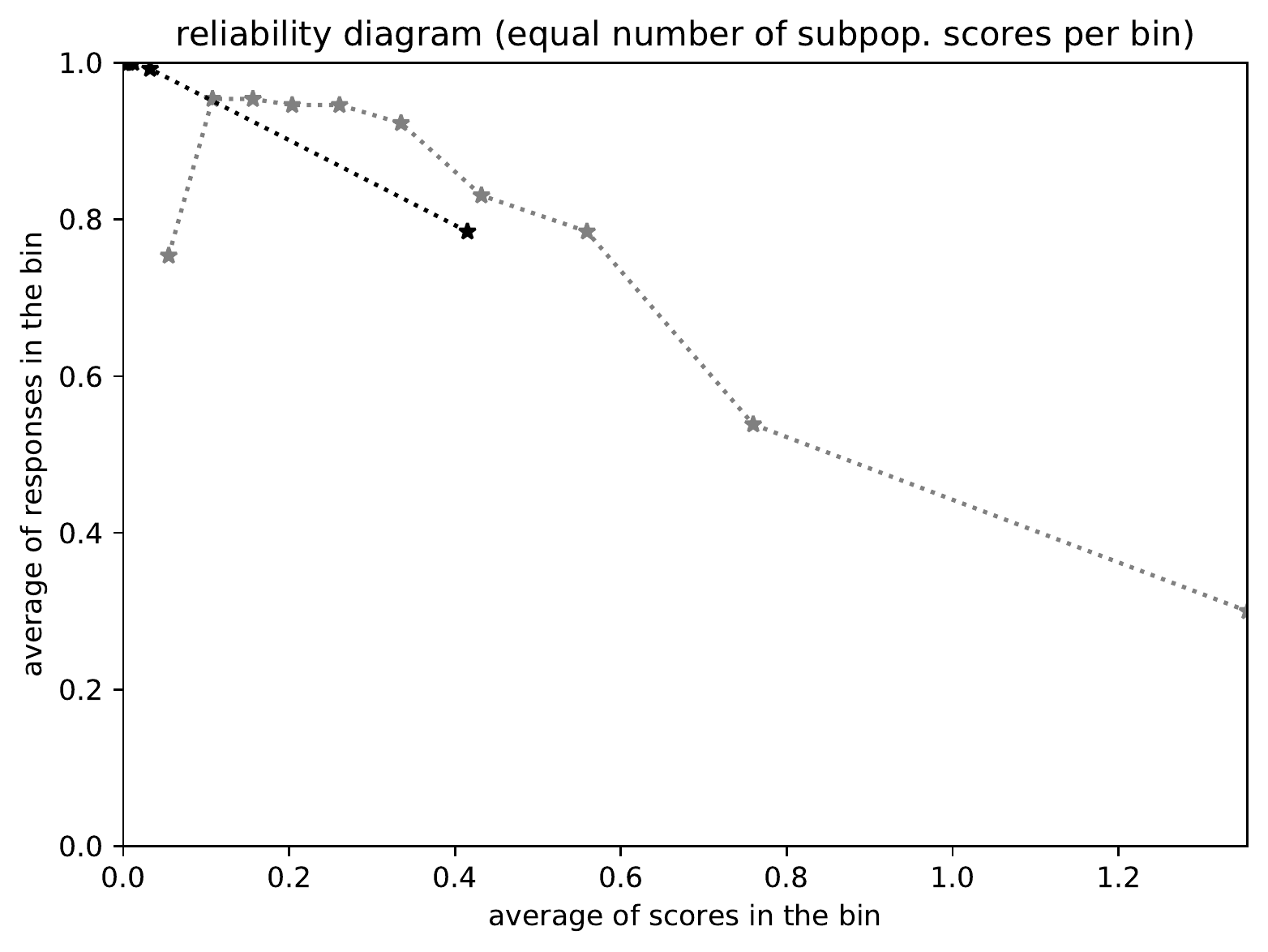}}
\quad\quad
(c)
\parbox{\imsize}{\includegraphics[width=\imsize]
{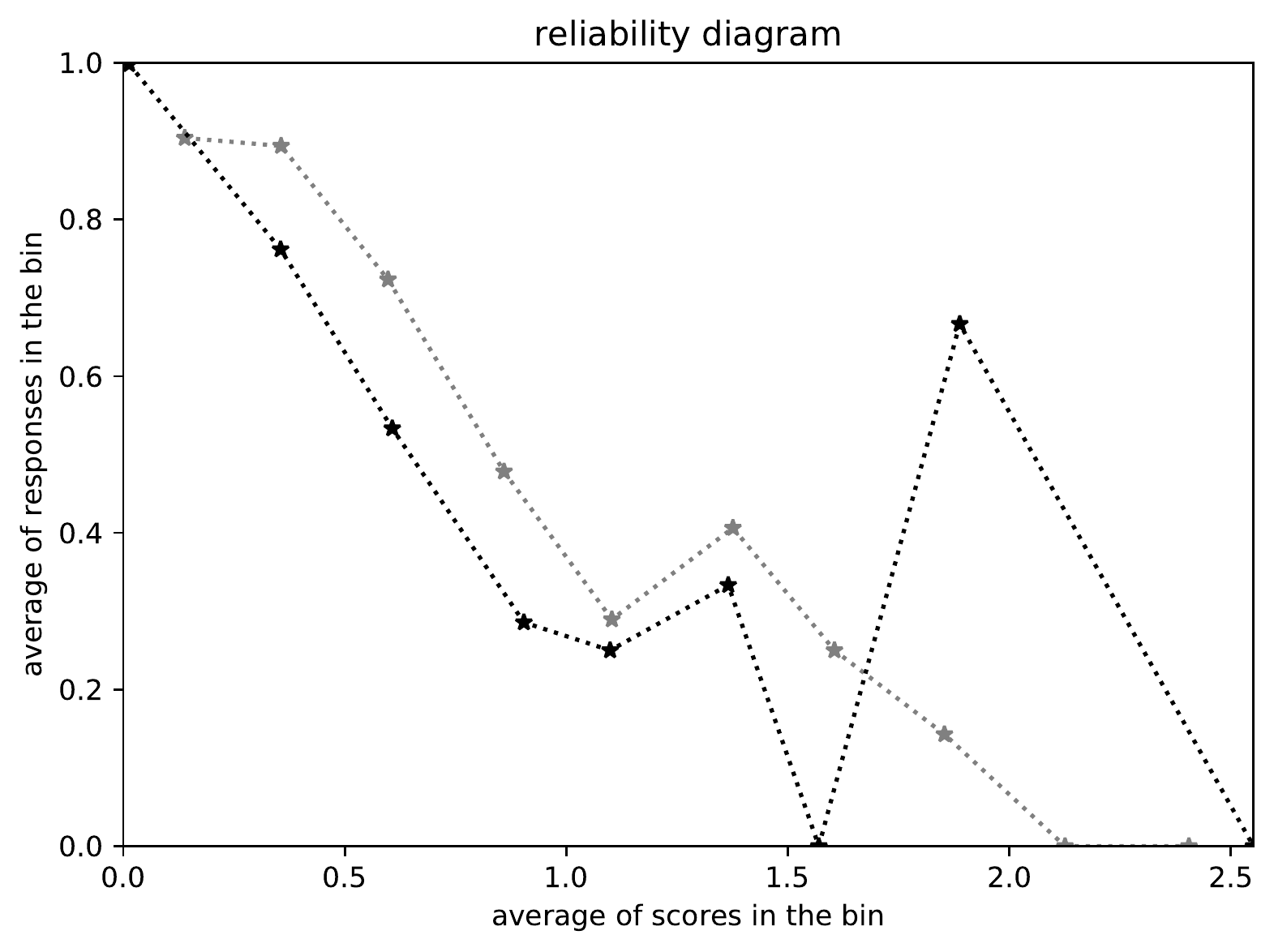}}

\vspace{\vertsep}

(d)
\parbox{\imsize}{\includegraphics[width=\imsize]
{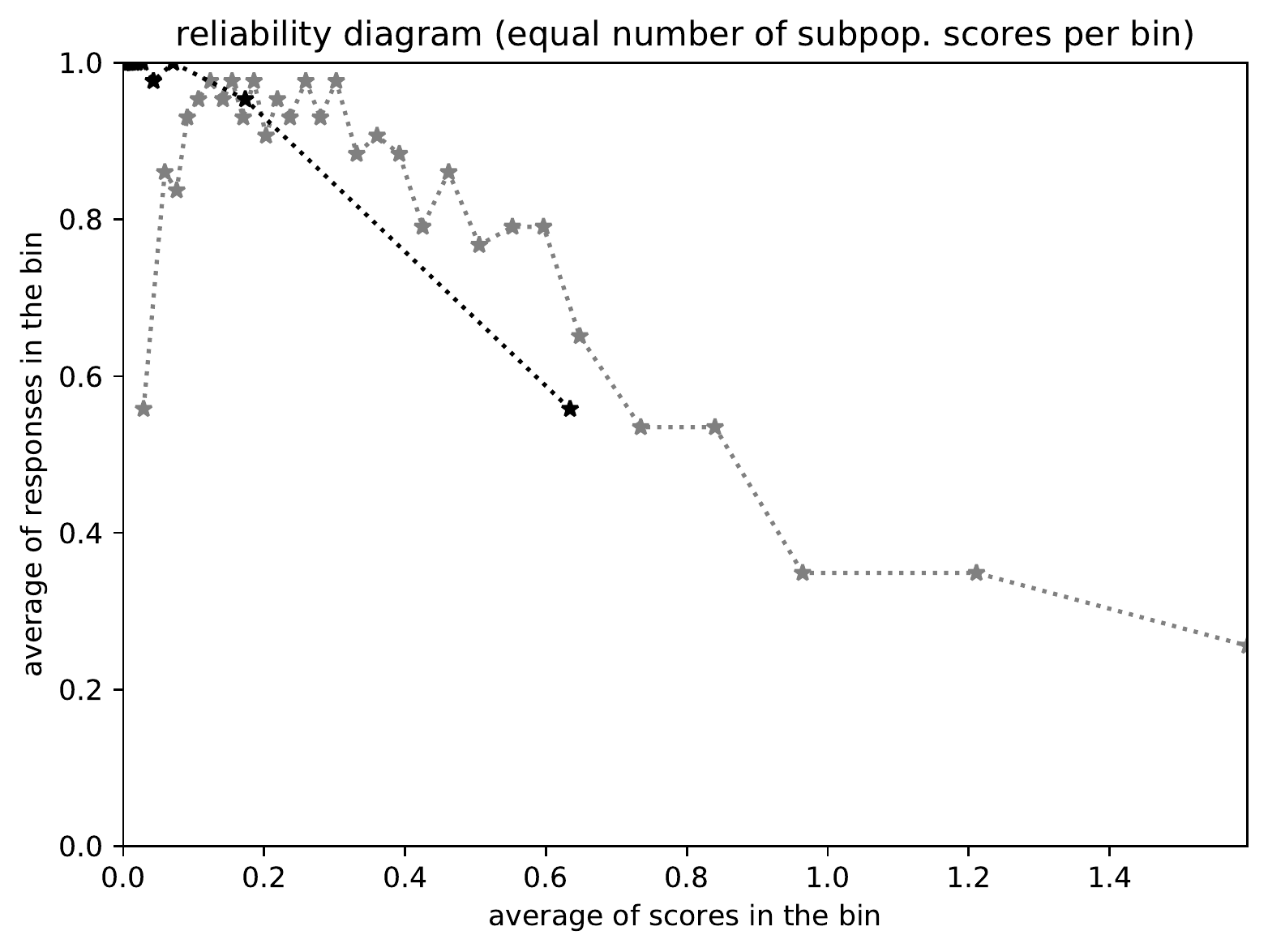}}
\quad\quad
(e)
\parbox{\imsize}{\includegraphics[width=\imsize]
{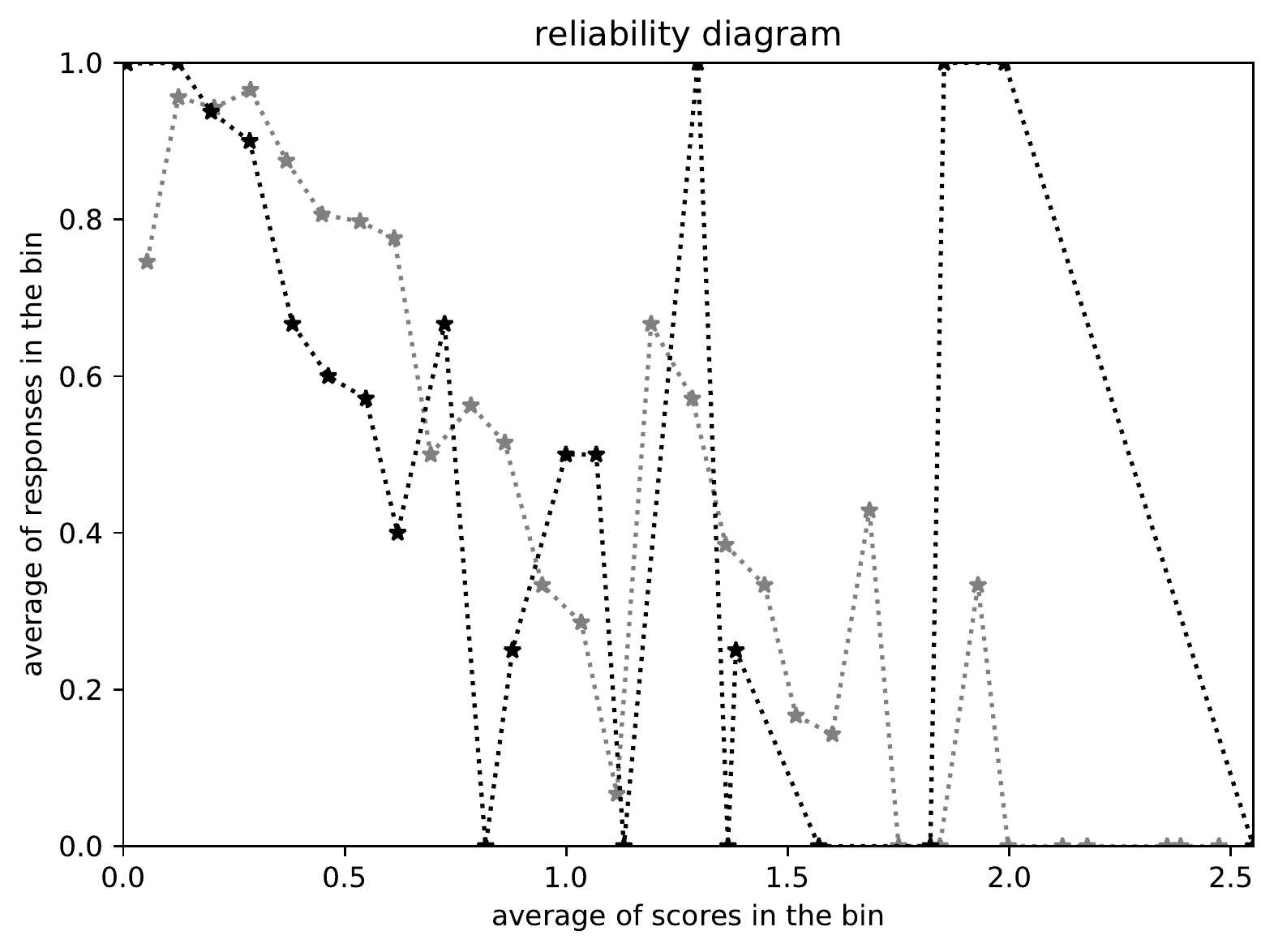}}

\vspace{\vertsep}

(f)
\parbox{\imsize}{\includegraphics[width=\imsize]
{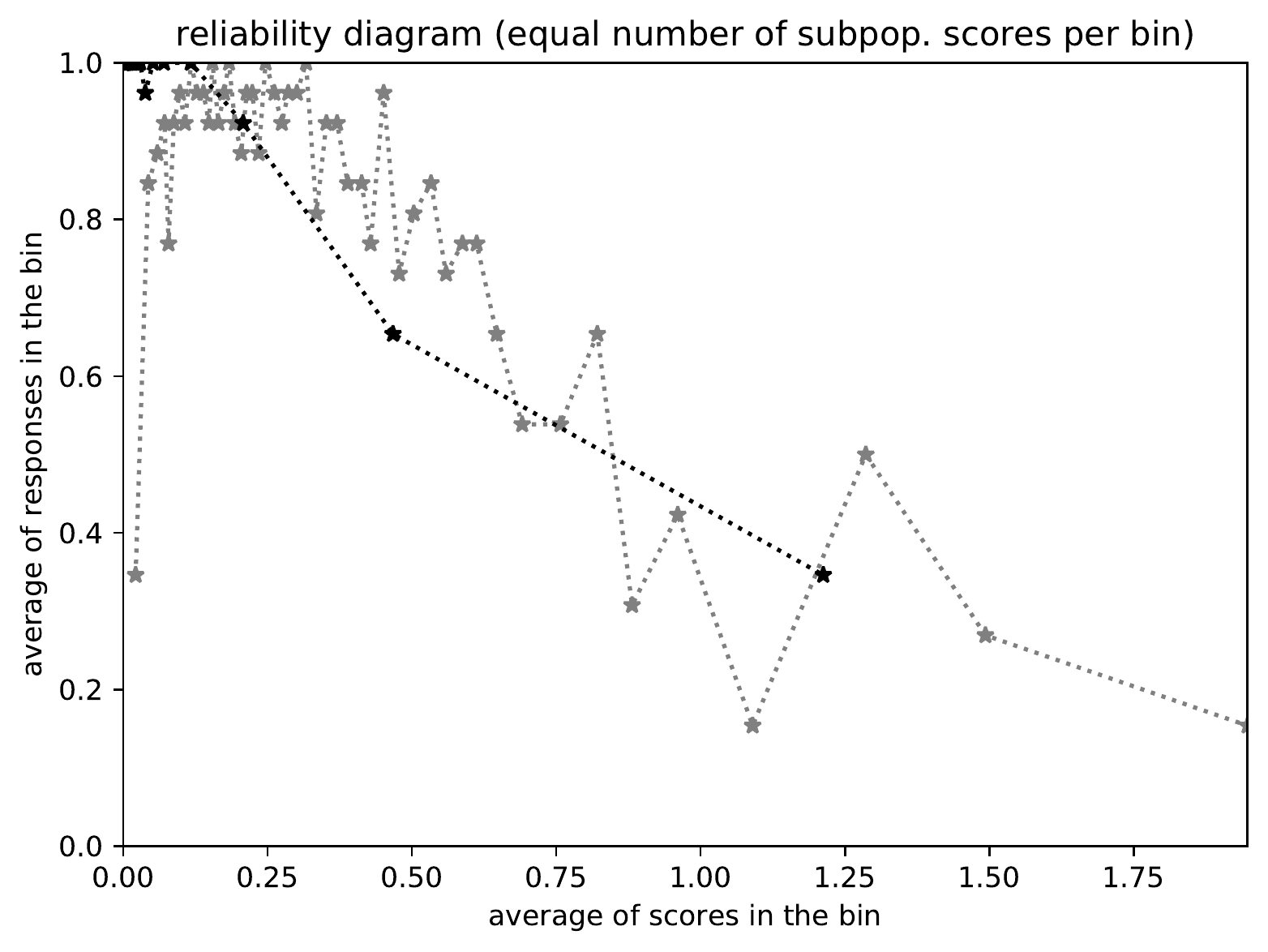}}
\quad\quad
(g)
\parbox{\imsize}{\includegraphics[width=\imsize]
{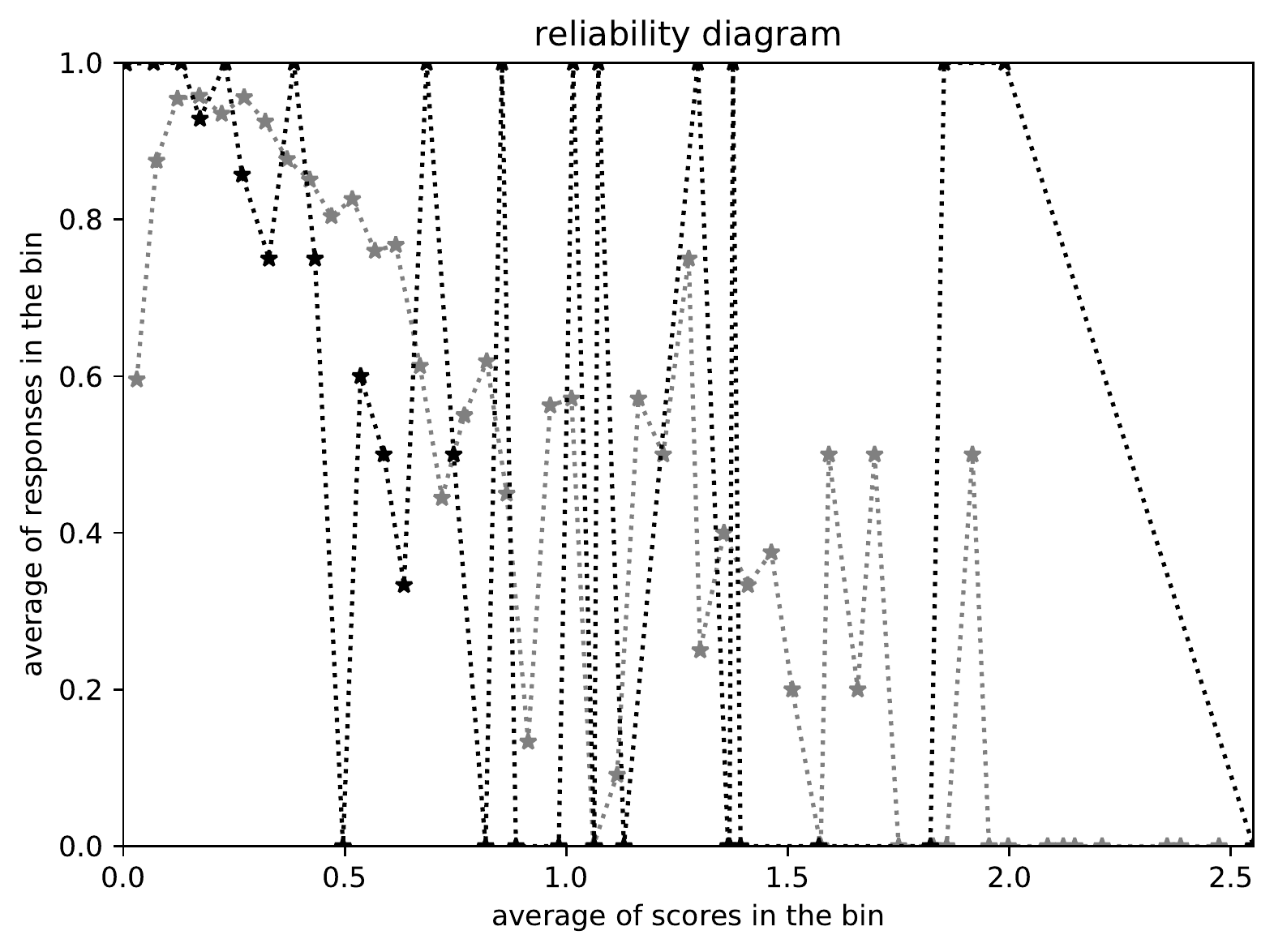}}

\end{centering}
\caption{Monarch (or milkweed) butterfly (Danaus plexippus) vs.
         Wild boar (Sus scrofa); $n =$ 315;
         Kuiper's statistic is $0.1292 / \sigma = 2.294$,
         Kolmogorov's and Smirnov's is $0.1292 / \sigma = 2.294$;
         the reliability diagrams with 30 bins or less (b, c, d, and e)
         underestimate (or fail to resolve) the extreme deviation
         at the lowest scores, whereas the diagrams with 50 bins (f and g)
         are far too noisy for the other scores.
         The graph of cumulative differences (a) resolves
         all these behaviors clearly.
         The metrics of Kuiper and of Kolmogorov and Smirnov both report
         somewhat statistically significant deviations
         between the subpopulations, though much less extreme than in
         Figures~\ref{Eskimo-dog-husky_cheetah}
         and~\ref{night-snake_monarch-butterfly}.
}
\label{monarch-butterfly_wild-boar}
\end{figure}

This subsection applies the methods of Section~\ref{methods}
to the training data set ``ImageNet-1000'' of~\cite{imagenet},
which contains a thousand labeled classes.
Each class forms a natural subpopulation to consider,
with each class considered consisting of 1,300 images of a particular noun
(such as a ``cheetah,'' a ``night snake,'' or an ``Eskimo Dog or Husky'').
The total number of members of the data set over all classes is 1,281,167,
as some classes in the data set contain fewer than 1,300 images,
but each subpopulation considered below comes from a class with 1,300 images.
The images are unweighted (or, equivalently, uniformly or equally weighted),
not requiring the methods of Subsection~\ref{weighted} above.
We calculate the scores using the pretrained ResNet18 classifier
of~\cite{he-zhang-ren-sun} from the computer-vision module, ``torchvision,''
in the PyTorch software library of~\cite{pytorch};
the score for an image is the negative of the natural logarithm
of the probability assigned by the classifier
to the class predicted to be most likely,
with the scores randomly perturbed by about one part in $10^8$ to guarantee
their uniqueness.
The response (also known as ``result'' or ``outcome'') corresponding
to a given score takes the value 1 when the class predicted to be most likely
is the correct class; the response takes the value 0 otherwise.
Figures~\ref{Eskimo-dog-husky_cheetah}--\ref{monarch-butterfly_wild-boar}
present three examples; the captions first list the names of the classes
for the subpopulations and then compare the different kinds of plots.

\clearpage

\subsection{American Community Survey of the U.S. Census Bureau}
\label{census}

\begin{figure}
\begin{centering}

(a)
\parbox{\imsize}{\includegraphics[width=\imsize]
{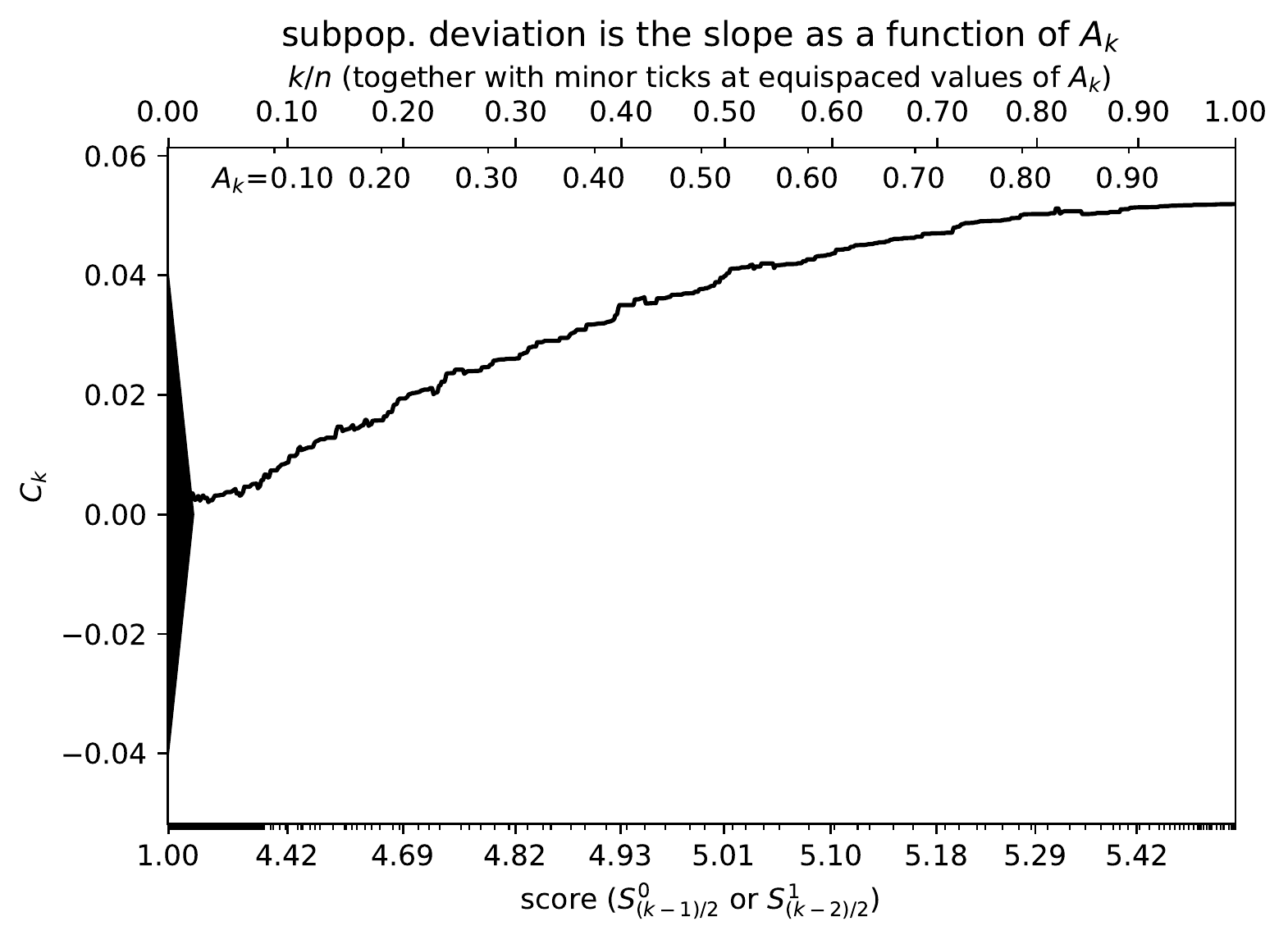}}

\vspace{\vertsep}

(b)
\parbox{\imsize}{\includegraphics[width=\imsize]
{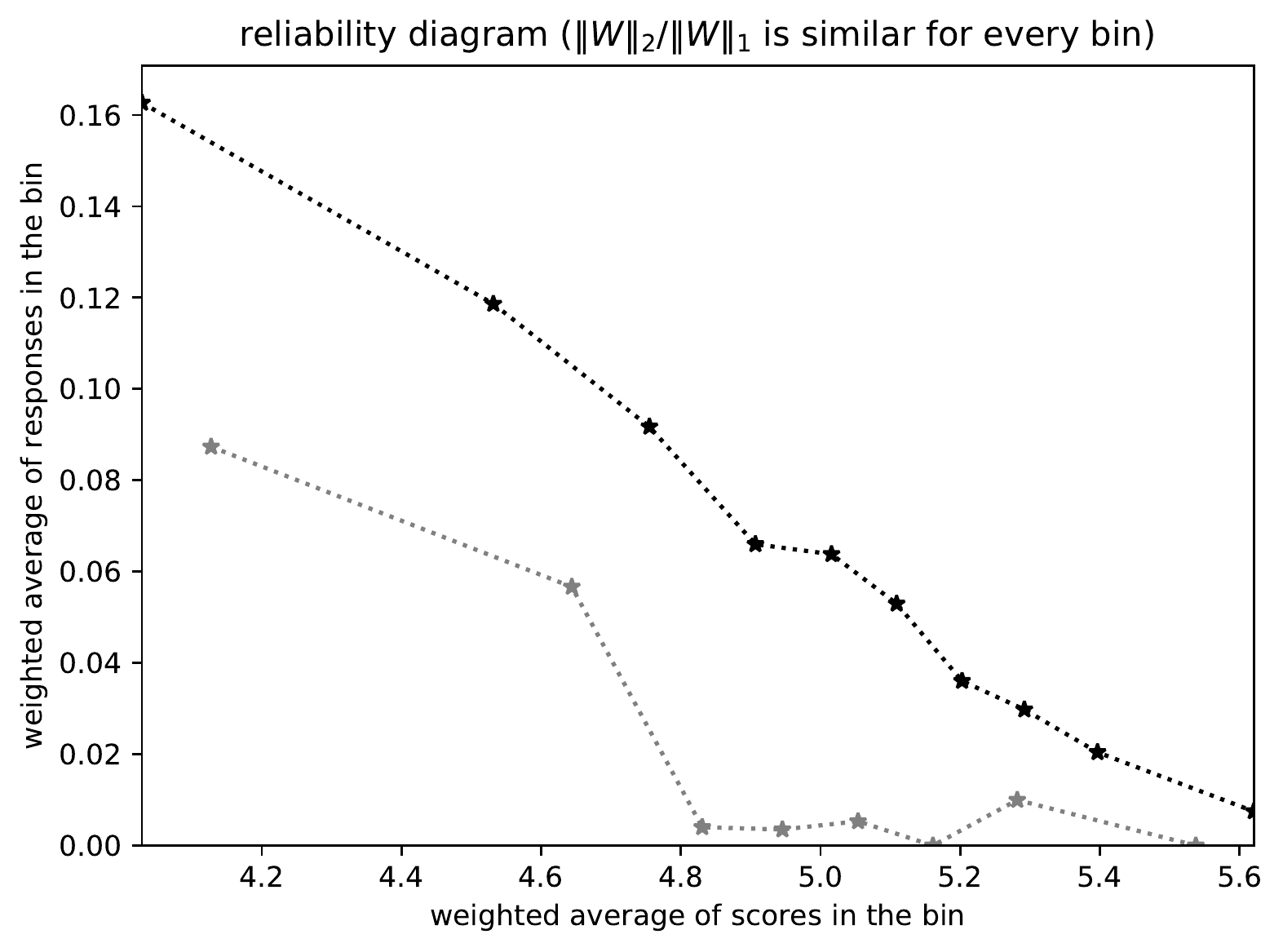}}
\quad\quad
(c)
\parbox{\imsize}{\includegraphics[width=\imsize]
{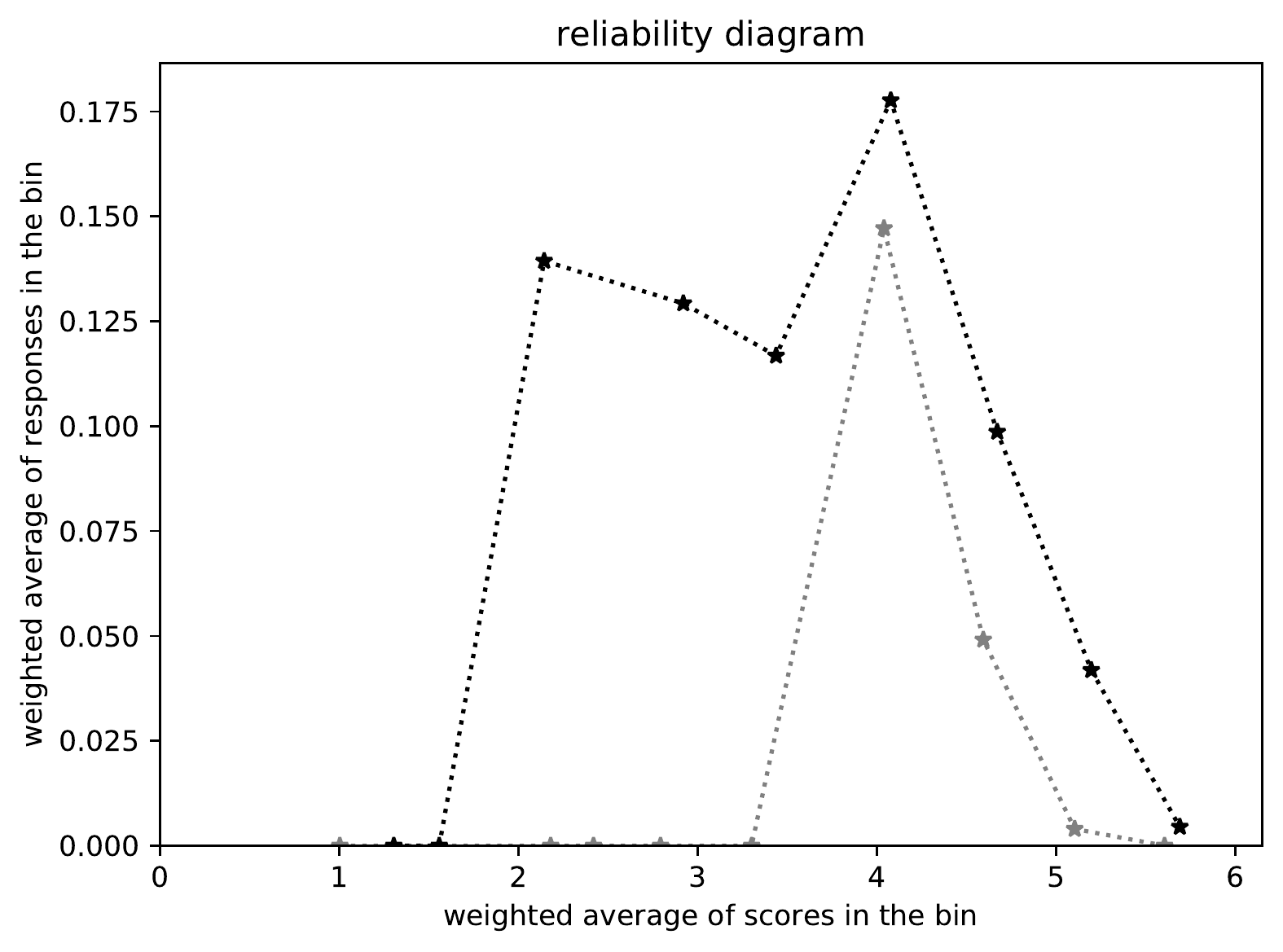}}

\vspace{\vertsep}

(d)
\parbox{\imsize}{\includegraphics[width=\imsize]
{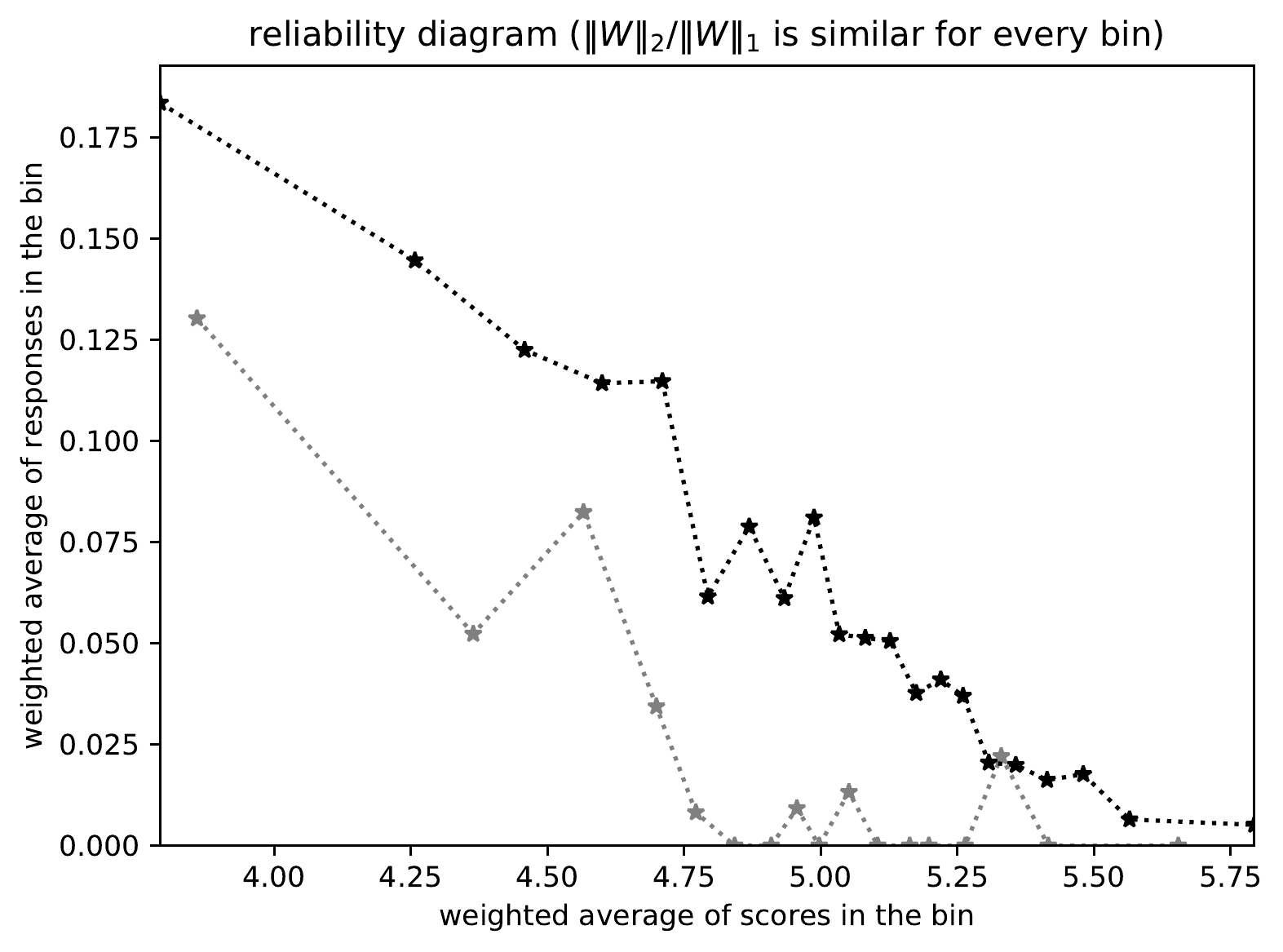}}
\quad\quad
(e)
\parbox{\imsize}{\includegraphics[width=\imsize]
{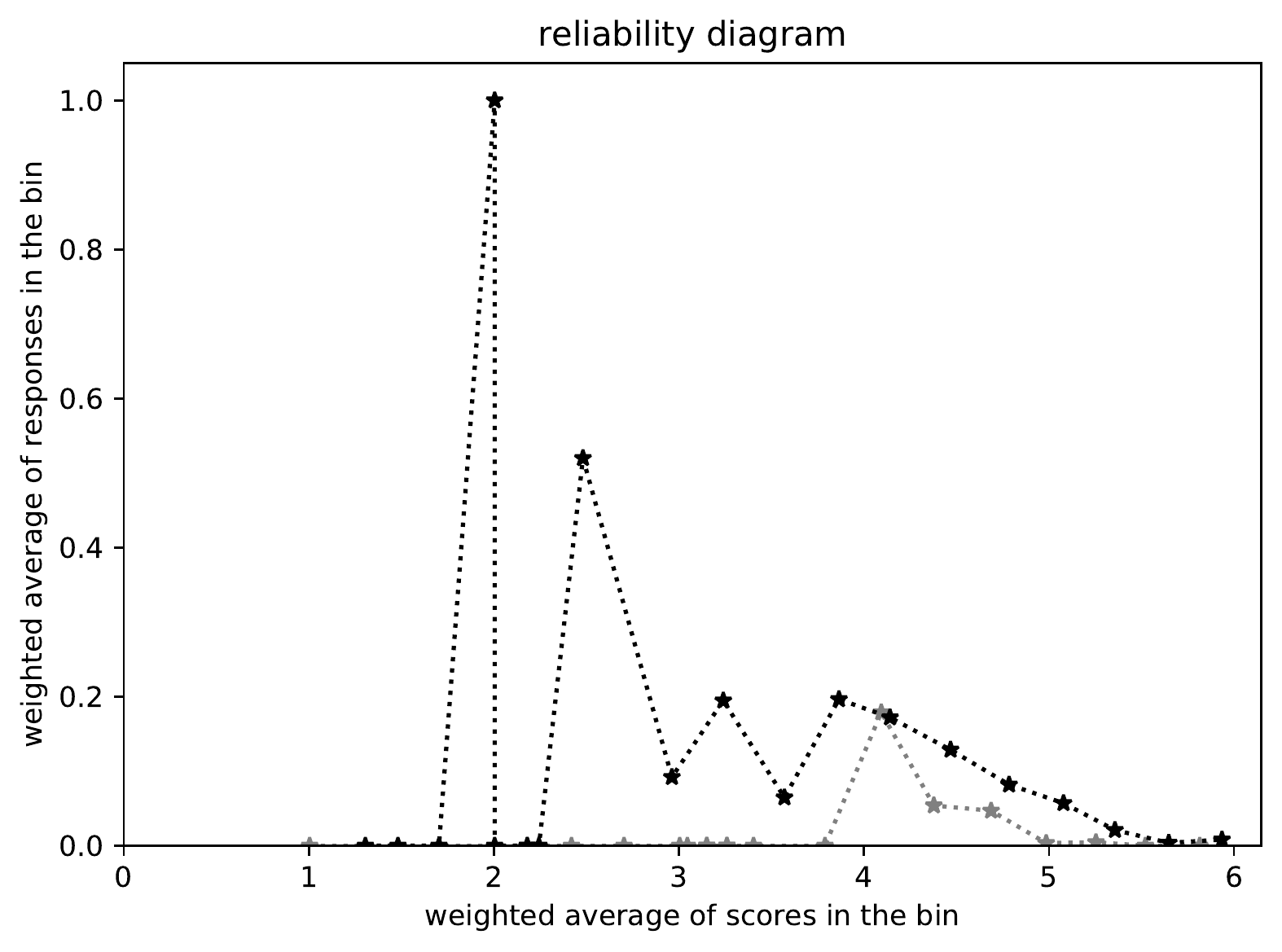}}

\vspace{\vertsep}

(f)
\parbox{\imsize}{\includegraphics[width=\imsize]
{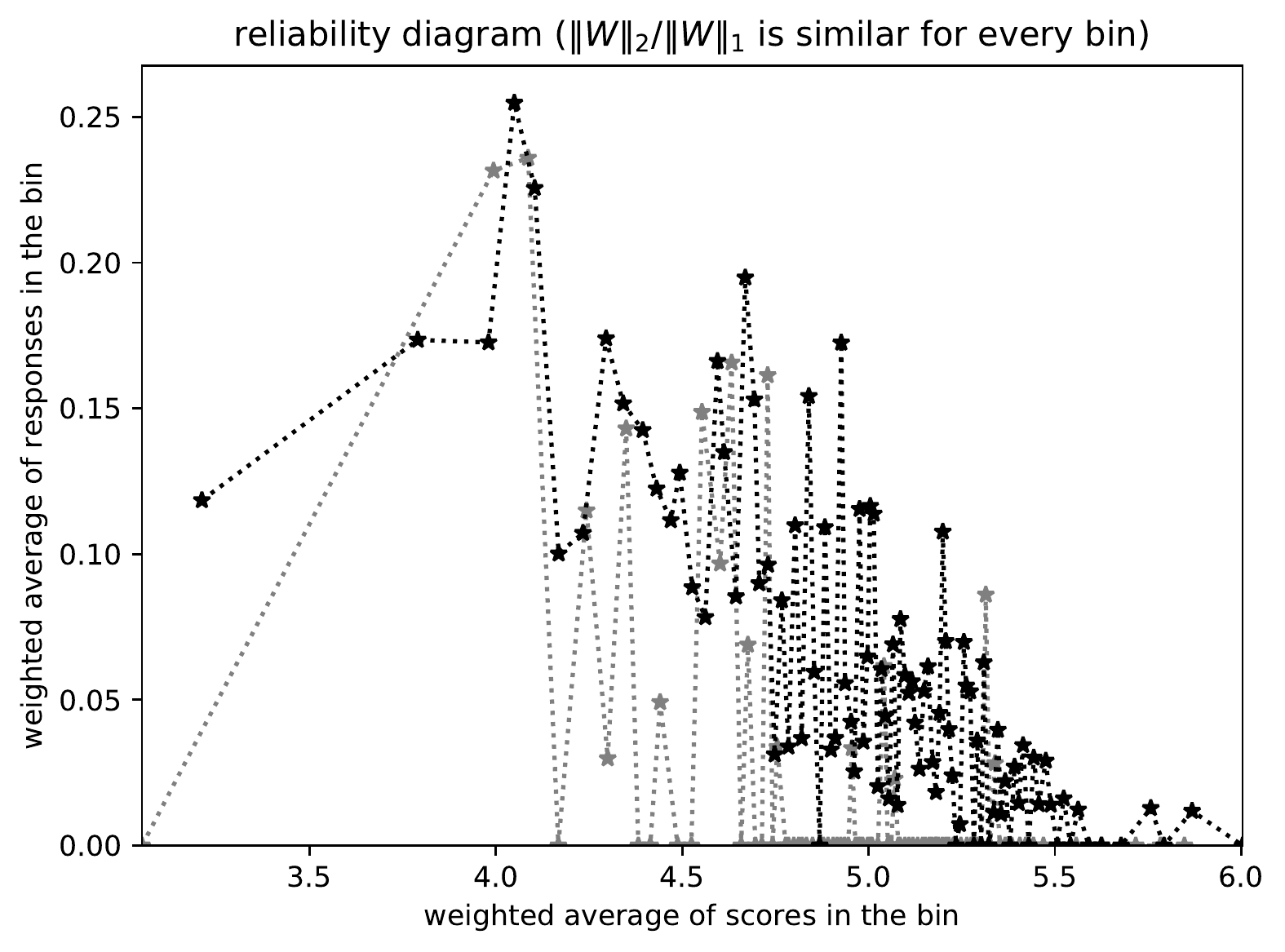}}
\quad\quad
(g)
\parbox{\imsize}{\includegraphics[width=\imsize]
{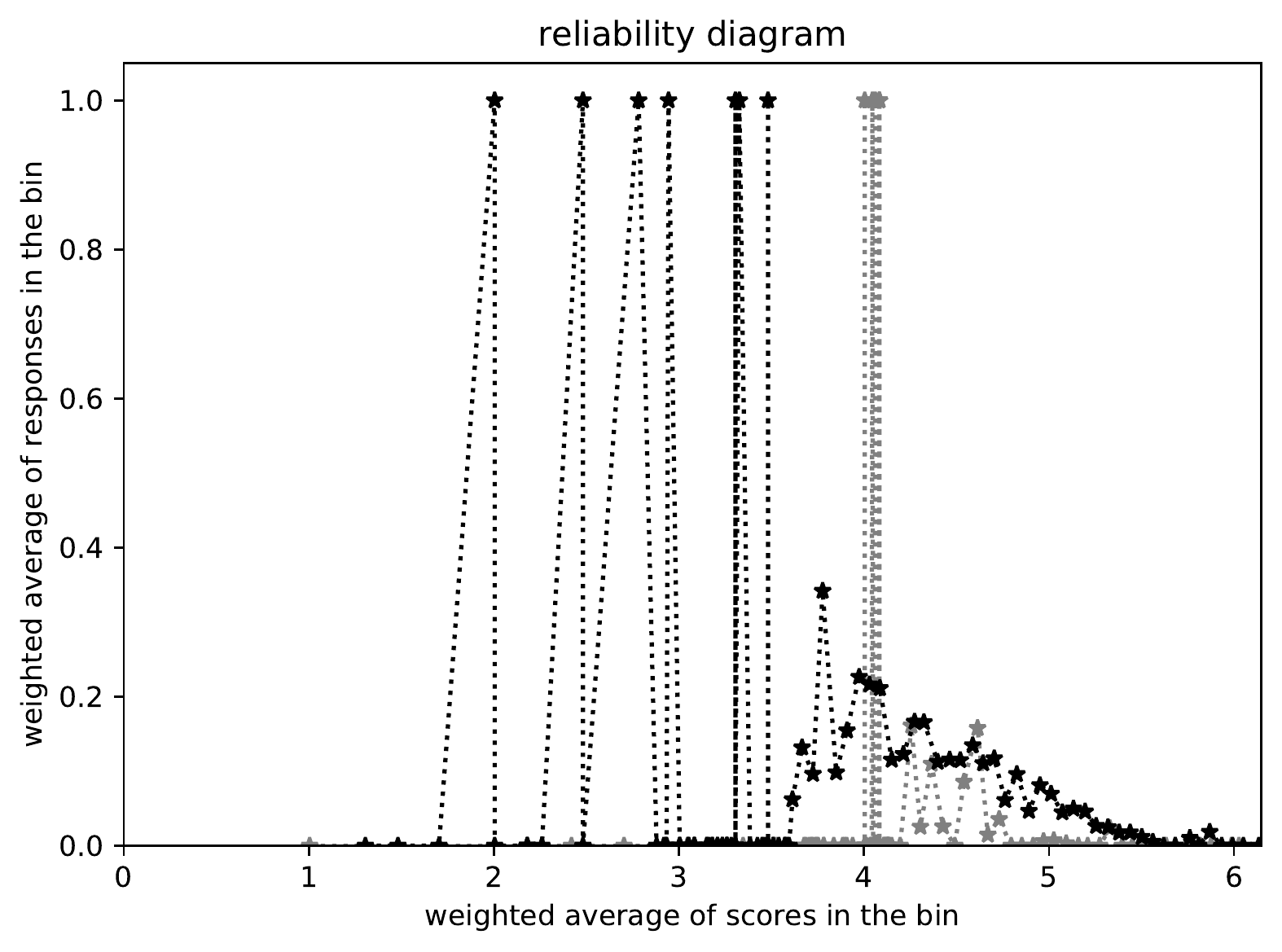}}

\end{centering}
\caption{Alameda County vs.\ Placer County; $n =$ 2,536;
         Kuiper's statistic is $0.05192 / \sigma = 2.450$,
         Kolmogorov's and Smirnov's is $0.05192 / \sigma = 2.450$;
         the behavior for small scores is interesting,
         as the cumulative graph (a) shows a big spike
         at the very lowest scores and then a very flat part, and
         only the reliability diagrams with 100 bins (f and g) reflect those.
         Yet the latter reliability diagrams are very, very noisy
         for the other scores. 
         The metrics of Kuiper and of Kolmogorov and Smirnov report
         mildly statistically significant deviation between the subpopulations.
}
\label{Alameda-Placer}
\end{figure}

\begin{figure}
\begin{centering}

(a)
\parbox{\imsize}{\includegraphics[width=\imsize]
{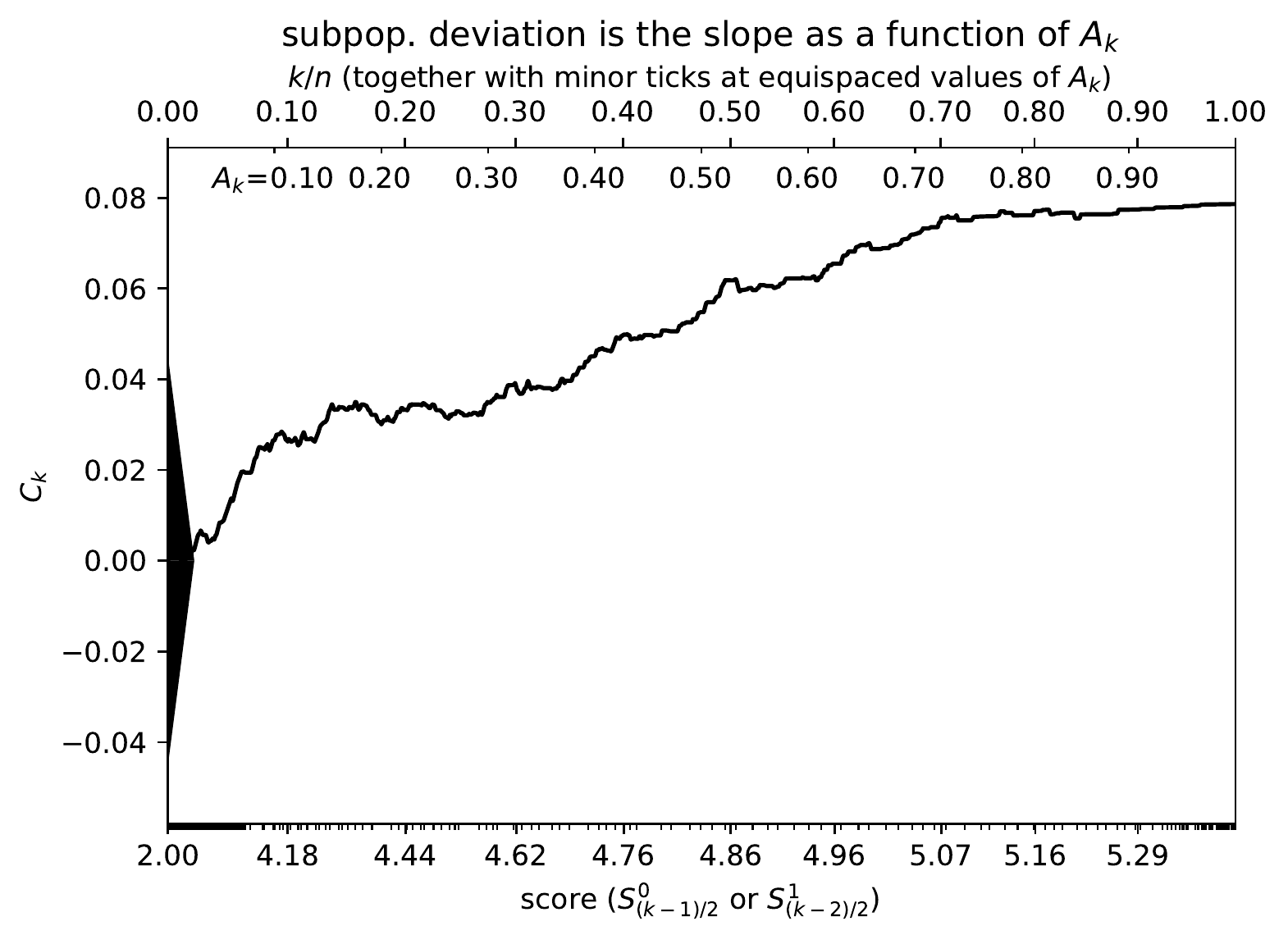}}

\vspace{\vertsep}

(b)
\parbox{\imsize}{\includegraphics[width=\imsize]
{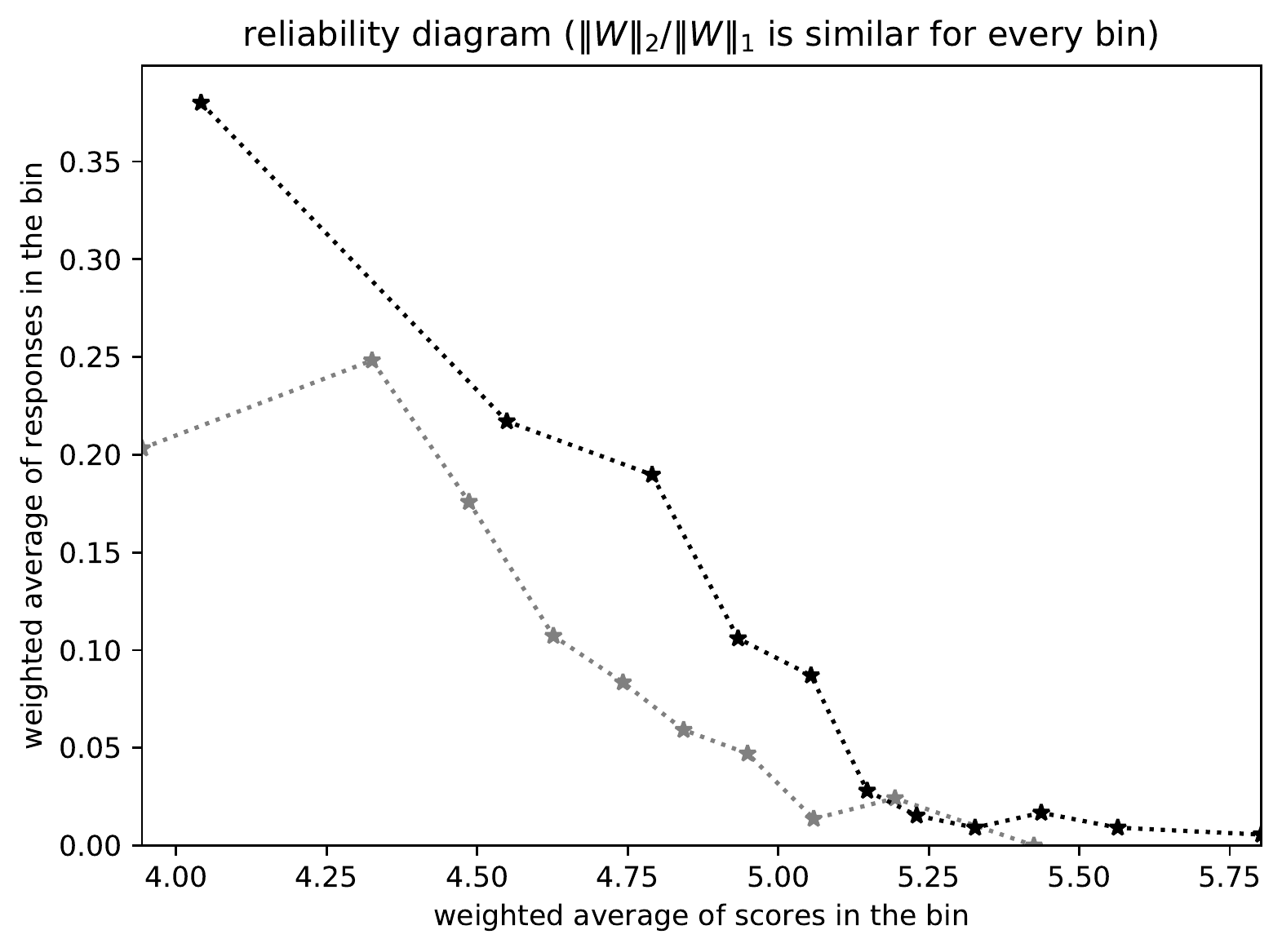}}
\quad\quad
(c)
\parbox{\imsize}{\includegraphics[width=\imsize]
{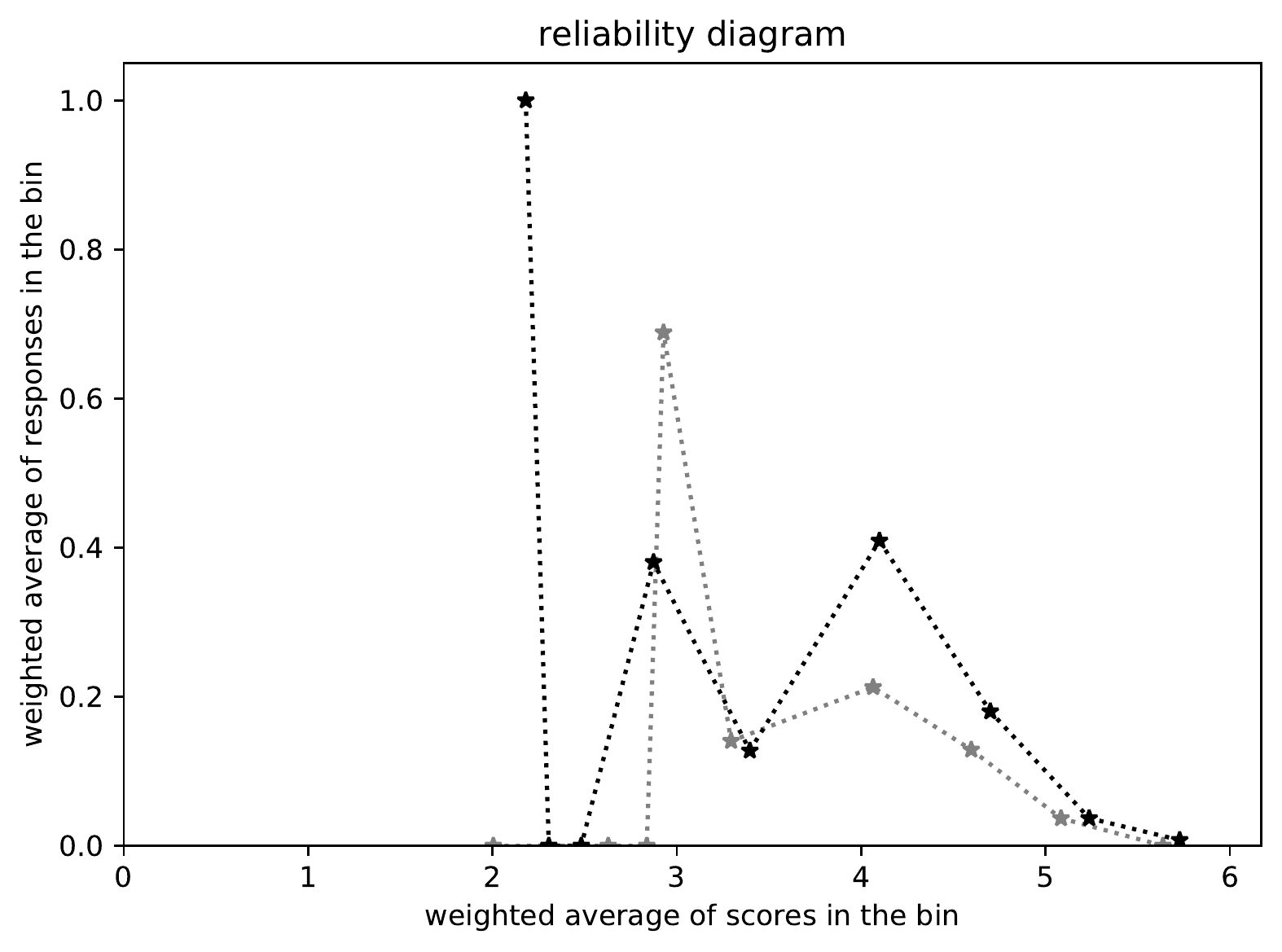}}

\vspace{\vertsep}

(d)
\parbox{\imsize}{\includegraphics[width=\imsize]
{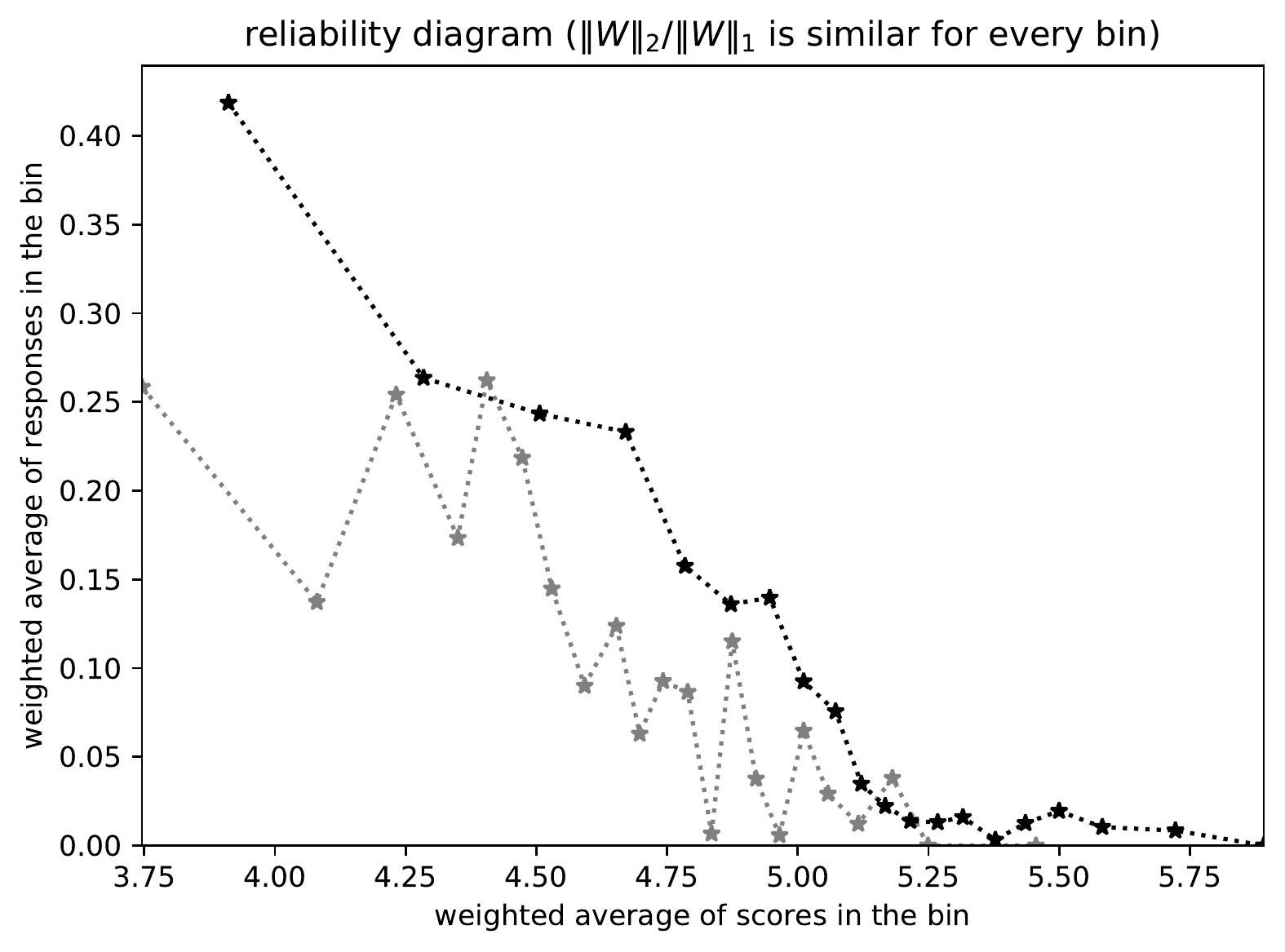}}
\quad\quad
(e)
\parbox{\imsize}{\includegraphics[width=\imsize]
{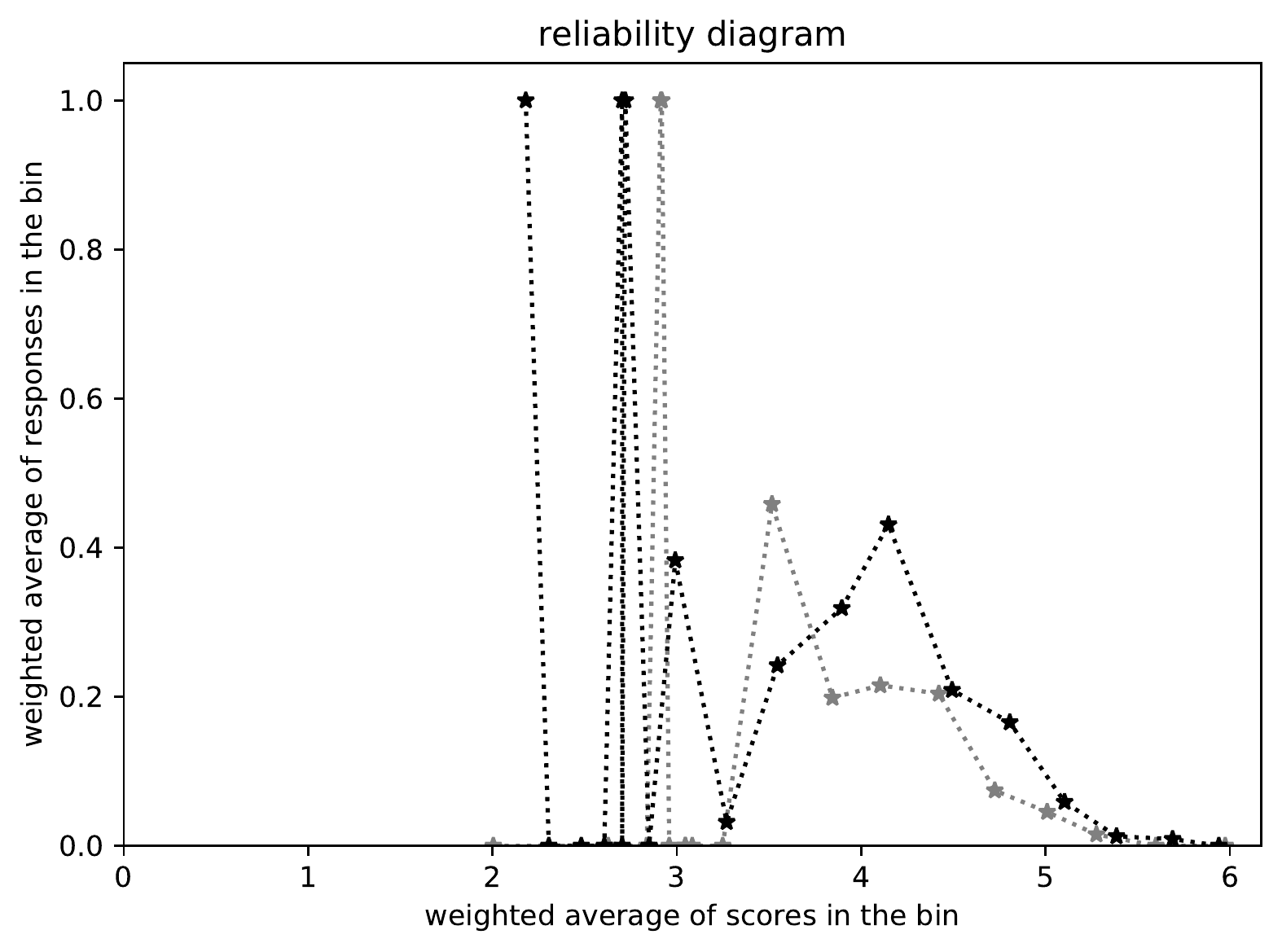}}

\vspace{\vertsep}

(f)
\parbox{\imsize}{\includegraphics[width=\imsize]
{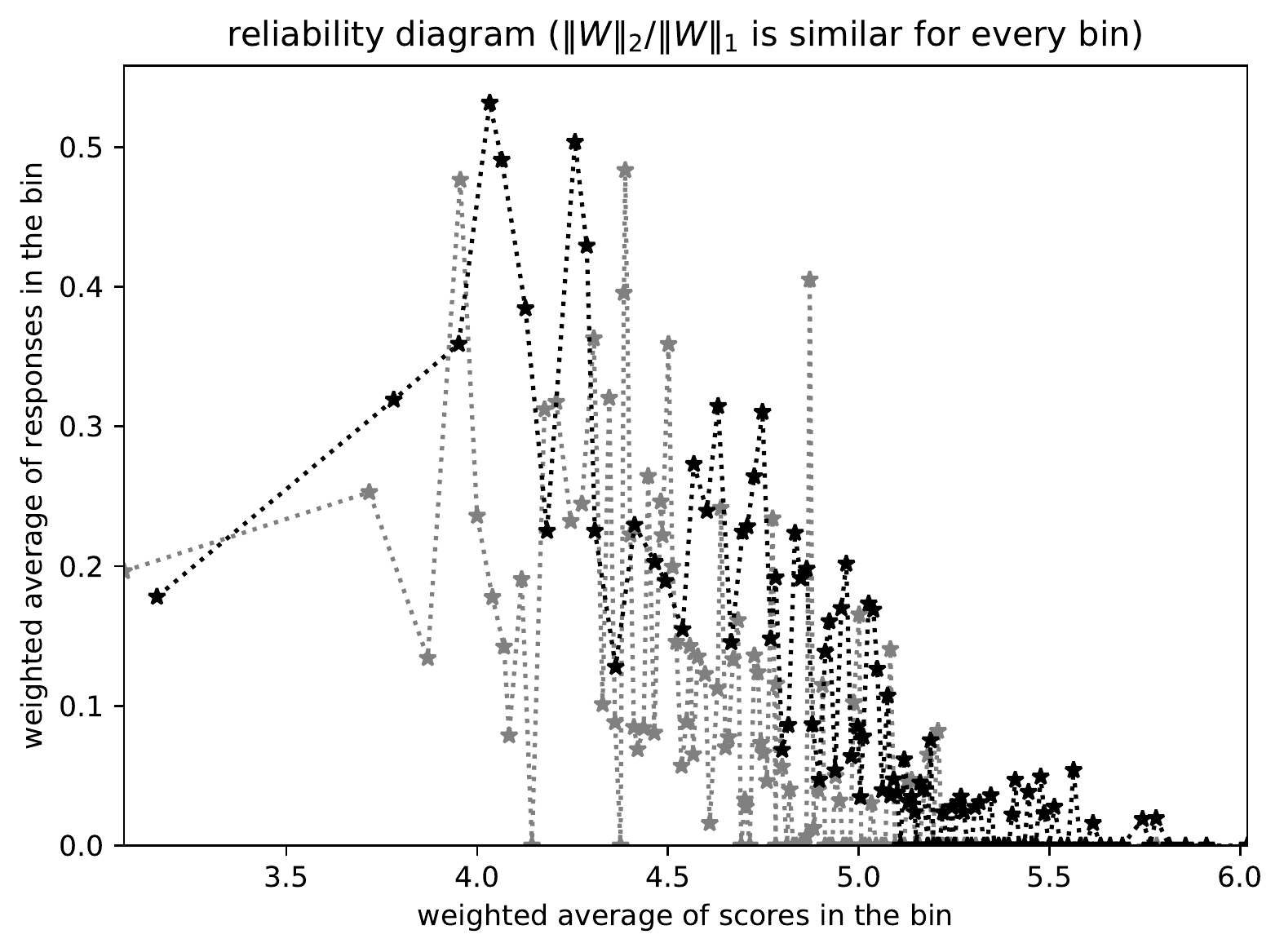}}
\quad\quad
(g)
\parbox{\imsize}{\includegraphics[width=\imsize]
{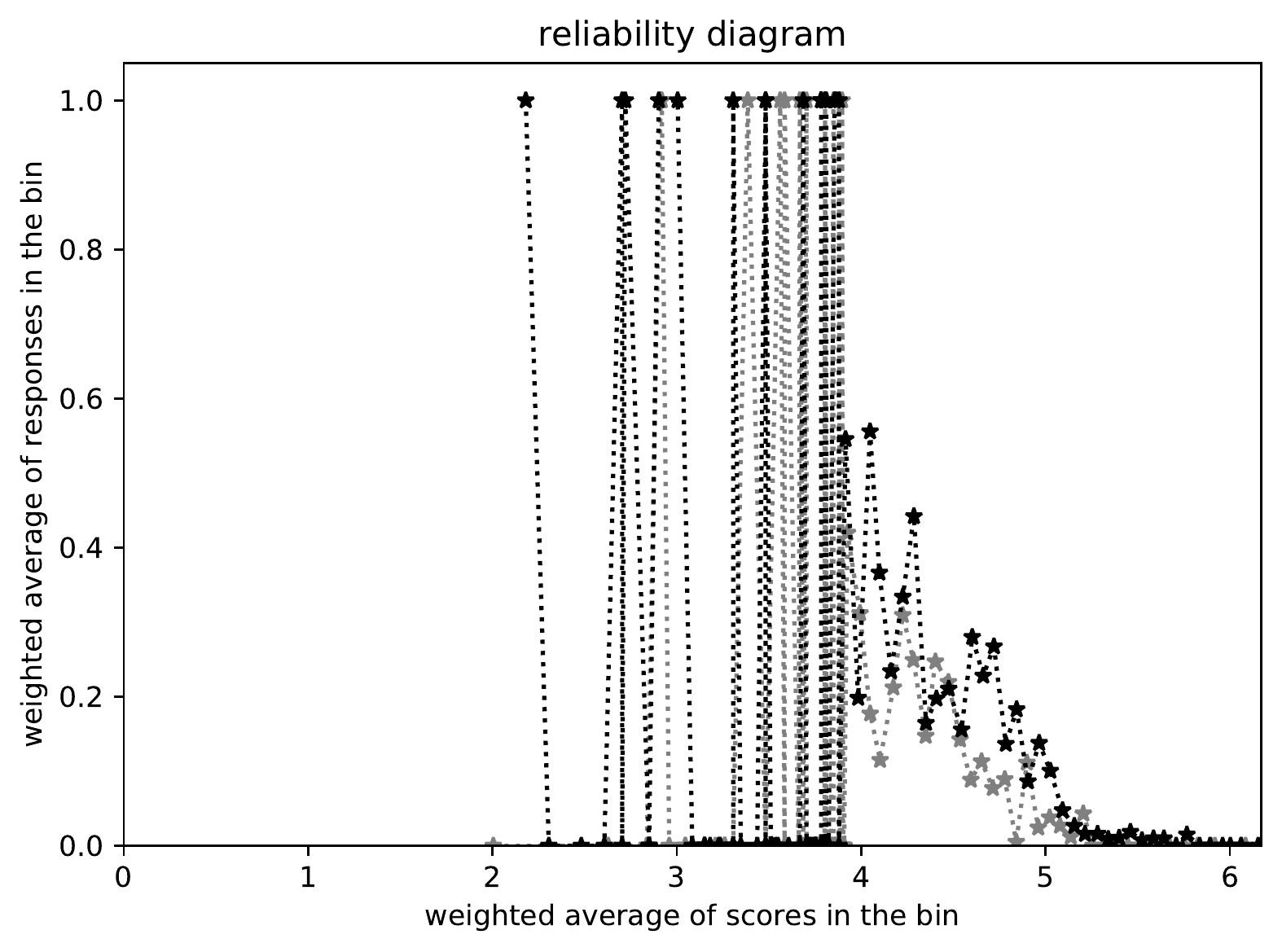}}

\end{centering}
\caption{San Francisco County vs.\ Kern County; $n =$ 2,260;
         Kuiper's statistic is $0.07882 / \sigma = 3.454$,
         Kolmogorov's and Smirnov's is $0.07863 / \sigma = 3.445$;
         only the cumulative graph (a) and the reliability diagrams
         with 100 bins (f and g) resolve both the extreme deviation
         for many low scores and the relatively small deviation
         for the very lowest scores, whereas 100 bins (f and g) produce
         far too much noise for most scores.
         The statistics of Kuiper and of Kolmogorov and Smirnov report
         statistically significant deviation between the subpopulations.
}
\label{San_Francisco-Kern}
\end{figure}

\begin{figure}
\begin{centering}

(a)
\parbox{\imsize}{\includegraphics[width=\imsize]
{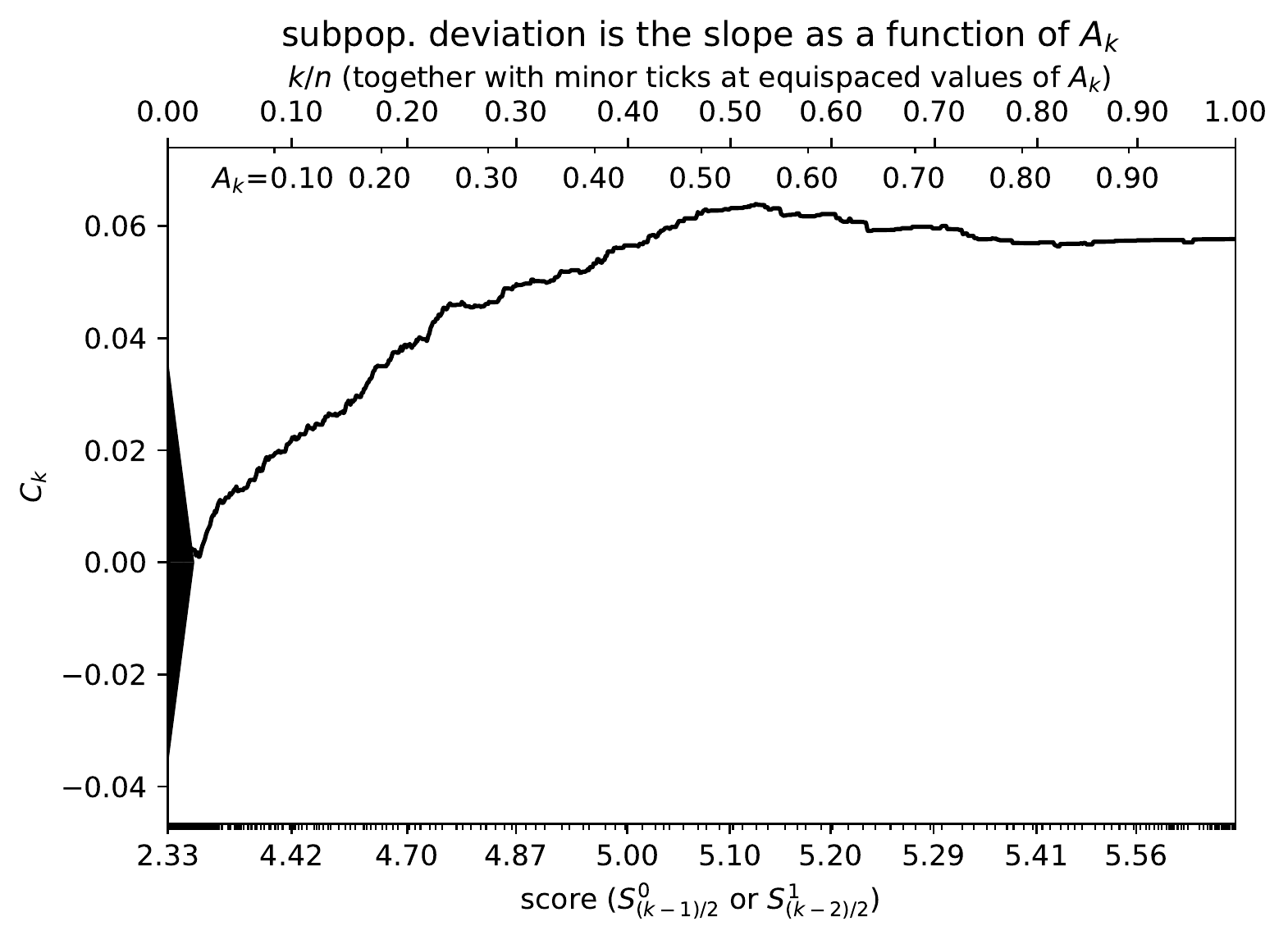}}

\vspace{\vertsep}

(b)
\parbox{\imsize}{\includegraphics[width=\imsize]
{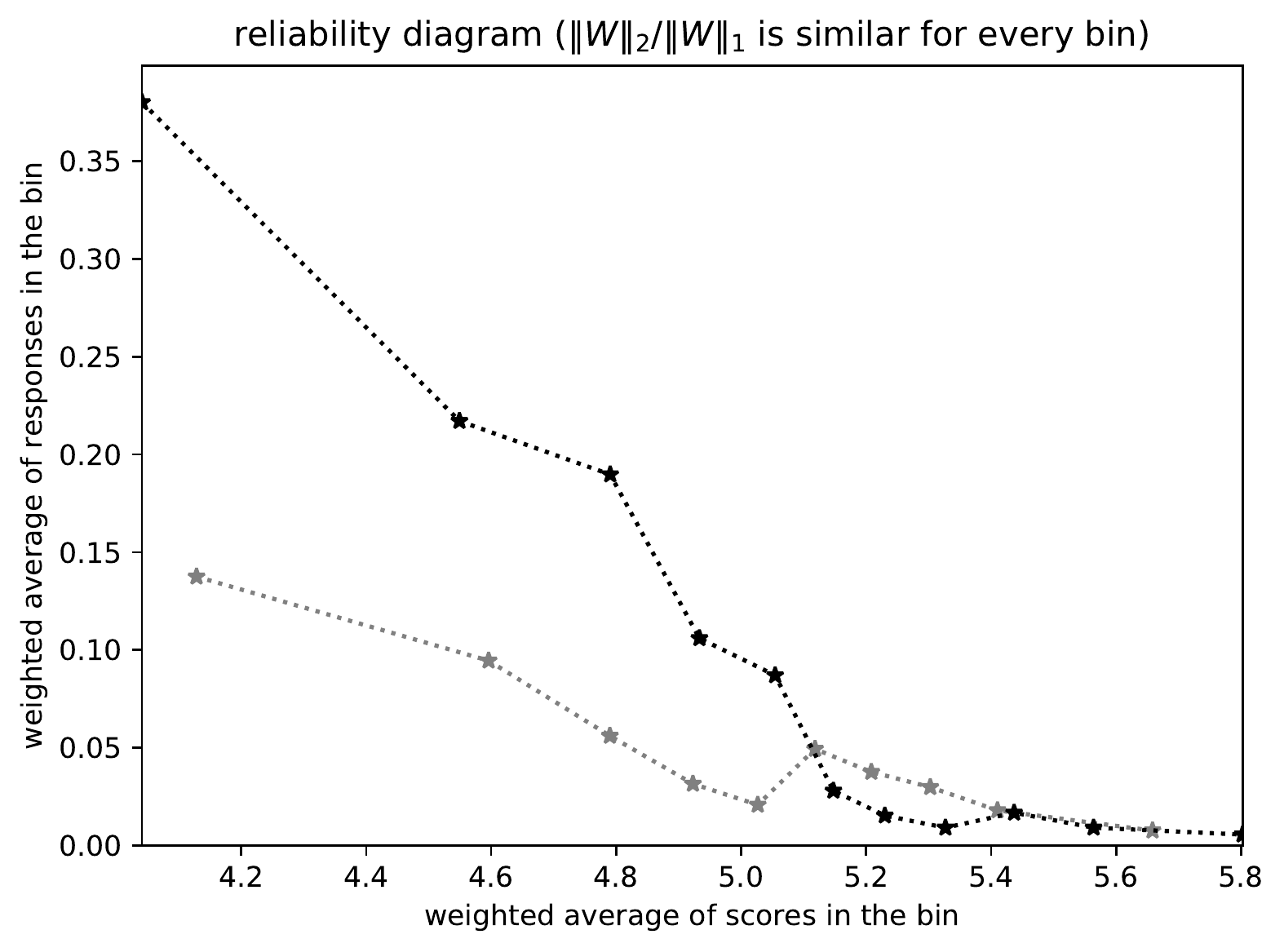}}
\quad\quad
(c)
\parbox{\imsize}{\includegraphics[width=\imsize]
{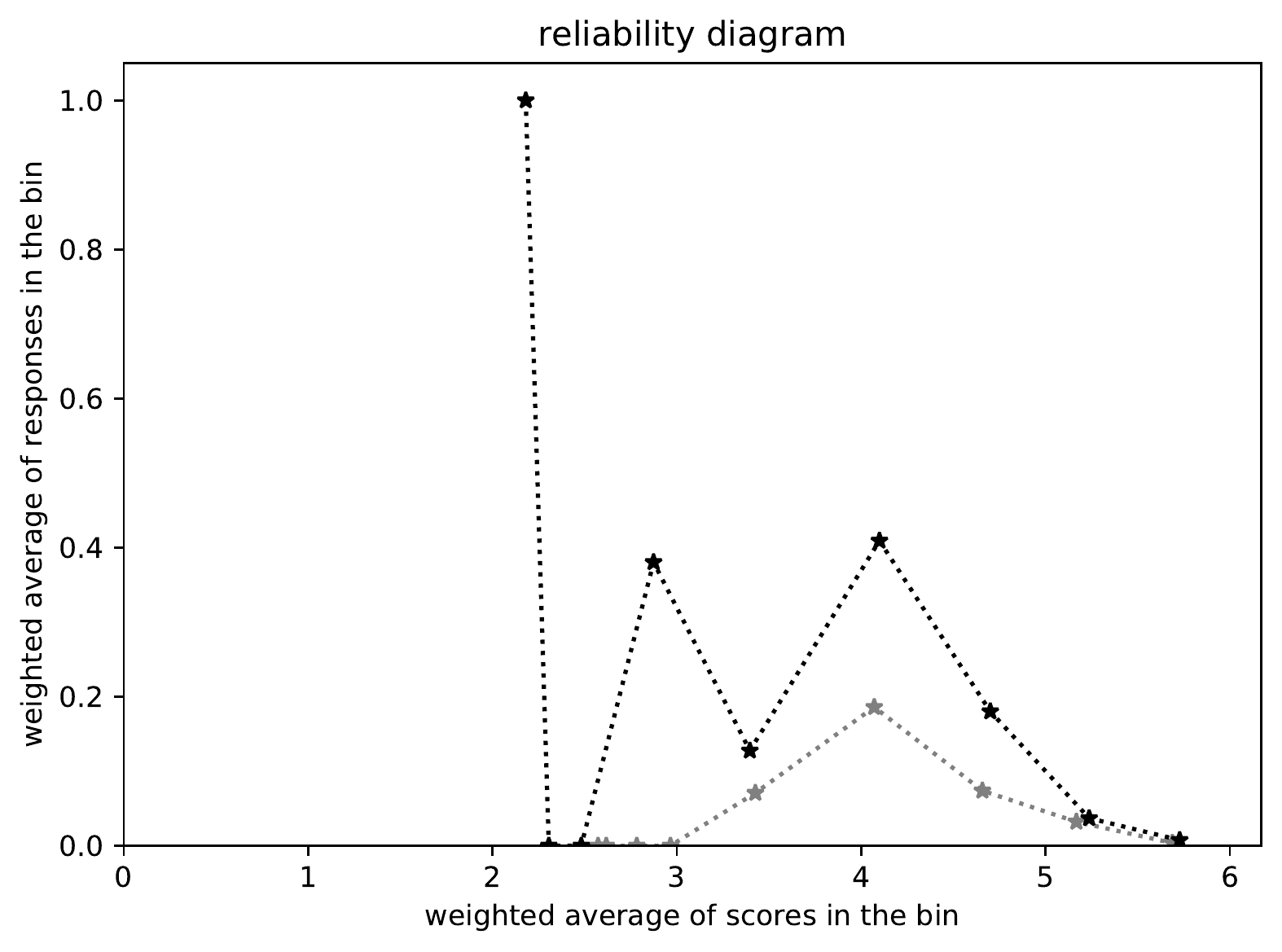}}

\vspace{\vertsep}

(d)
\parbox{\imsize}{\includegraphics[width=\imsize]
{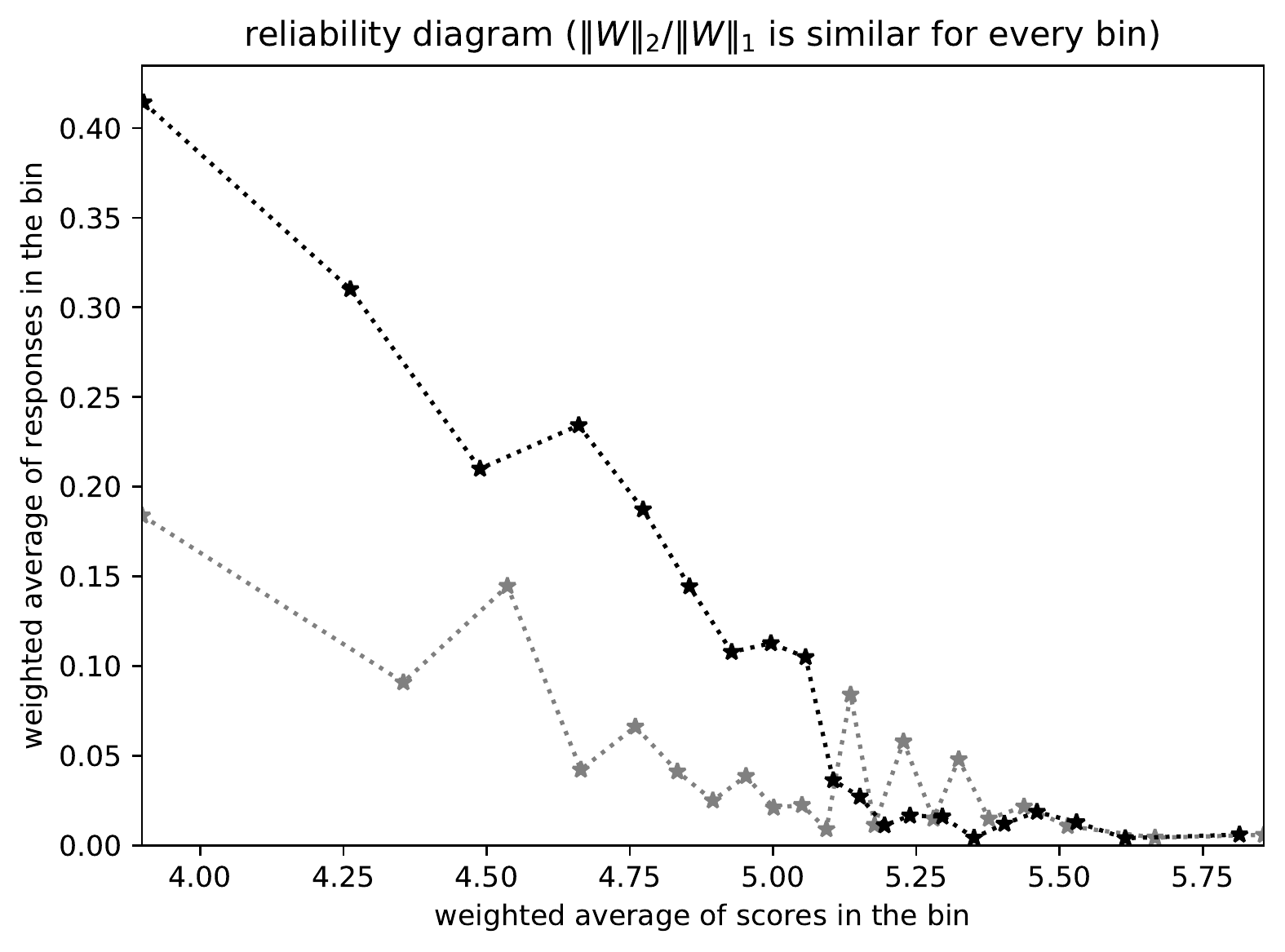}}
\quad\quad
(e)
\parbox{\imsize}{\includegraphics[width=\imsize]
{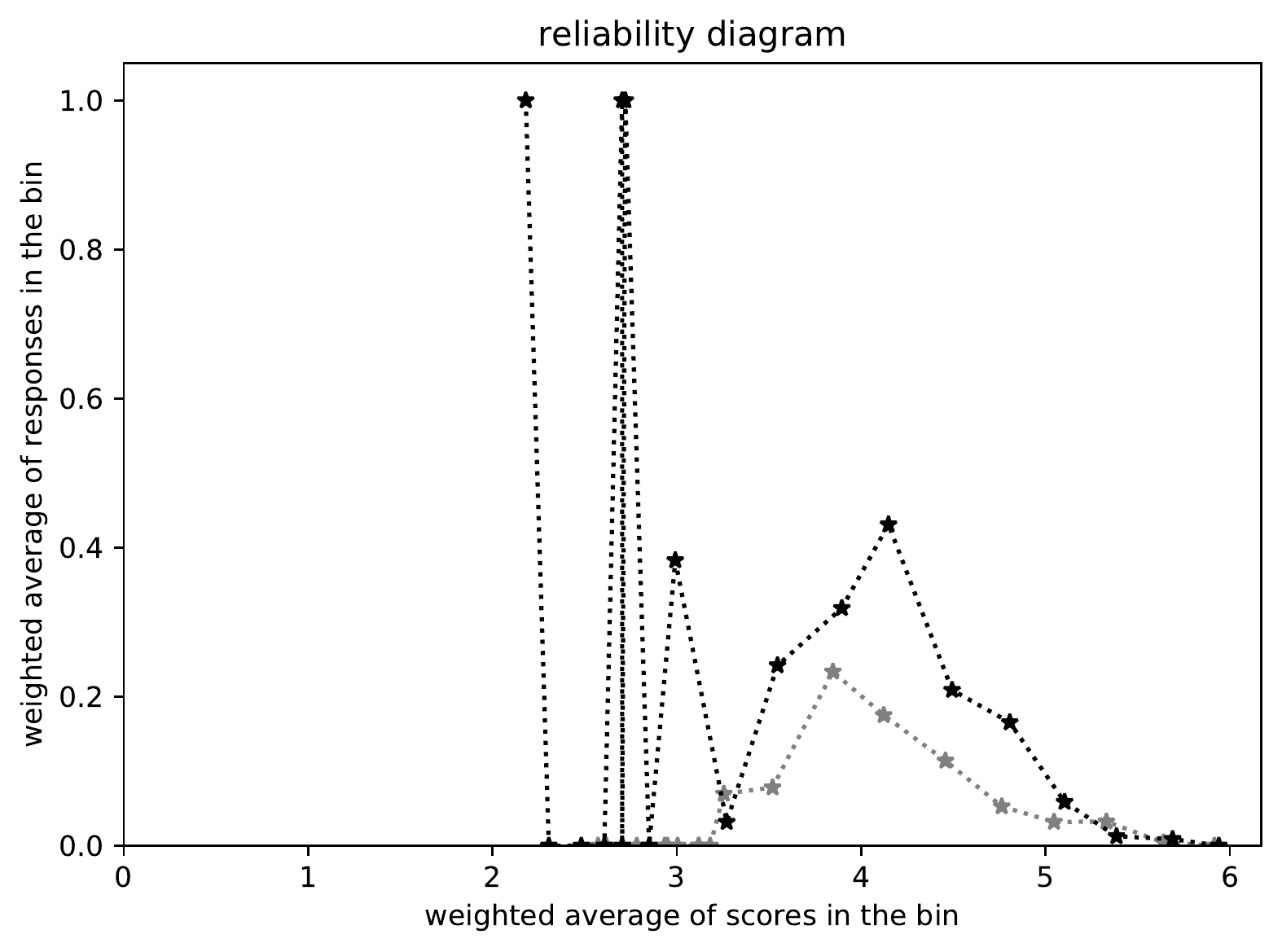}}

\vspace{\vertsep}

(f)
\parbox{\imsize}{\includegraphics[width=\imsize]
{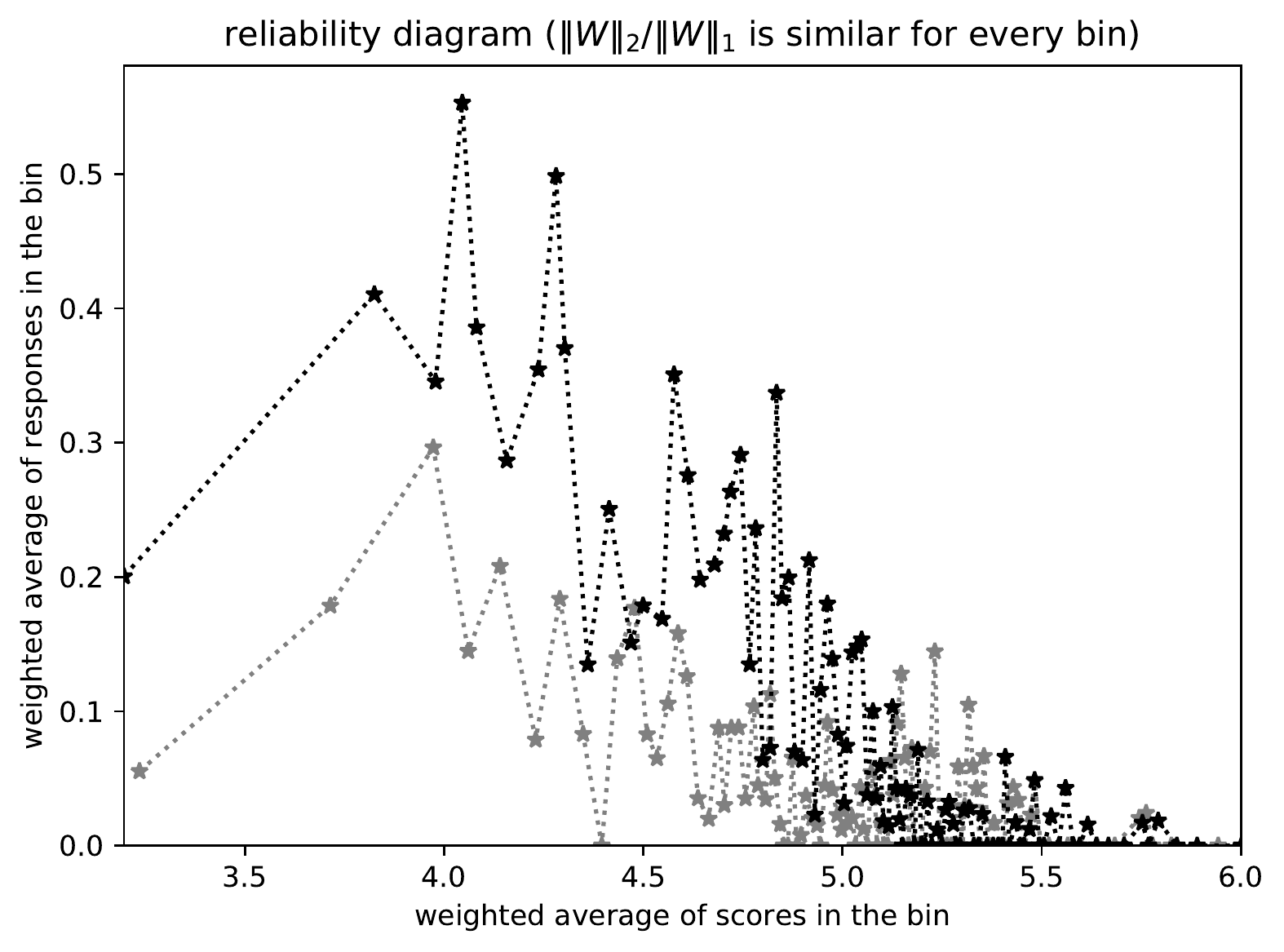}}
\quad\quad
(g)
\parbox{\imsize}{\includegraphics[width=\imsize]
{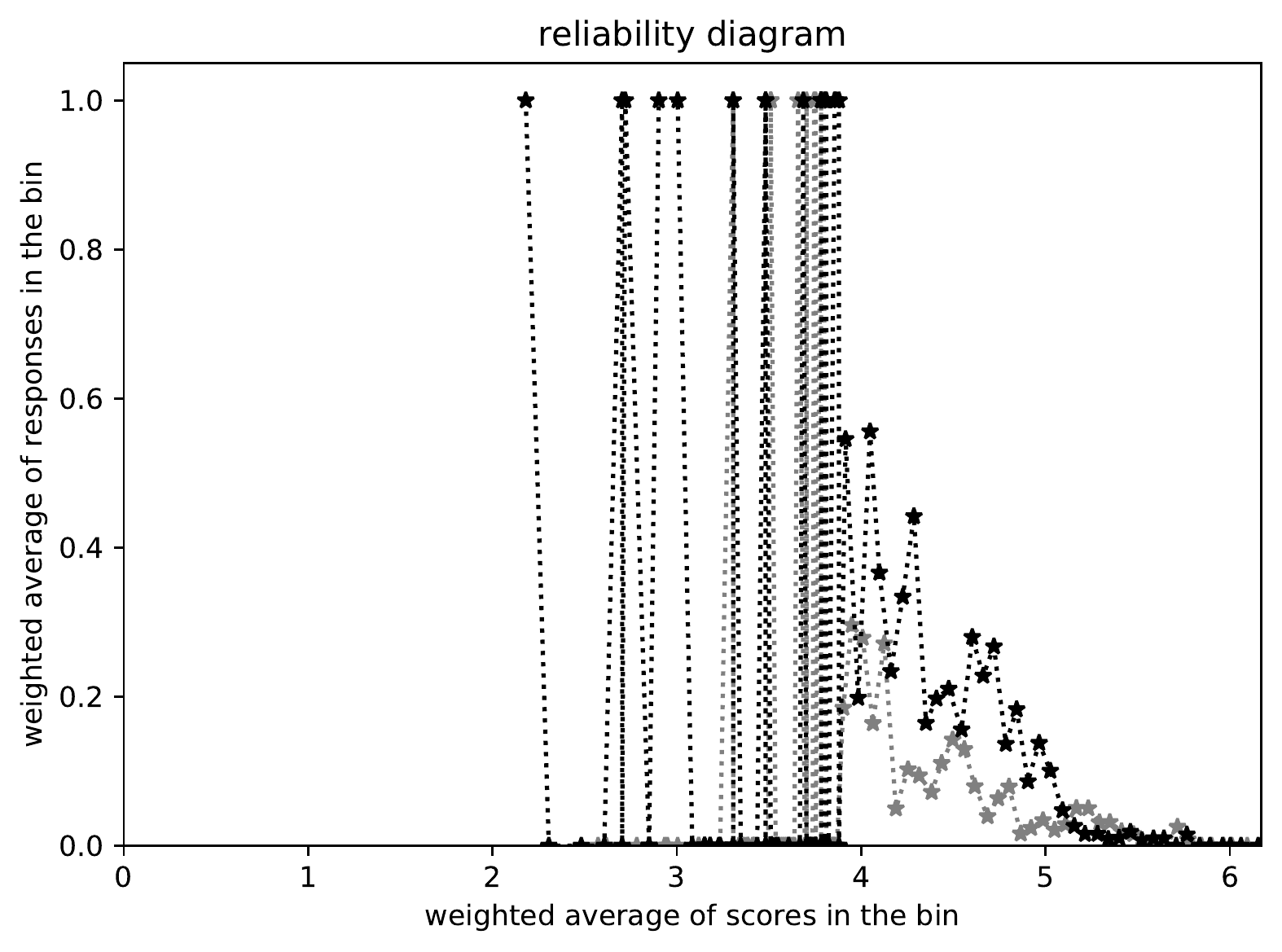}}

\end{centering}
\caption{San Francisco County vs.\ Contra Costa County; $n =$ 3,407;
         Kuiper's statistic is $0.06395 / \sigma = 3.488$,
         Kolmogorov's and Smirnov's is $0.06395 / \sigma = 3.488$;
         only the cumulative graph (a) fully captures the relatively small
         deviation for the very lowest scores, and having even just 100 bins
         in a reliability diagram (f and g) already produces far too much noise
         for most scores.
         The metrics of Kuiper and of Kolmogorov and Smirnov report
         statistically significant deviation between the subpopulations.
}
\label{San_Francisco-Contra_Costa}
\end{figure}

\begin{figure}
\begin{centering}

(a)
\parbox{\imsize}{\includegraphics[width=\imsize]
{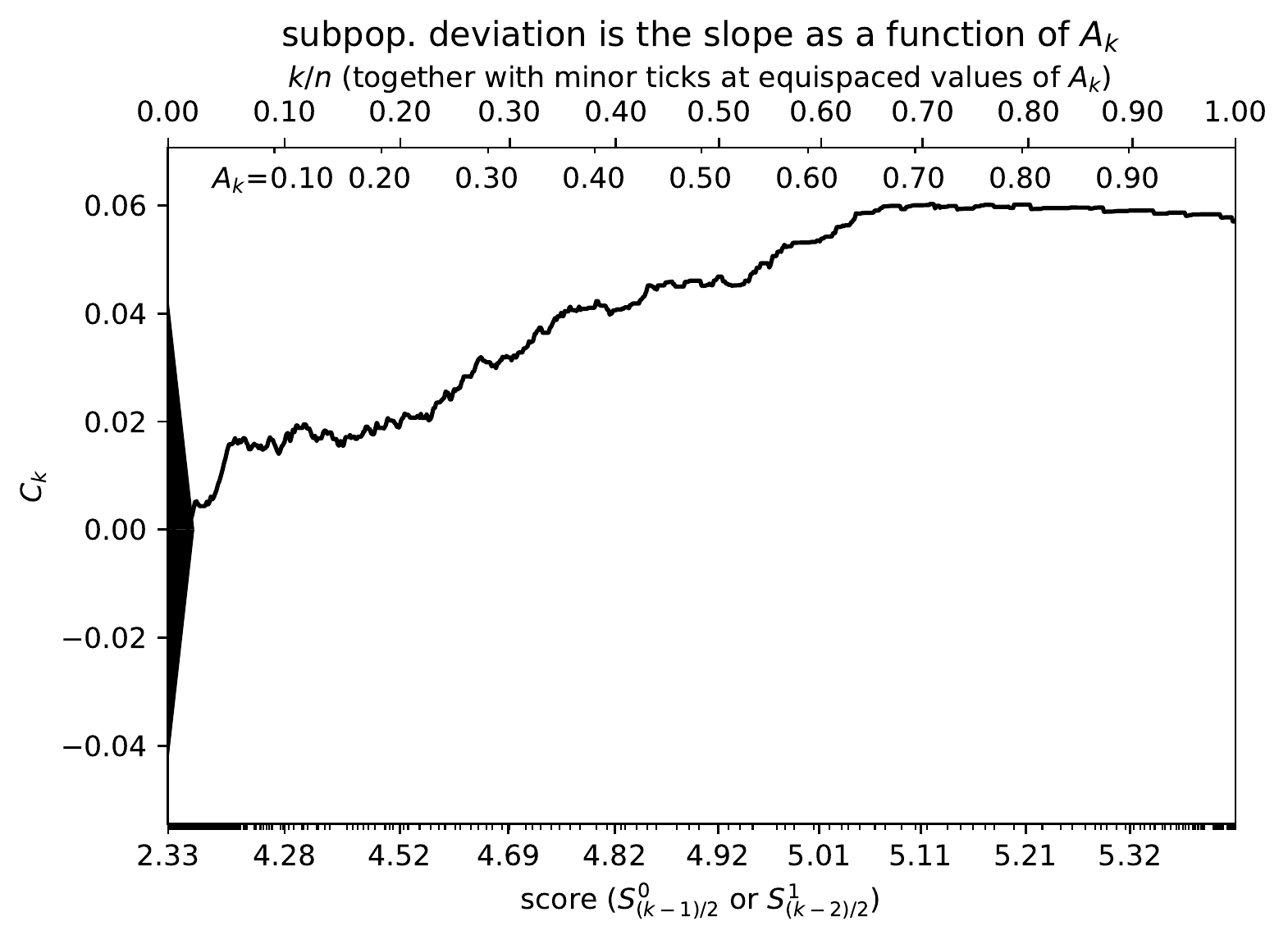}}

\vspace{\vertsep}

(b)
\parbox{\imsize}{\includegraphics[width=\imsize]
{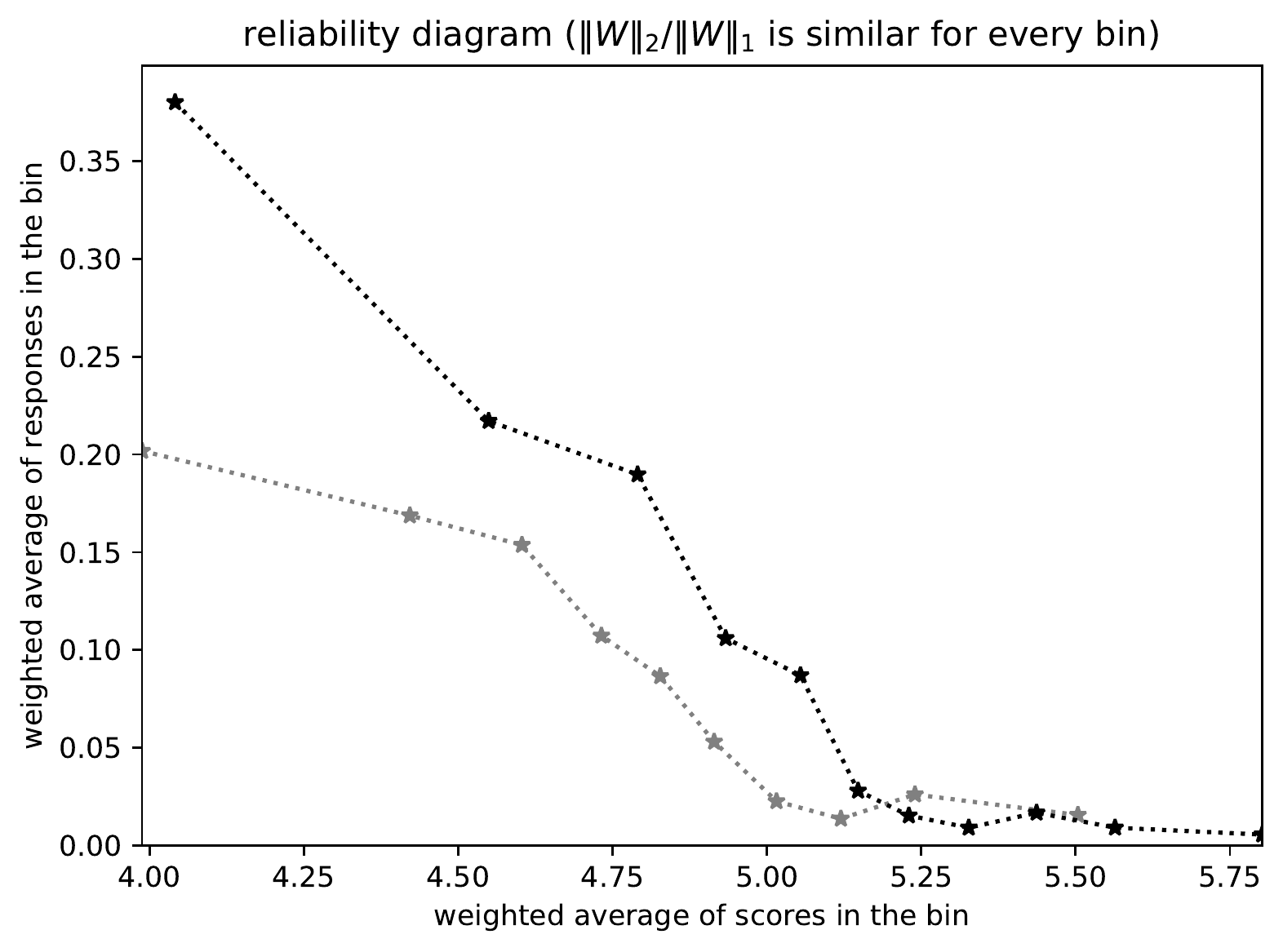}}
\quad\quad
(c)
\parbox{\imsize}{\includegraphics[width=\imsize]
{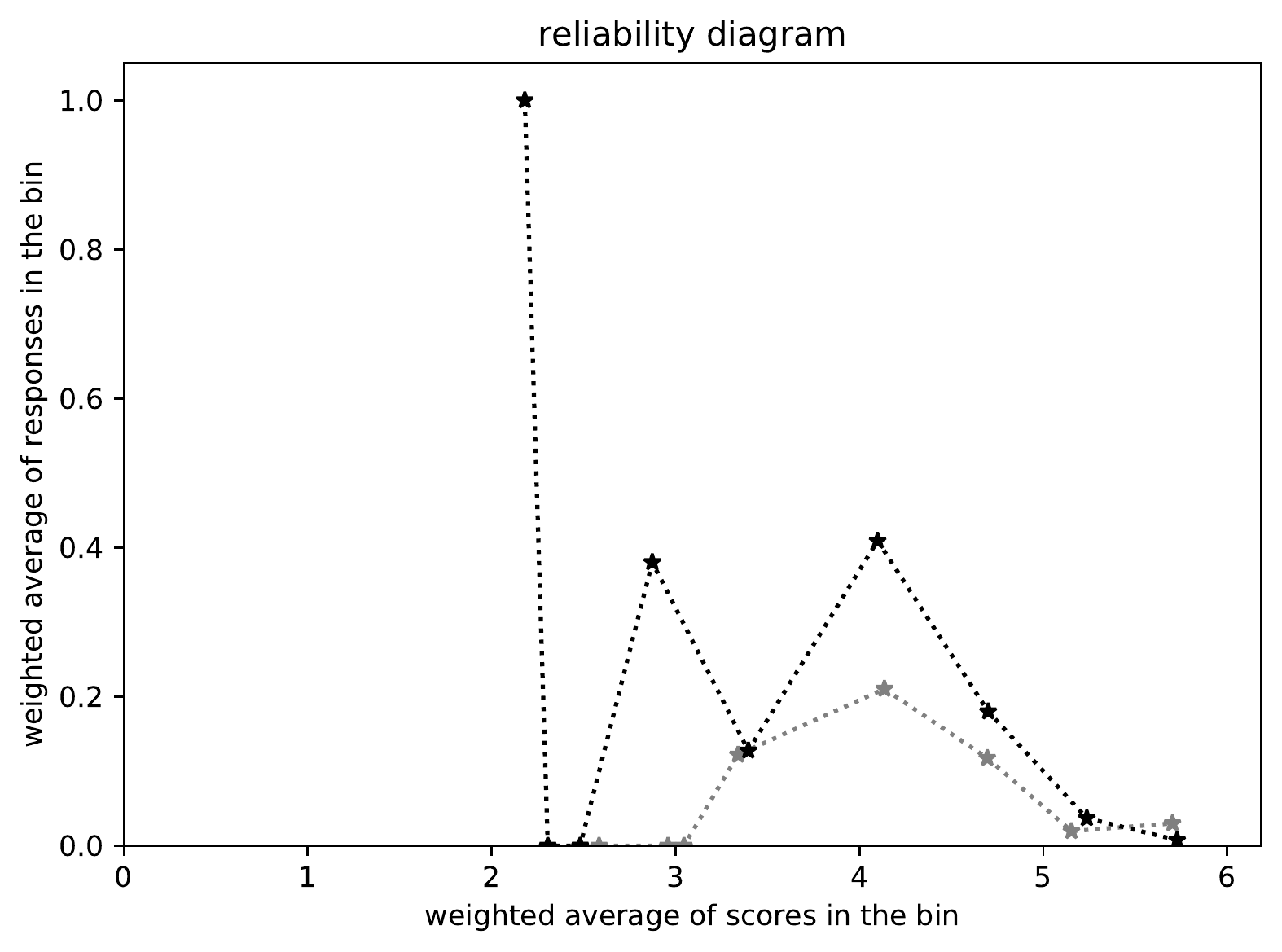}}

\vspace{\vertsep}

(d)
\parbox{\imsize}{\includegraphics[width=\imsize]
{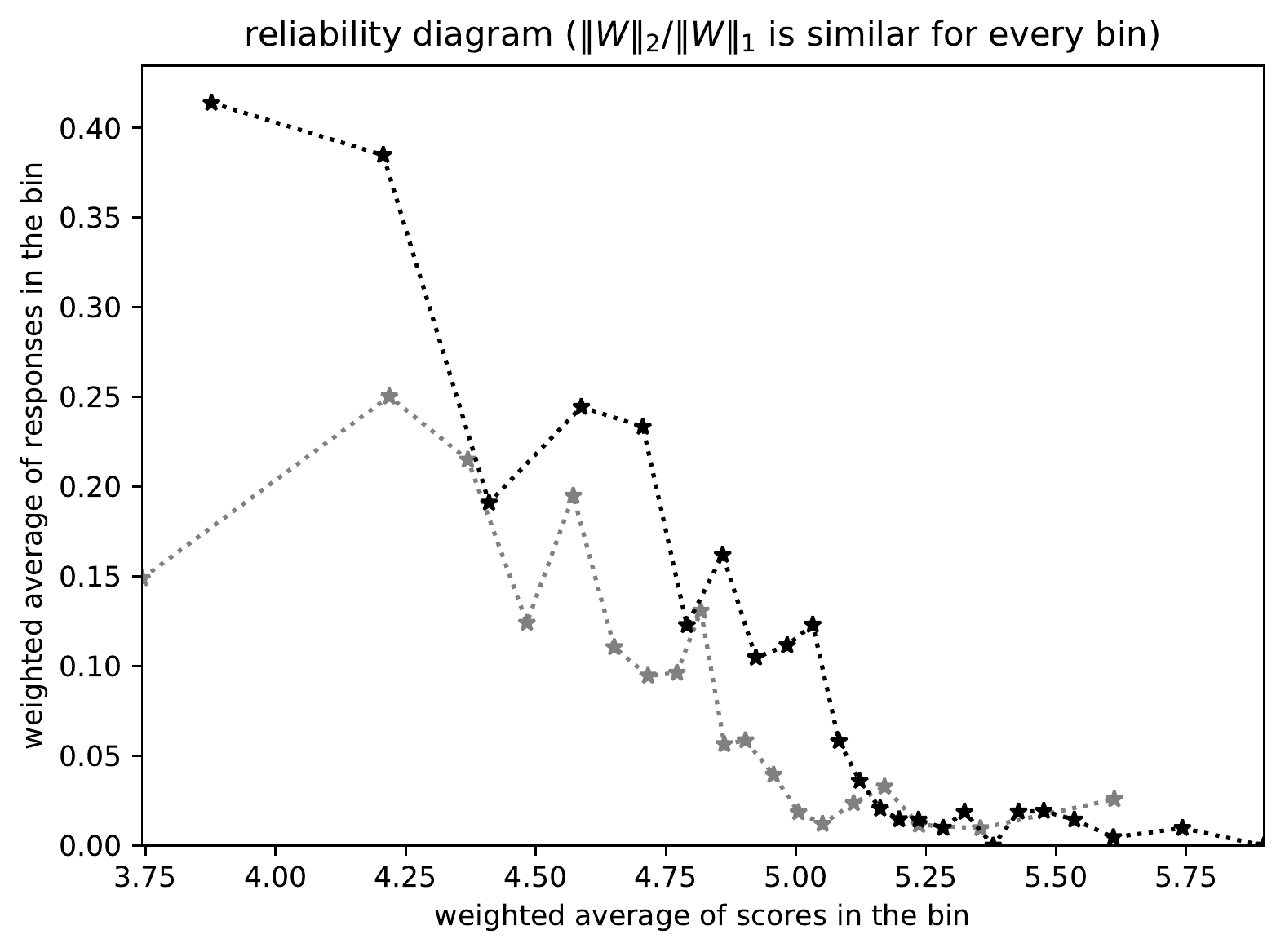}}
\quad\quad
(e)
\parbox{\imsize}{\includegraphics[width=\imsize]
{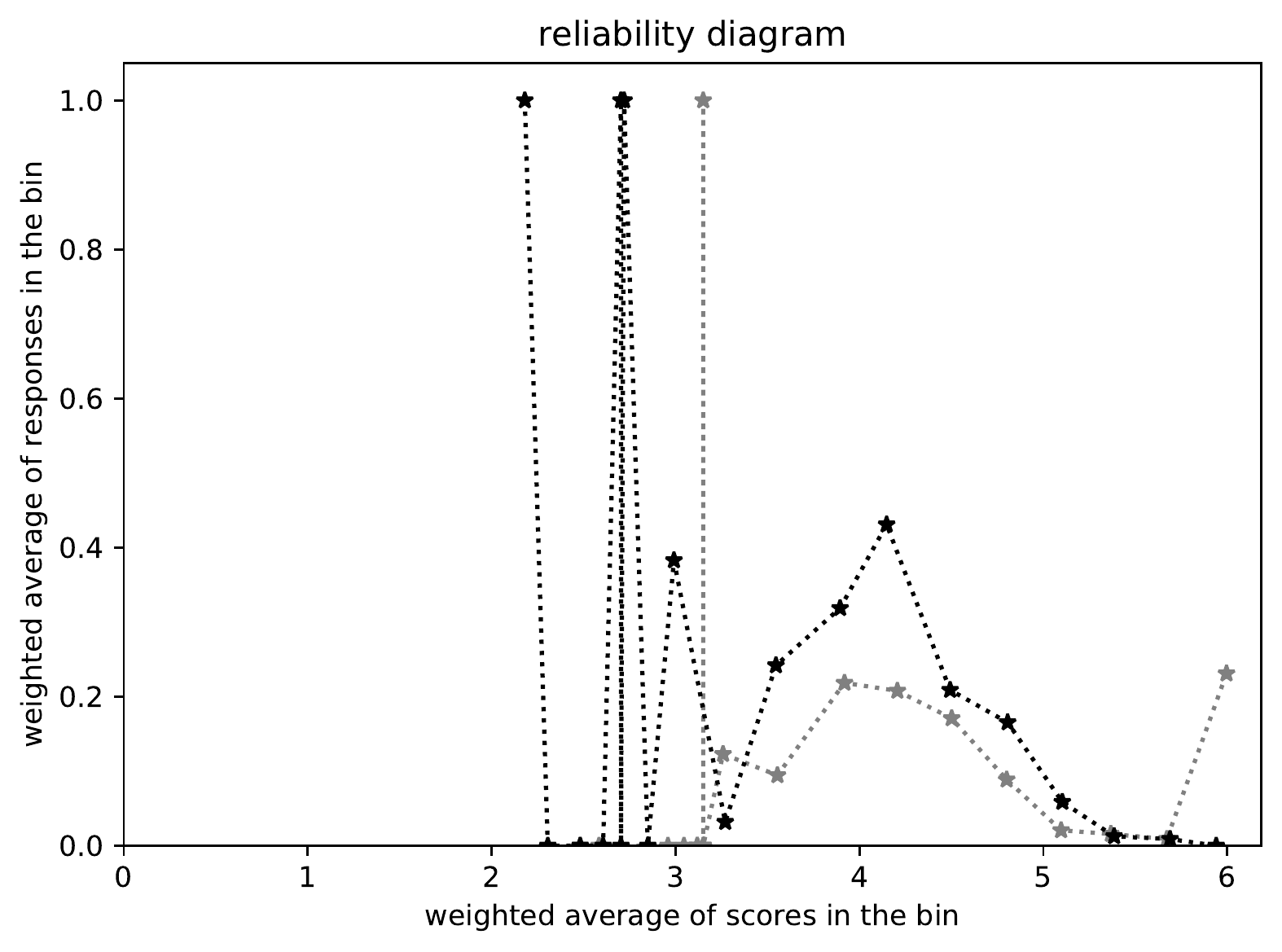}}

\vspace{\vertsep}

(f)
\parbox{\imsize}{\includegraphics[width=\imsize]
{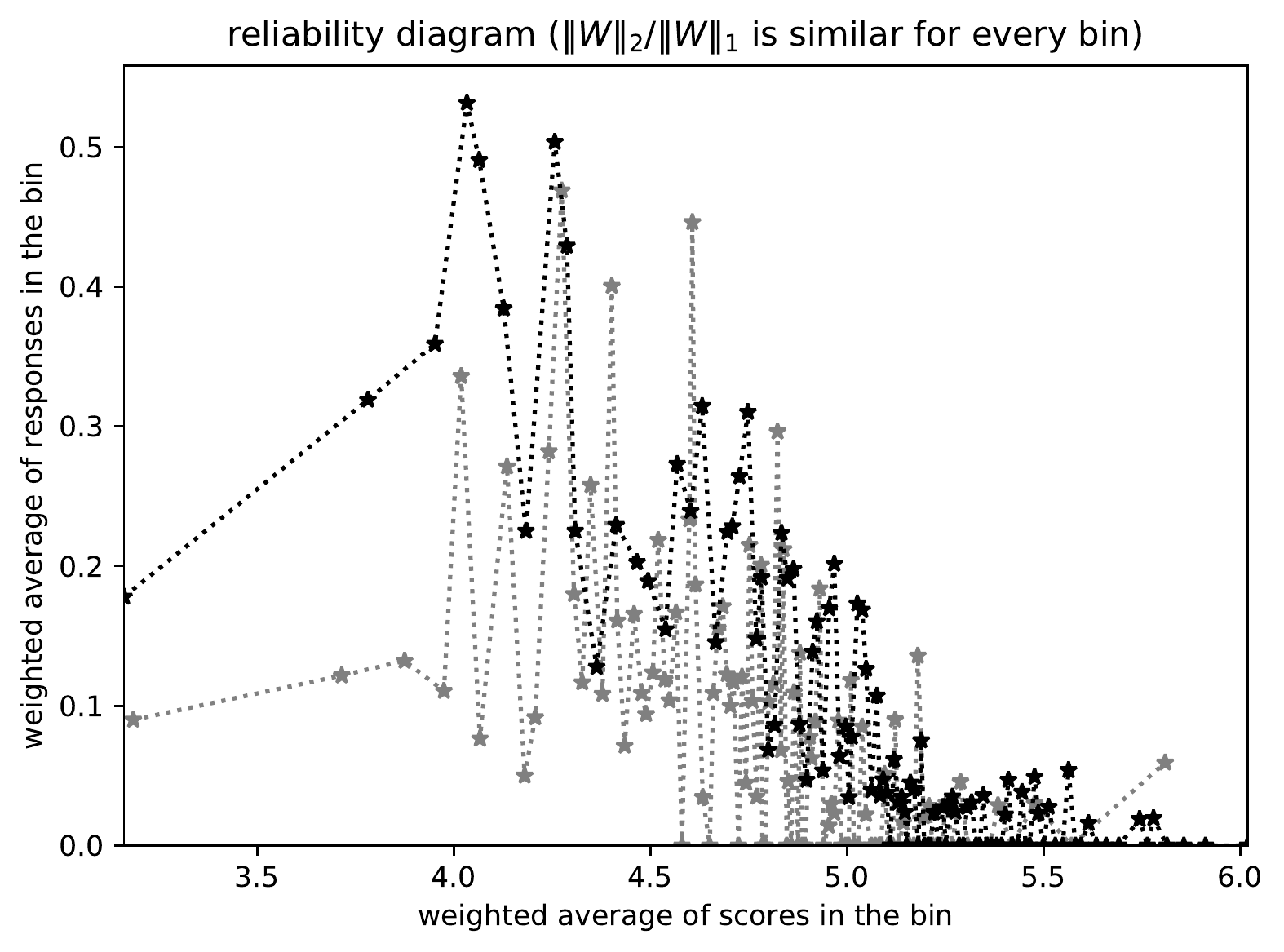}}
\quad\quad
(g)
\parbox{\imsize}{\includegraphics[width=\imsize]
{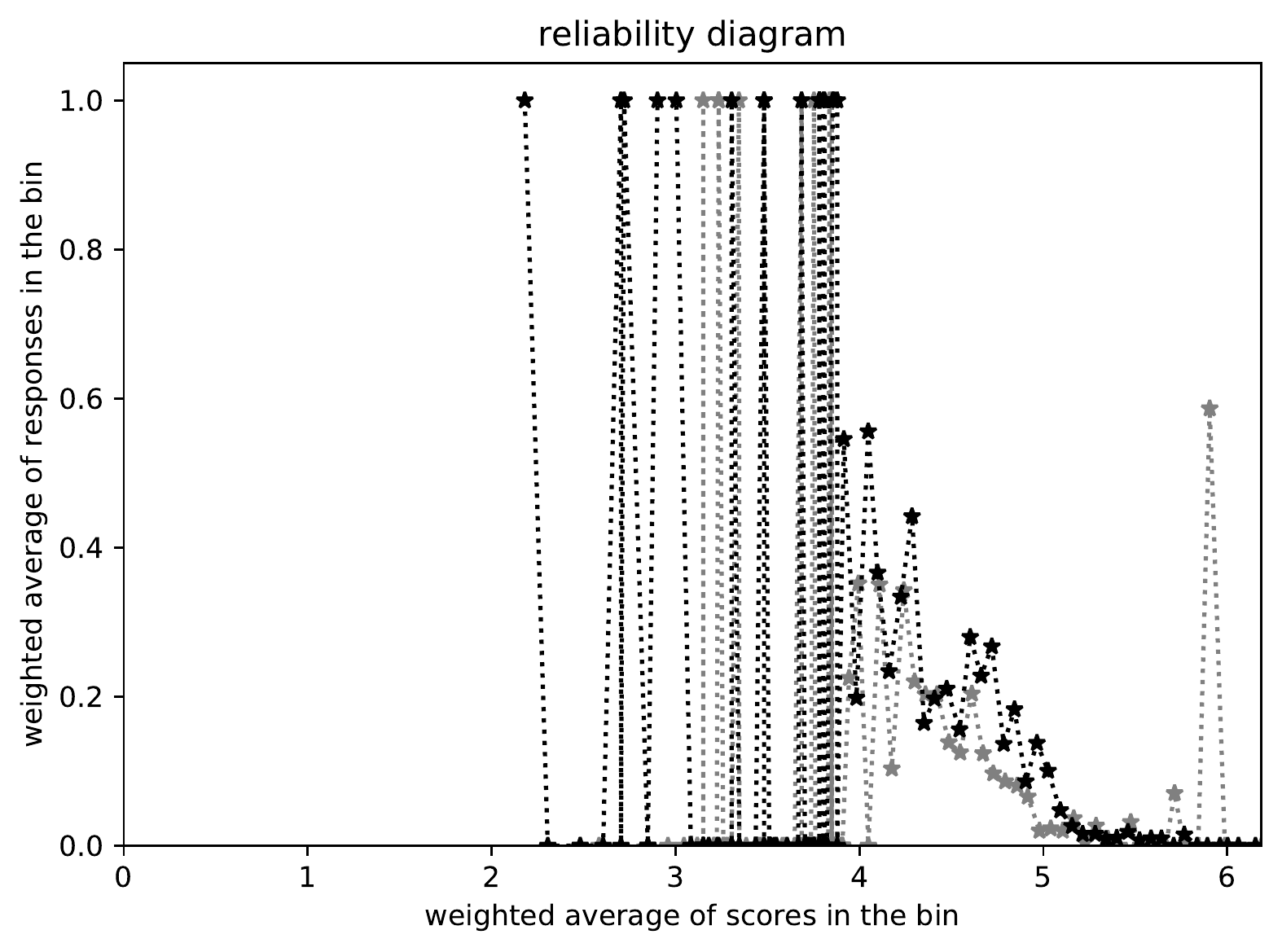}}

\end{centering}
\caption{San Francisco County vs.\ San Joaquin County; $n =$ 2,358;
         Kuiper's statistic is $0.06160 / \sigma = 2.794$,
         Kolmogorov's and Smirnov's is $0.06025 / \sigma = 2.733$;
         only the cumulative graph (a) and the otherwise extremely noisy
         reliability diagrams each with 100 bins (f and g) fully detail
         the sharp spike at scores just slightly greater than 4.
         The metrics of Kuiper and of Kolmogorov and Smirnov report
         some statistically significant deviation between the subpopulations.
}
\label{San_Francisco-San_Joaquin}
\end{figure}

\begin{figure}
\begin{centering}

(a)
\parbox{\imsize}{\includegraphics[width=\imsize]
{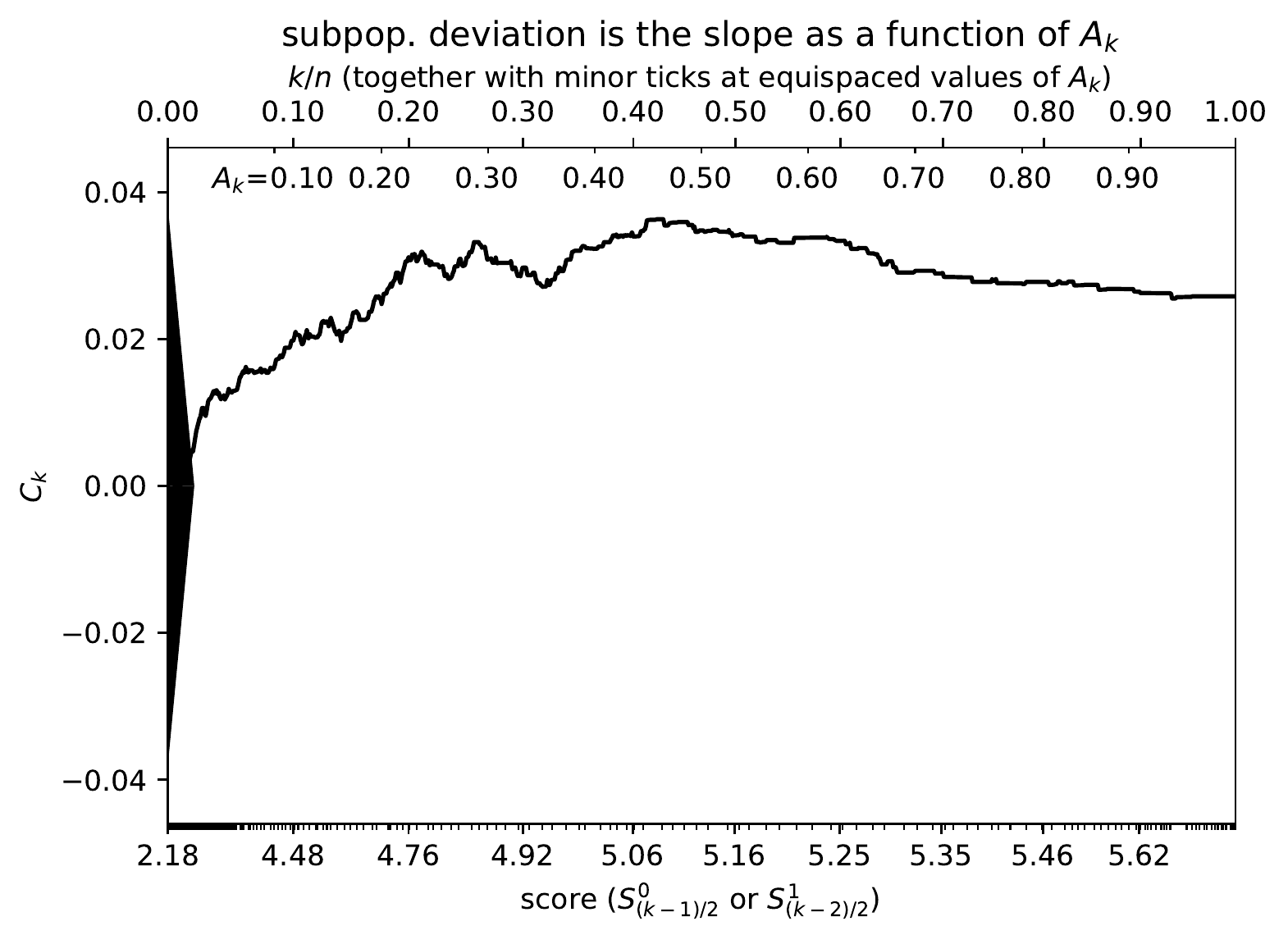}}

\vspace{\vertsep}

(b)
\parbox{\imsize}{\includegraphics[width=\imsize]
{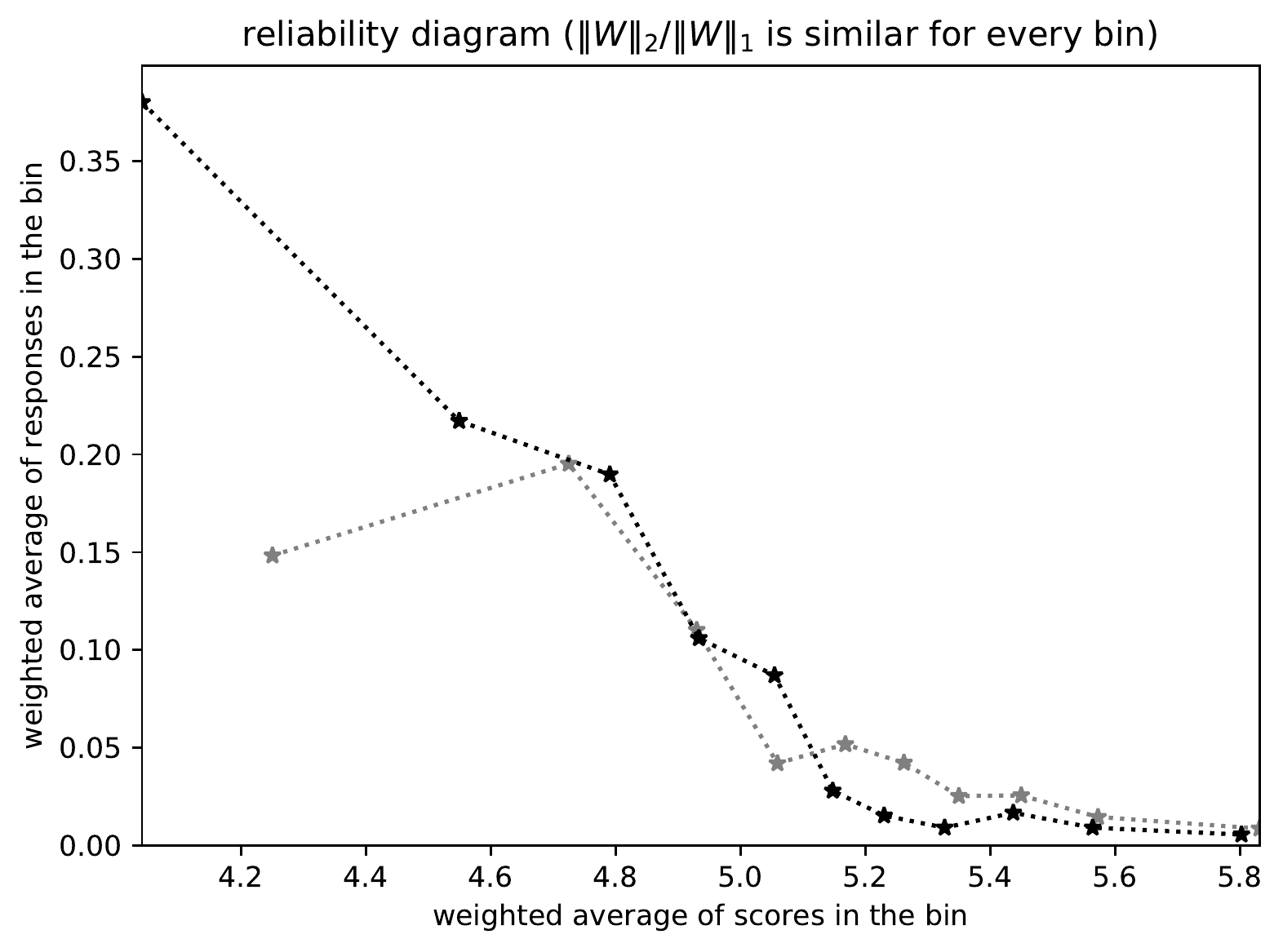}}
\quad\quad
(c)
\parbox{\imsize}{\includegraphics[width=\imsize]
{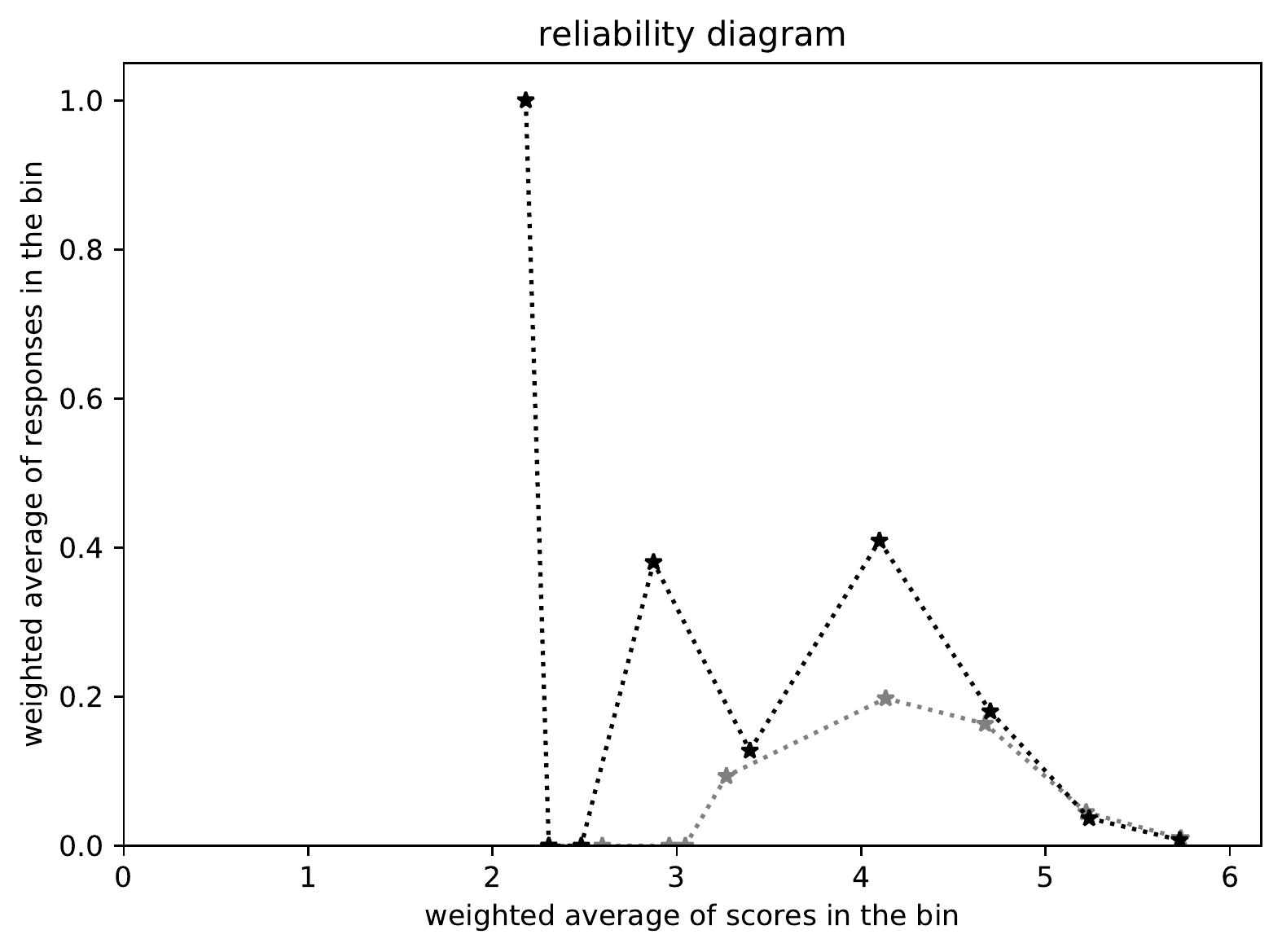}}

\vspace{\vertsep}

(d)
\parbox{\imsize}{\includegraphics[width=\imsize]
{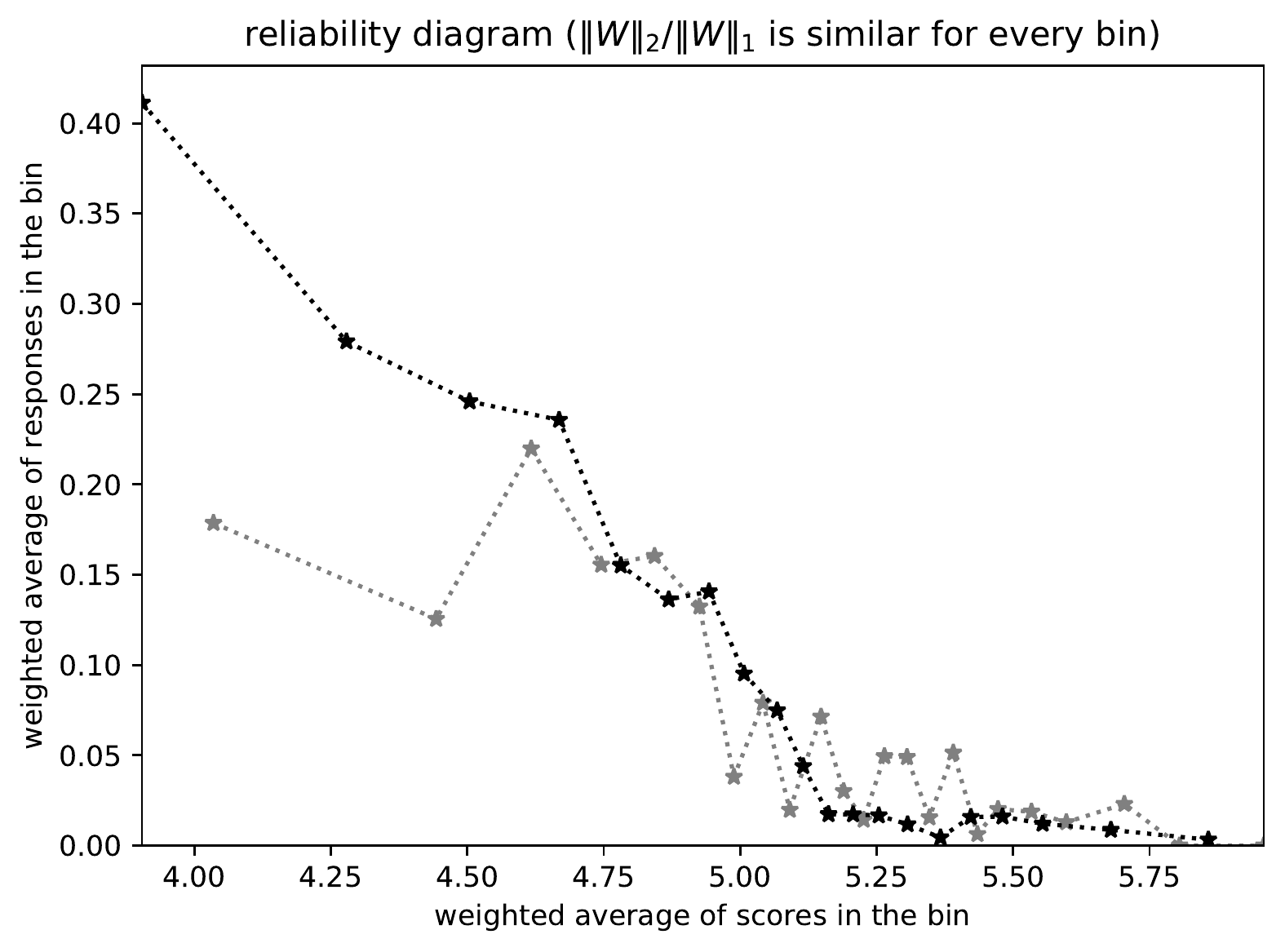}}
\quad\quad
(e)
\parbox{\imsize}{\includegraphics[width=\imsize]
{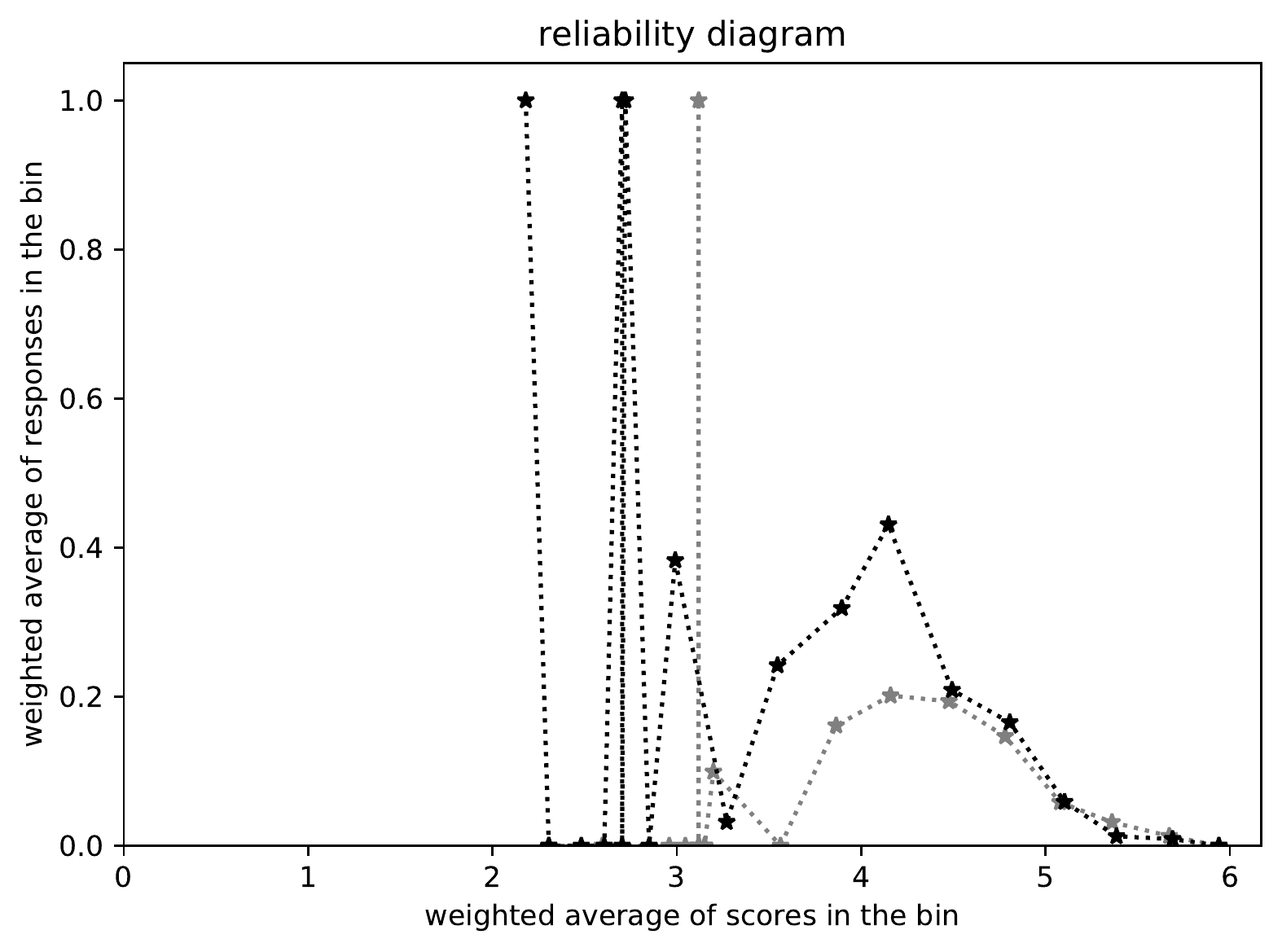}}

\vspace{\vertsep}

(f)
\parbox{\imsize}{\includegraphics[width=\imsize]
{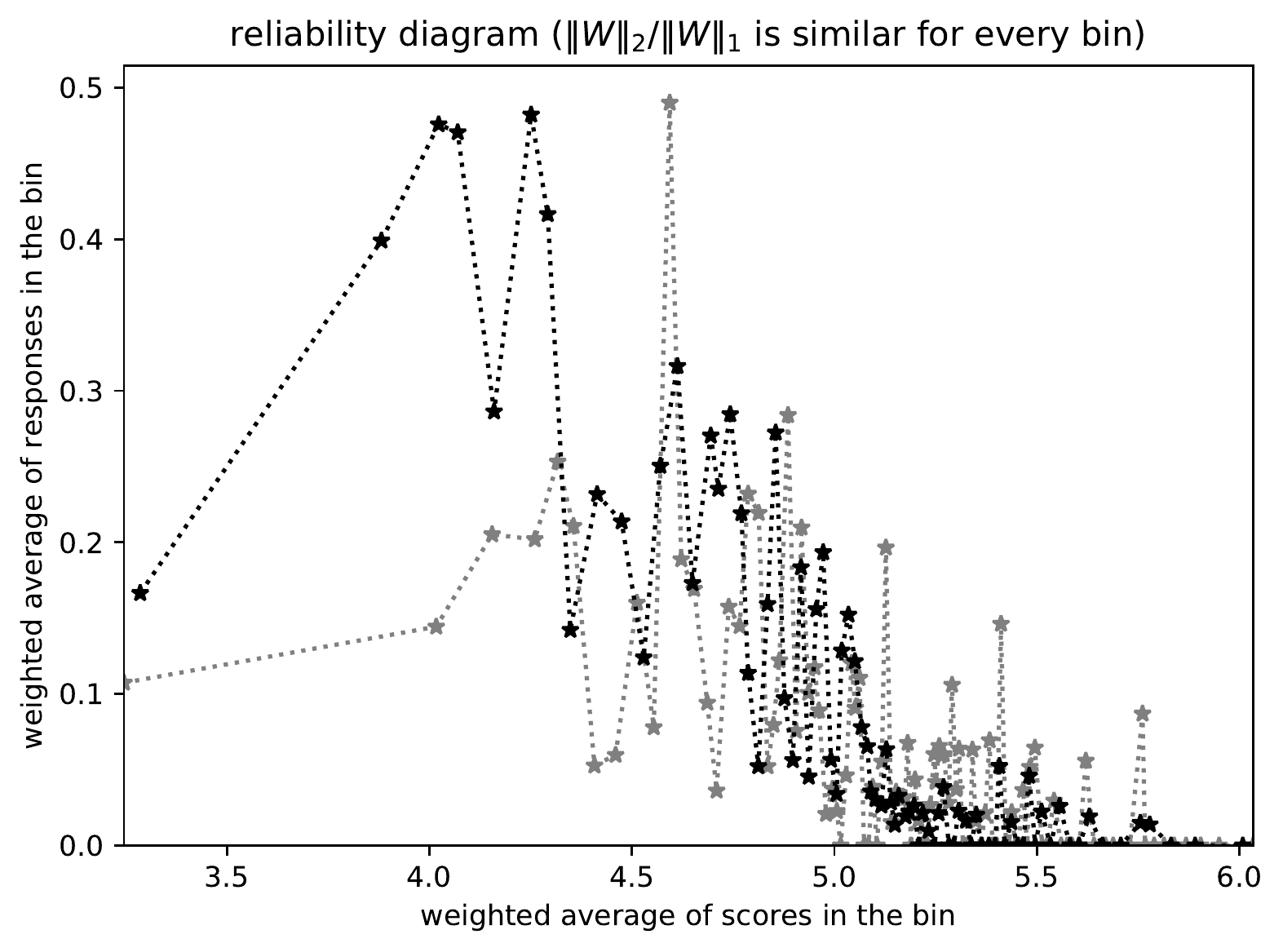}}
\quad\quad
(g)
\parbox{\imsize}{\includegraphics[width=\imsize]
{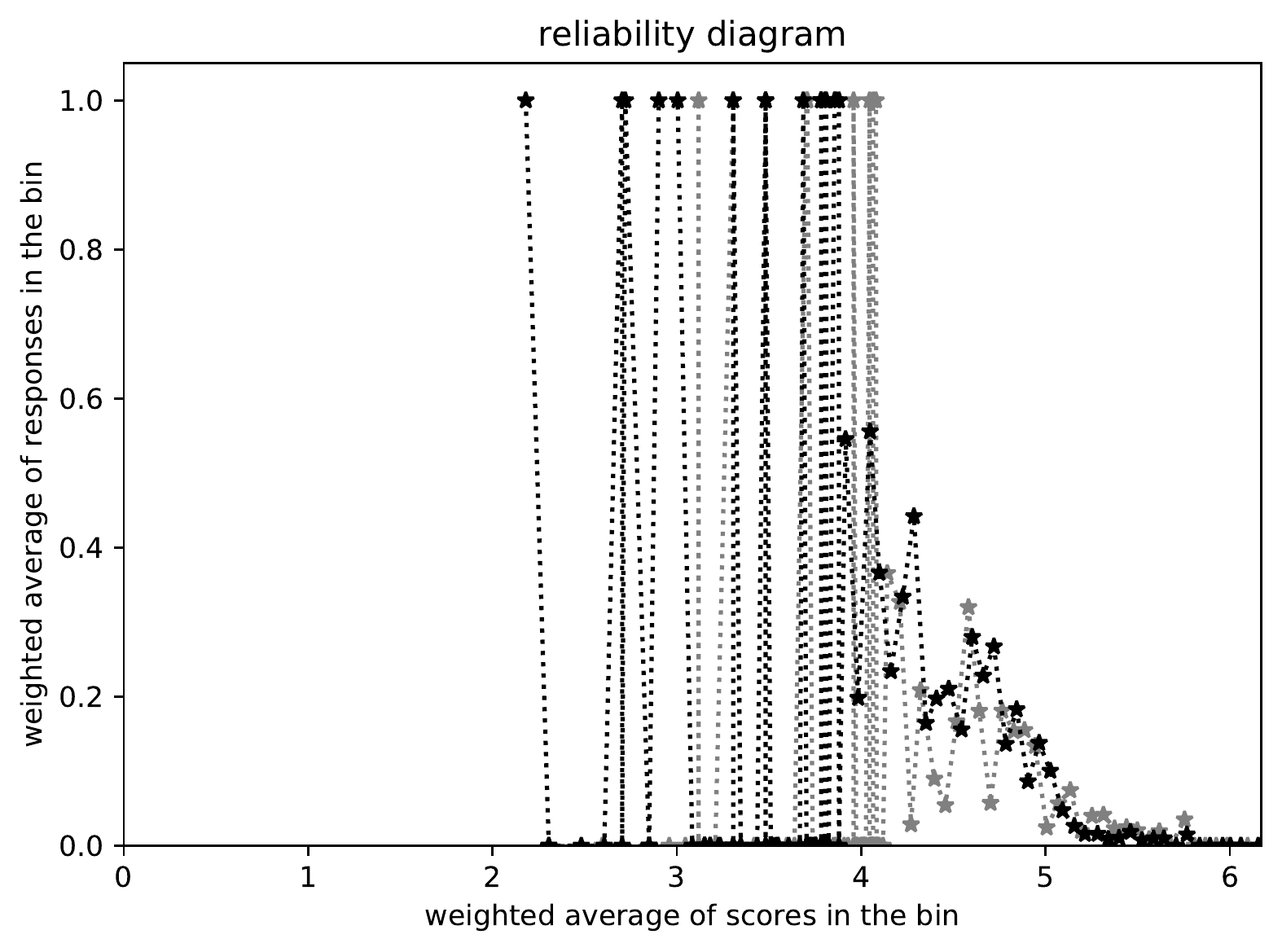}}

\end{centering}
\caption{San Francisco County vs.\ San Mateo County; $n =$ 3,147;
         Kuiper's statistic is $0.03688 / \sigma = 1.923$,
         Kolmogorov's and Smirnov's is $0.03631 / \sigma = 1.893$;
         resolving the full extent of the spike
         at some of the lowest scores in the cumulative graph (a) requires
         at least 100 bins in the reliability diagrams (f and g),
         but then the reliability diagrams are too noisy at the other scores.
         The statistics of Kuiper and of Kolmogorov and Smirnov do not report
         very statistically significant deviation between the subpopulations.
}
\label{San_Francisco-San_Mateo}
\end{figure}

\begin{figure}
\begin{centering}

(a)
\parbox{\imsize}{\includegraphics[width=\imsize]
{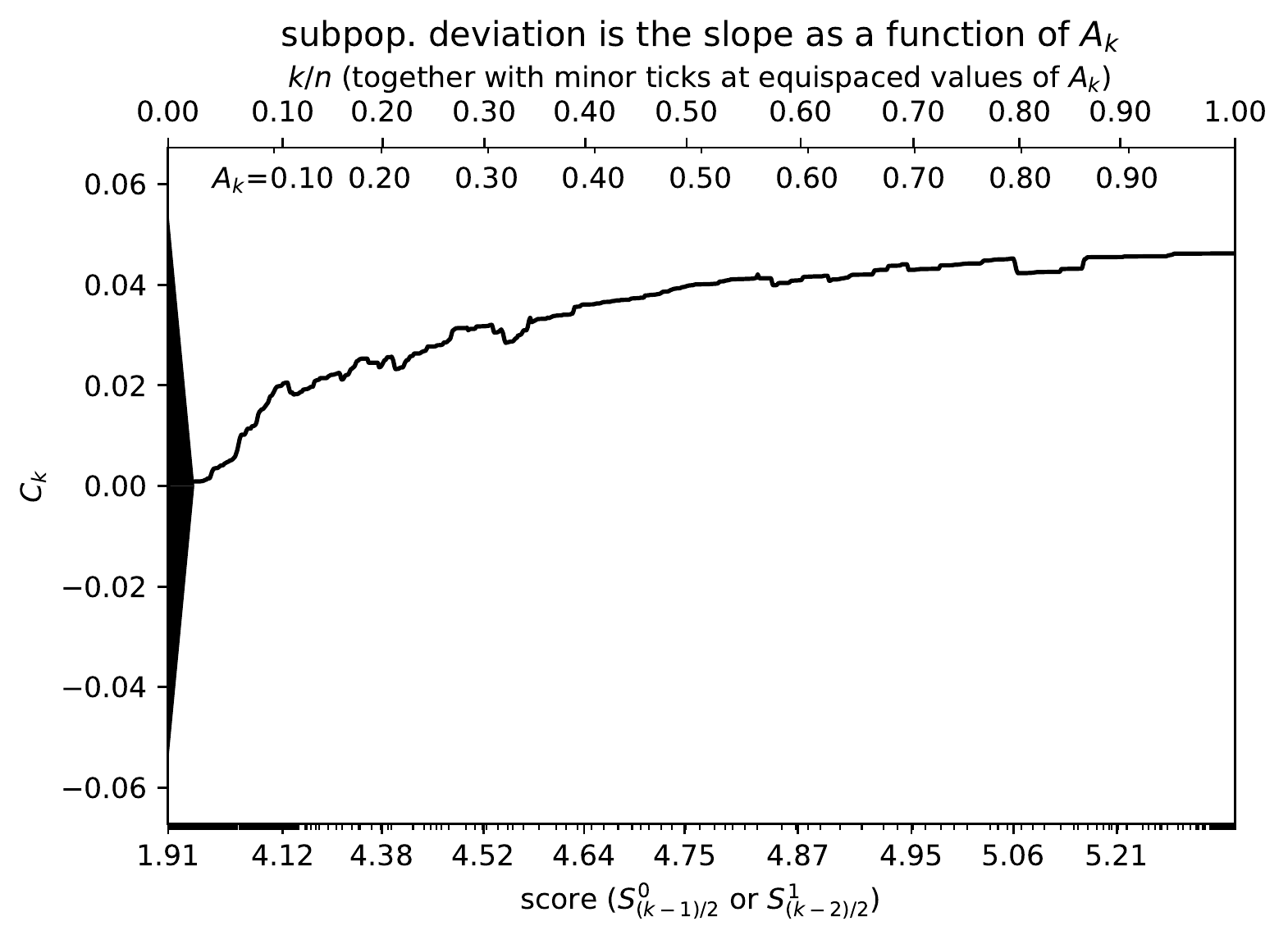}}

\vspace{\vertsep}

(b)
\parbox{\imsize}{\includegraphics[width=\imsize]
{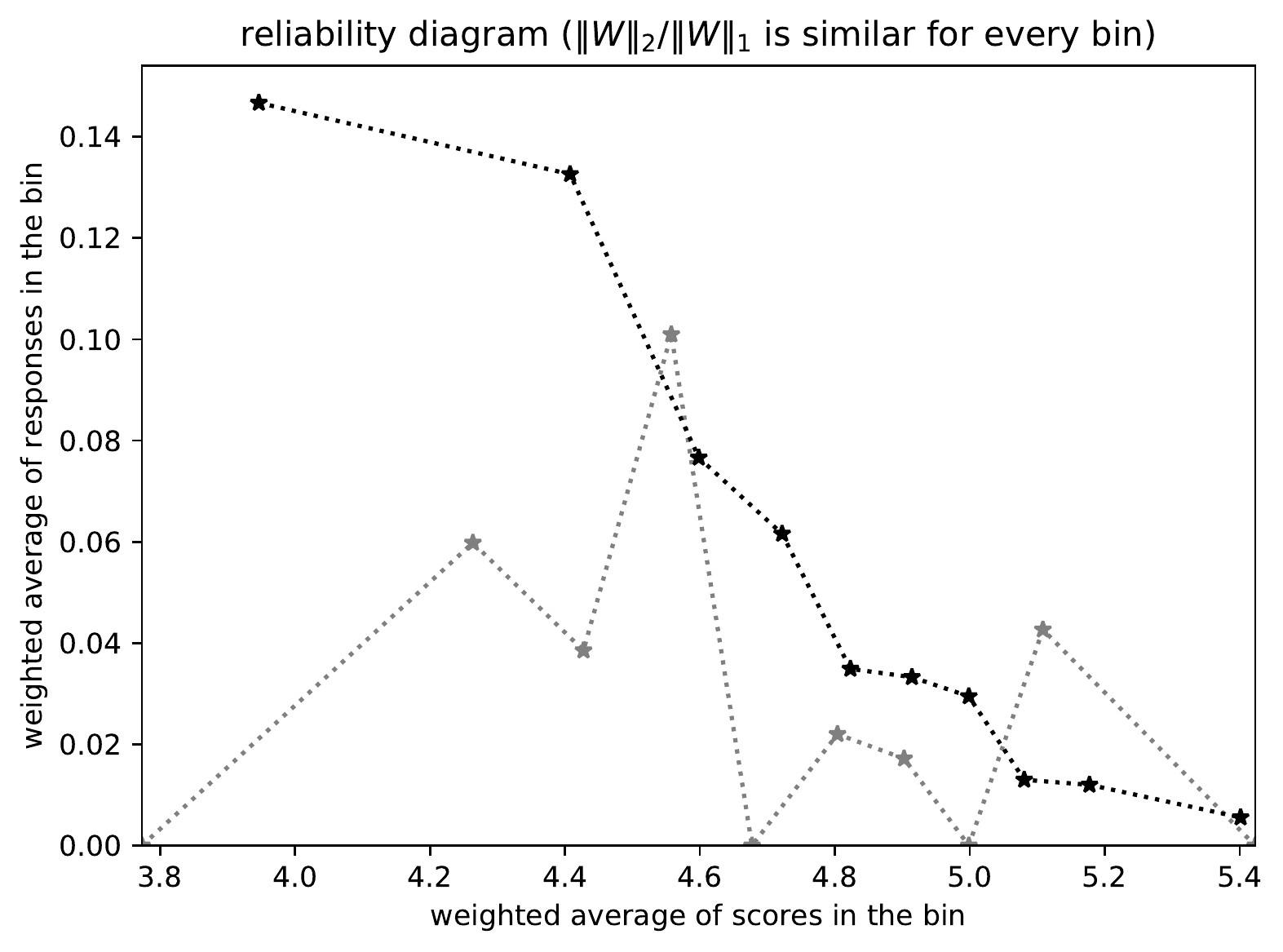}}
\quad\quad
(c)
\parbox{\imsize}{\includegraphics[width=\imsize]
{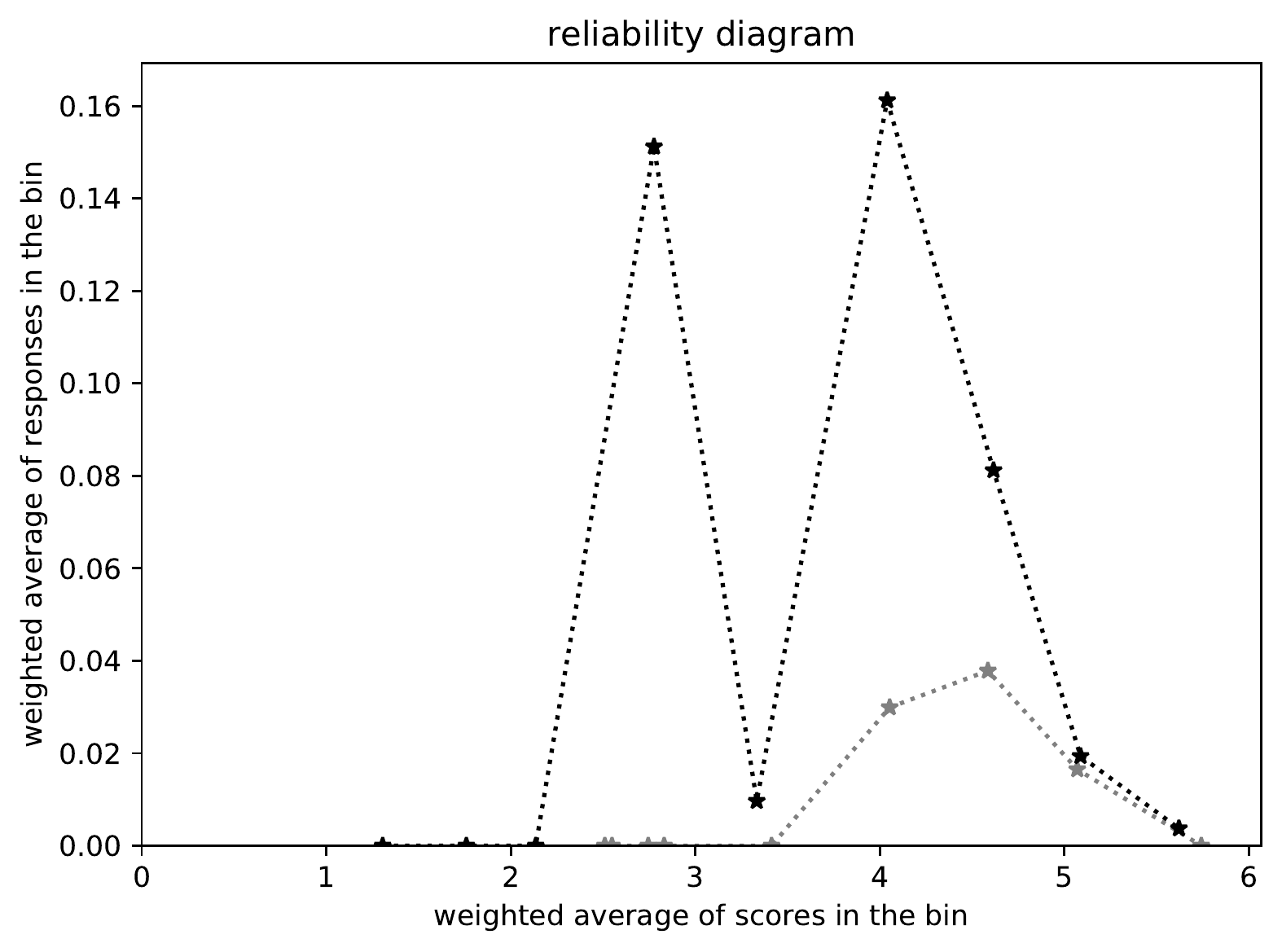}}

\vspace{\vertsep}

(d)
\parbox{\imsize}{\includegraphics[width=\imsize]
{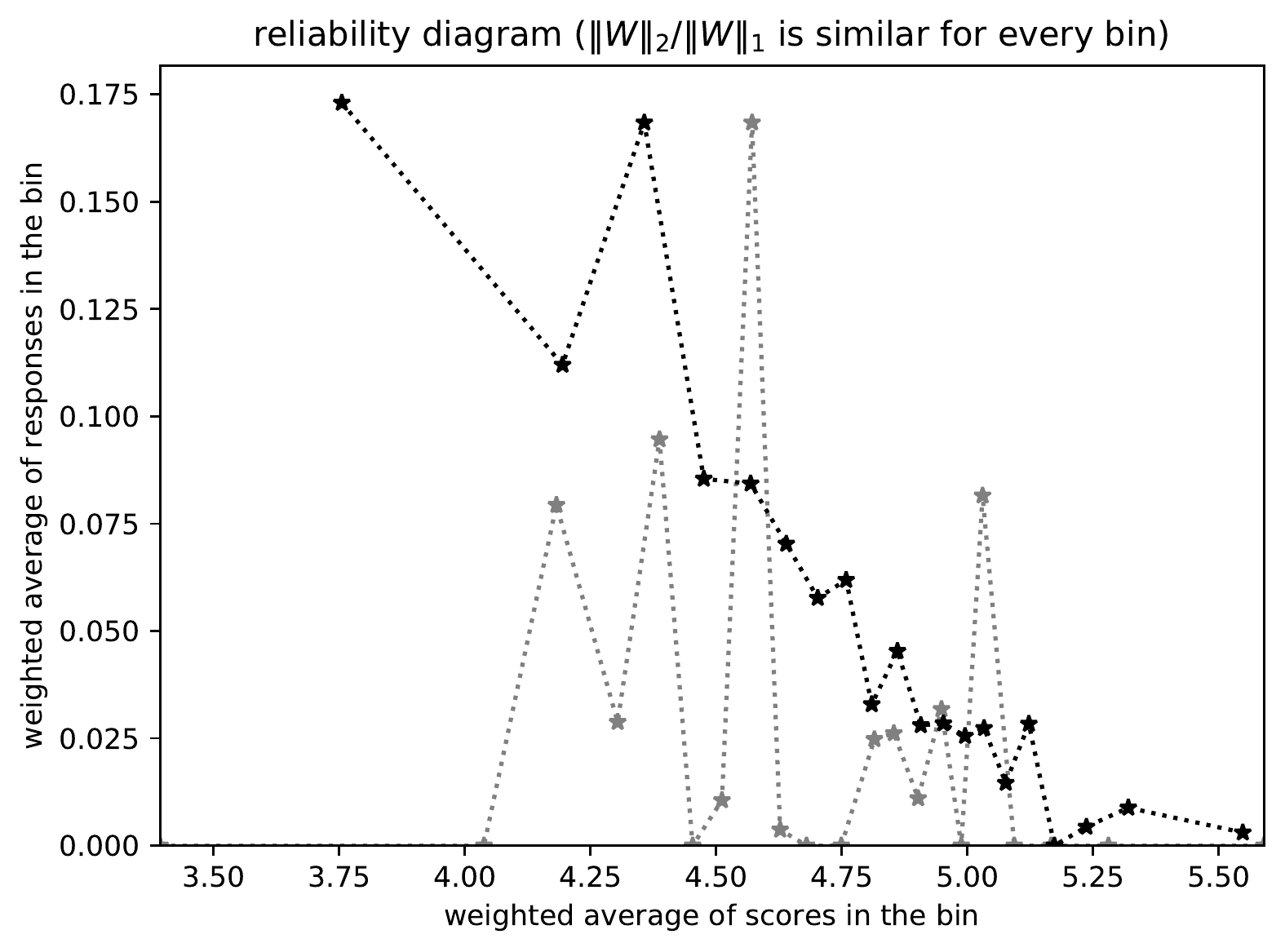}}
\quad\quad
(e)
\parbox{\imsize}{\includegraphics[width=\imsize]
{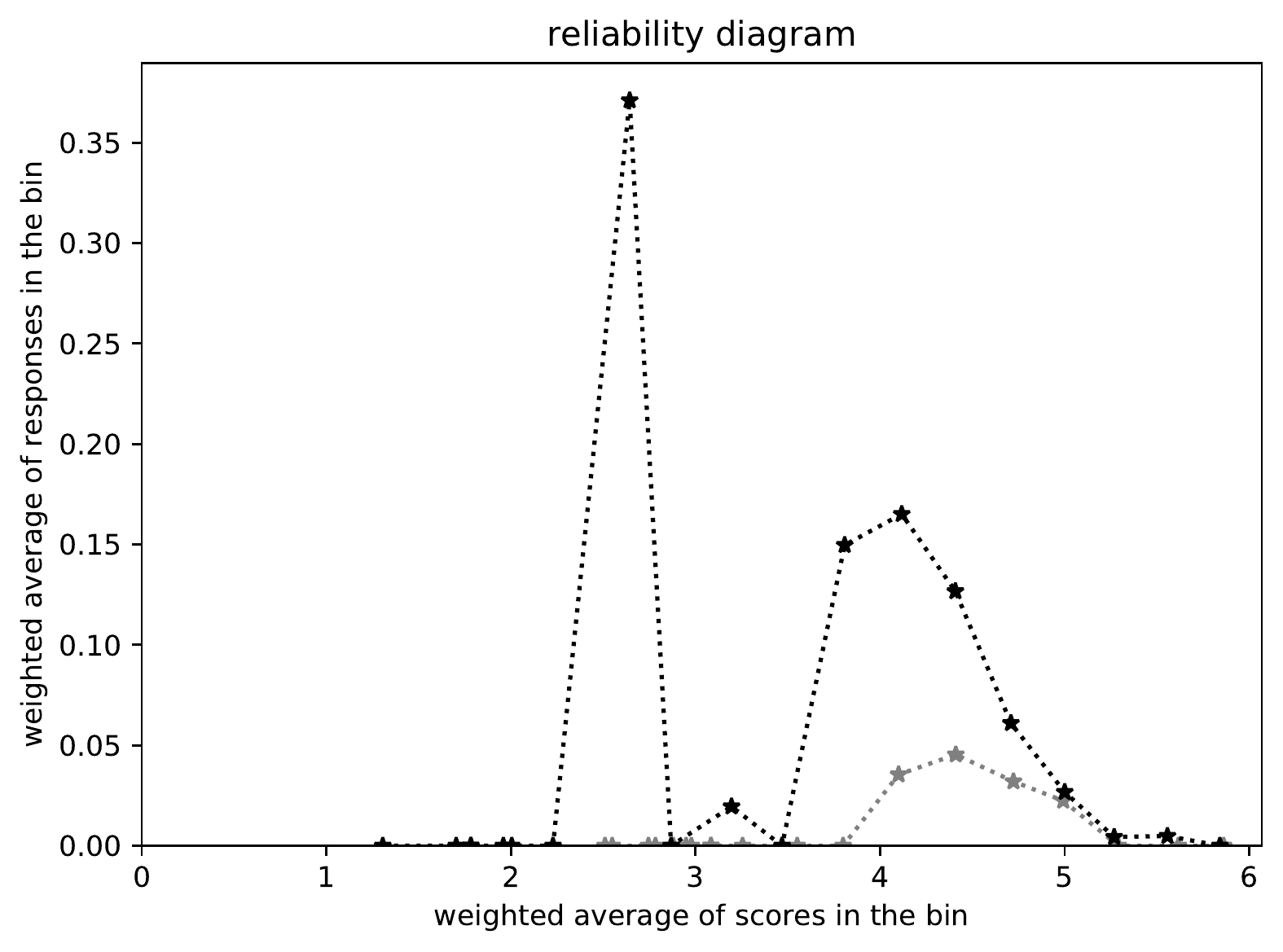}}

\vspace{\vertsep}

(f)
\parbox{\imsize}{\includegraphics[width=\imsize]
{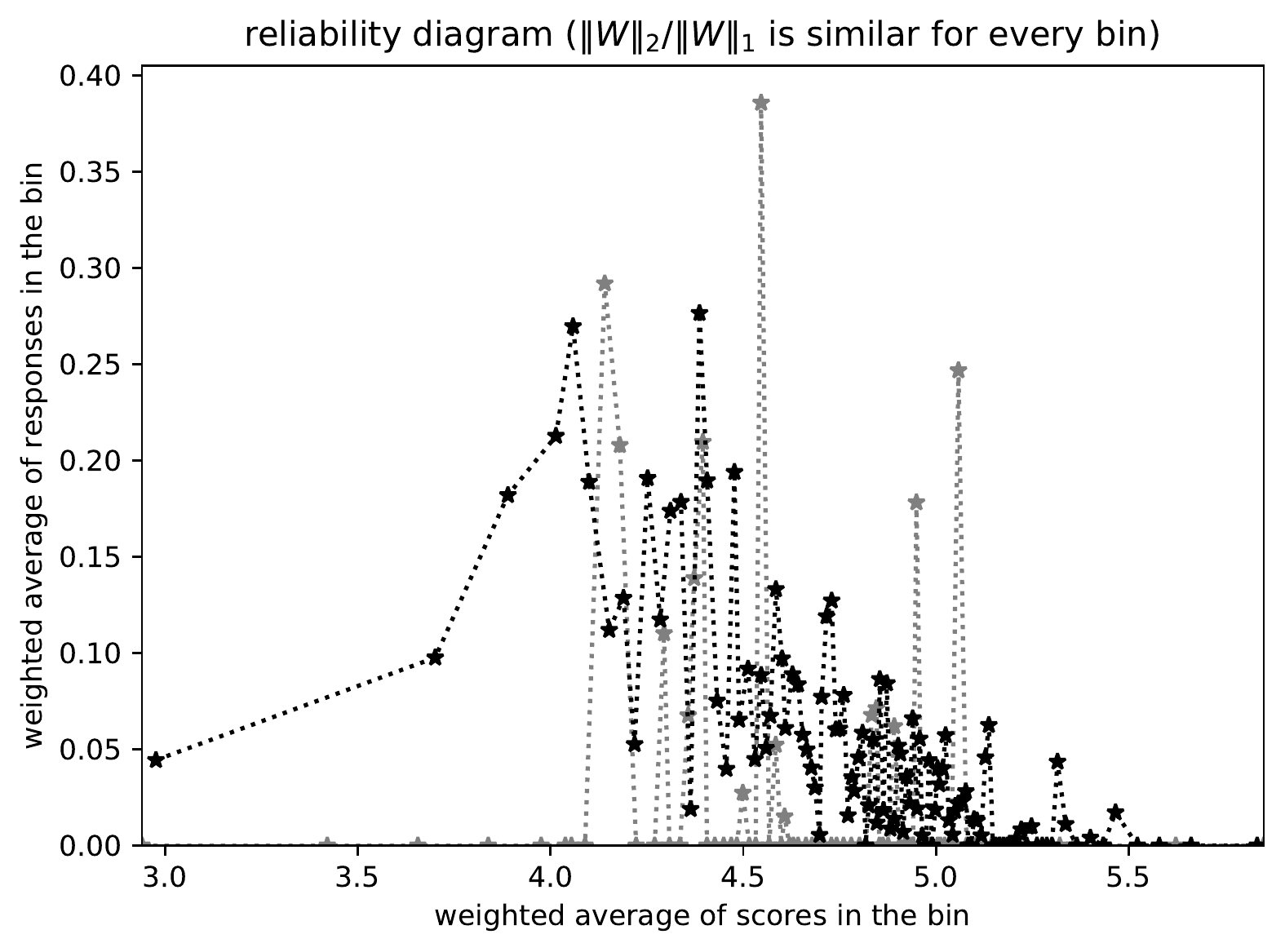}}
\quad\quad
(g)
\parbox{\imsize}{\includegraphics[width=\imsize]
{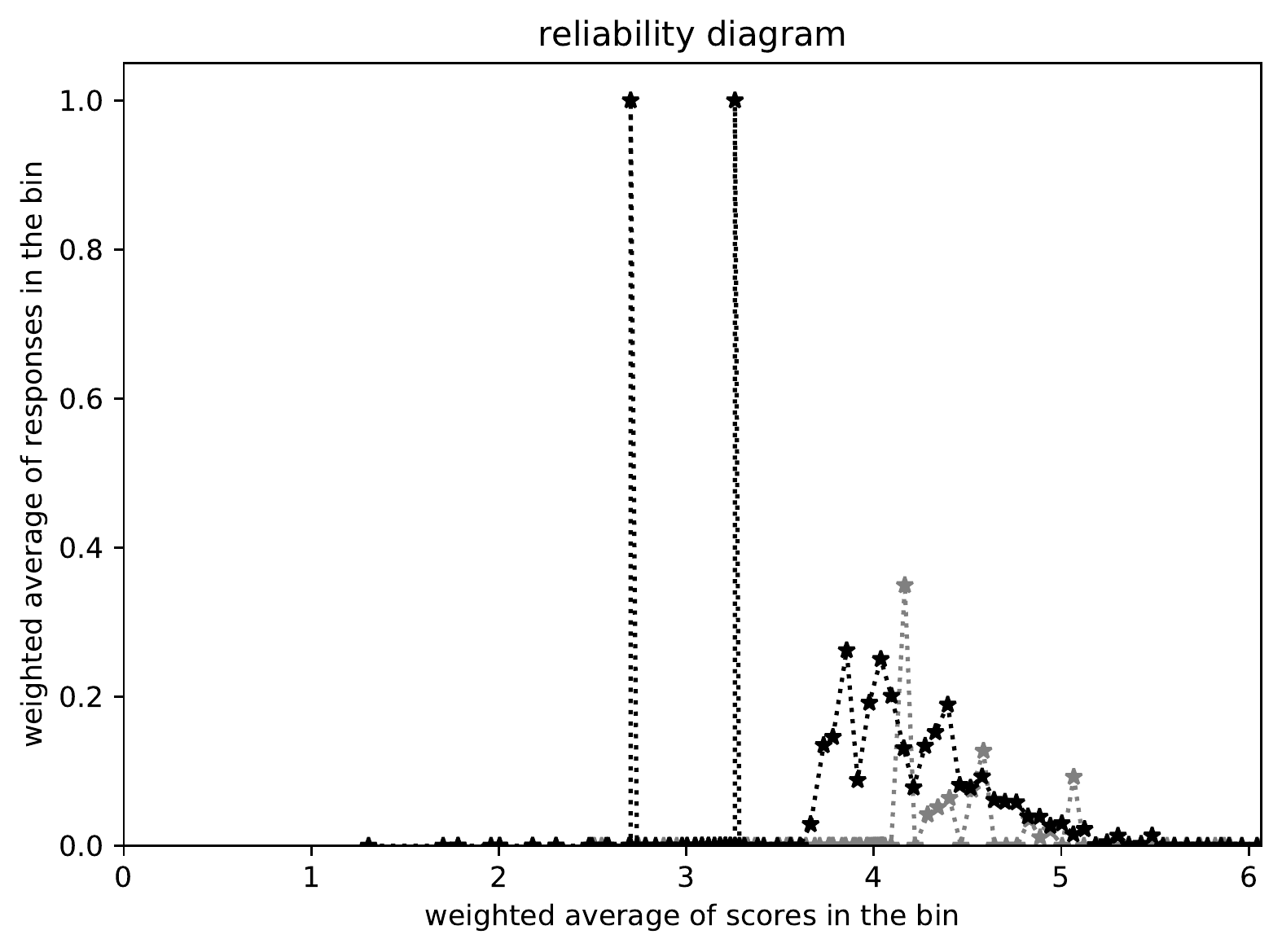}}

\end{centering}
\caption{Riverside County vs.\ Butte County; $n =$ 1,478;
         Kuiper's statistic is $0.04624 / \sigma = 1.650$,
         Kolmogorov's and Smirnov's is $0.04624 / \sigma = 1.650$;
         resolving both the phenomena corresponding to the fairly flat part
         and the phenomena corresponding to the very steep part
         of the cumulative graph (a) for the lowest scores requires
         at least 100 bins in the reliability diagrams (f and g),
         but then the rest of the diagrams is very noisy.
         The statistics of Kuiper and of Kolmogorov and Smirnov do not report
         much statistically significant deviation between the subpopulations.
}
\label{Riverside-Butte}
\end{figure}

This subsection applies the methods of Subsection~\ref{weighted}
to the latest (year 2019) microdata from the American Community Survey
of the United States Census Bureau;\footnote{All microdata
from the United States Census Bureau's American Community Survey of 2019
is available for download at
\url{https://www.census.gov/programs-surveys/acs/microdata.html}}
specifically, we consider each subpopulation to be the observations
from a county in California. The sampling in this survey is weighted,
and we retain only those members whose weights (``WGTP'' in the microdata)
are nonzero, omitting any member whose household personal income
(``HINCP'') is zero or for which the adjustment factor to income (``ADJINC'')
is missing. The scores are the logarithm to base 10
of the adjusted household personal income
(the adjusted income is ``HINCP'' times ``ADJINC,'' divided by one million
when ``ADJINC'' omits its decimal point in the integer-valued microdata),
and we randomly perturb the scores by about one part in $10^8$ to guarantee
their uniqueness.
The response (also known as ``result'' or ``outcome'') for a given score
takes the value 1 when the corresponding household has limited English speaking
(limited English speaking refers to a household in which every member
strictly older than 13 has some difficulty speaking English);
the response takes the value 0
when the corresponding household is fully English speaking.
Table~\ref{sizes} lists the numbers of scores in the subpopulations
prior to any binning.
Figures~\ref{Alameda-Placer}--\ref{Riverside-Butte} present several examples;
the captions first list the names of the counties corresponding
to the subpopulations considered and then compare the reliability diagrams
with the cumulative graph.

\begin{table}
\caption{Numbers of observations in the original data sets}
\label{sizes}
\begin{center}
\begin{tabular}{rrr}
\hline
number of & number of scores for & number of scores for \\
the figure & the first subpop. & the second subpop. \\\hline
\ref{ex0}, \ref{ex1}, \ref{ex2}, \ref{ex3} & 10,000 & 7,000 \\
\ref{Eskimo-dog-husky_cheetah}, \ref{night-snake_monarch-butterfly},
\ref{monarch-butterfly_wild-boar} & 1,300 & 1,300 \\
\ref{Alameda-Placer} & 6,415 & 1,616 \\
\ref{San_Francisco-Kern} & 3,440 & 2,276 \\
\ref{San_Francisco-Contra_Costa} & 3,440 & 3,697 \\
\ref{San_Francisco-San_Joaquin} & 3,440 & 2,282 \\
\ref{San_Francisco-San_Mateo} & 3,440 & 2,888 \\
\ref{Riverside-Butte} & 7,826 & 843 \\
\hline
\end{tabular}
\end{center}
\vspace{-1.5em}
\end{table}

\subsection{Cautions}
\label{caution}

This subsection warns about some limitations of both the methods
of the present paper and the conventional reliability diagrams.

The fourth example from Subsection~\ref{synthetic},
with its corresponding Figure~\ref{ex3}, emphasizes a cautionary note:
avoid hallucinating deviations between the subpopulations
on account of statistically insignificant random fluctuations!
The indicators such as $\sigma$ and the triangle at the origin
discussed in Subsections~\ref{scalarstats},
\ref{significance}, and~\ref{weighted} are critical
for the proper interpretation of statistical significance.
(Note that similar questions of significance also arise
for the conventional reliability diagrams, on account of multiple testing:
error bars for each bin could report 95\% confidence intervals, for instance,
but then 1 out of every 20 such bins would be expected to report results
exceeding its error bar.)

A chief drawback of the approach of the present paper
is the limitation highlighted in the abstract, in the introduction,
and in an italicized sentence of Section~\ref{methods}, too:
the score for every observation in either subpopulation
must not be exactly equal to the score for any other observation
from the subpopulations. Of course, one way to enforce the required uniqueness
of scores is to perturb them at random slightly.
Another drawback is that the observations
from one subpopulation get compared to observations
from the other subpopulation at slightly different scores;
although the bias that this introduces in the cumulative approach
is less than in the classical reliability diagrams, the bias is still there
and potentially worrisome.
An ideal means of circumventing such drawbacks is to compare
a subpopulation to the full population as detailed by~\cite{tygert}.
The approach of~\cite{tygert} is effectively ideal
and should be the method of choice whenever applicable.
The approach of the present paper is only relevant
when comparing subpopulations directly is necessary.

\section{Conclusion}
\label{conclusion}

The plot of cumulative differences between the two subpopulations
is easy to interpret --- the slope of a secant line for the graph over
a long range becomes the average difference between the two subpopulations,
and slope is easy to gauge irrespective of any constant offset
of the secant line. The plots for the examples of Section~\ref{results}
clearly demonstrate many advantages of the cumulative approach
over the classical reliability diagrams, and the scalar summary statistics
of Kuiper and of Kolmogorov and Smirnov usually faithfully reflect
significant differences between the subpopulations if any occur
across the full range of scores in the plots.
The graphs of cumulative differences avoid explicitly making
a trade-off between statistical confidence and resolution as a function
of score --- a trade-off that is inherent to the traditional binned diagrams.

\bibliography{disjoint}
\bibliographystyle{siam}

\end{document}